\documentclass[a4paper,fleqn,usenatbib]{mnras}

\usepackage{newtxtext,newtxmath}

\usepackage[T1]{fontenc}
\usepackage{ae,aecompl}

\usepackage{graphicx}	
\usepackage{amsmath}	
\usepackage{amssymb}	
\usepackage{gensymb}
\usepackage{color,soul}
\usepackage{xcolor}

\title[From small-scale AGN winds to galactic super-winds]{Powering galactic super-winds with small-scale AGN winds}

\author[Costa, Pakmor and Springel]{
Tiago Costa\thanks{e-mail: tcosta@mpa-garching.mpg.de}, R\"udiger Pakmor and Volker Springel
\\
Max-Planck-Institut f\"ur Astrophysik, Karl-Schwarzschild-Stra{\ss}e 1, D-85748 Garching b. M\"unchen, Germany
}

\date{Submitted June 2020}

\pubyear{2020}

\begin{document}
\label{firstpage}
\pagerange{\pageref{firstpage}--\pageref{lastpage}}
\maketitle
\begin{abstract}
We present a new implementation for active galactic nucleus (AGN) feedback through small-scale, ultra-fast winds in the moving-mesh hydrodynamic code {\sc AREPO}.
The wind is injected by prescribing mass, momentum and energy fluxes across a spherical boundary centred on a supermassive black hole according to available constraints for accretion disc winds. 
After sweeping-up a mass equal to their own, small-scale winds thermalise, powering energy-driven outflows with dynamics, structure and cooling properties in excellent agreement with those of analytic wind solutions.
Momentum-driven solutions do not easily occur, because the Compton cooling radius is usually much smaller than the free-expansion radius of the small-scale winds.
Through various convergence tests, we demonstrate that our implementation yields wind solutions which are well converged down to the typical resolution achieved in cosmological simulations.
We test our model in hydrodynamic simulations of isolated Milky Way \-- mass galaxies. 
Above a critical AGN luminosity, initially spherical, small-scale winds power bipolar, energy-driven super-winds that break out of the galactic nucleus, flowing at speeds $> 1000 \rm \, km \, s^{-1}$ out to $\sim 10 \, \rm kpc$.
These energy-driven outflows result in moderate, but long-term, reduction in star formation, which becomes more pronounced for higher AGN luminosities and faster small-scale winds. 
Suppression of star formation proceeds through a rapid mode that involves the removal of the highest-density, nuclear gas and through a slower mode that effectively halts halo gas accretion.
Our new implementation makes it possible to model AGN-driven winds in a physically meaningful and validated way in simulations of galaxy evolution, the interstellar medium and black hole accretion flows.
\end{abstract}


\section{Introduction}
\label{sec:Introduction}

Over the lifetime of a typical supermassive black hole, accretion releases a net energy hundreds of times greater than the binding energy of its host galaxy \citep[e.g.][]{Fabian:12, King:15}.
Since every massive galaxy is thought to harbour a supermassive black hole in its nucleus \citep{Gebhardt:00, Haring:04, McConnell:13, Kormendy:13}, the transfer of energy and momentum from active galactic nuclei (AGN) to the gaseous medium of their host galaxies (`AGN feedback') may profoundly influence the evolution of galaxy populations. 

The significance of AGN in galaxy evolution, however, depends on whether available energy and momentum are transferred to interstellar- and circumgalactic gas efficiently.
This efficiency is shaped by a myriad physical processes which operate over an extreme range of temporal and spatial scales.
Energy and momentum deposited in the form of radiation, jets or winds at accretion disc\-- ($\lesssim 10^{-2} \, \rm pc$) or dusty torus ($\lesssim 100 \, \rm pc$) scales first travel to galactic scales ($\sim \rm kpc$), where they can directly couple to star-forming gas.
The dominant channels through which star-forming clouds are disrupted, if at all, remain unconstrained. 
Questions such as the relative roles of ejection from the host galaxy versus in-situ cloud dissociation remain unanswered.
At larger scales ($\sim 100 \, \rm kpc$), AGN feedback acts on the gas reservoir that might, at some future time, accrete onto the AGN host galaxy and reignite star formation.
In this regime, the nature of the processes governing energy transfer across the gaseous halo, whether AGN feedback operates only on hot, tenuous gas or whether it also interrupts the more resilient, filamentary, cold gas streams \citep[e.g.][]{Dubois:13,Costa:14} all remain open questions. 

There is no shortage of candidate mechanisms for AGN feedback. The most conspicuous is the interaction between collimated jets and the gaseous haloes of galaxy clusters, where they inflate giant bubbles of hot, relativistic plasma on either side of the nucleus of the brightest cluster galaxy \citep[e.g.][]{McNamara:00, Churazov:01, Forman:07, Fabian:11}. These bubbles rise buoyantly in the cluster atmospheres, transferring energy into the intra-cluster medium via `PdV' work, mixing and a multitude of other processes such as turbulence, thermal conduction, shock- and sound waves \citep[e.g.][]{Zhuravleva:14, Yang:16, Soker:16, Prasad:17, Weinberger:17, Zhang:18, Bourne:19}.
In Seyferts, radio galaxies and quasars, compact jets are sometimes found to intercept a portion of the host's interstellar medium. Despite their collimation, the jets inflate hot bubbles that are capable of accelerating surrounding gas clouds \citep[e.g][]{Nesvadba:10, Morganti:13, Tadhunter:14, Jarvis:19}.  
Massive outflows comprising molecular, (atomic) neutral and ionised phases are often detected via spectral signatures in quasars and Seyfert galaxies. 
With velocities, which often exceed $1000 \, \rm km \, s^{-1}$ \citep[e.g.][]{Sturm:11, Zakamska:16}, sizes and inferred mass-, momentum- and kinetic energy outflow rates which appear to scale with the AGN luminosity $L_{\rm AGN}$, such outflows potentially extract gas from the host galaxy at a rate fast enough to put an end to star formation \citep[e.g.][]{Veilleux:13, Cicone:14, Fiore:17, Perna:17, Fluetsch:19, Wylezalek:20}. 

If indeed driven by AGN, such outflows must be launched via long-range forces such as radiation pressure \citep{Fabian:99, Murray:05} or through interaction with a smaller-scale wind emanating from the nucleus  \citep{Schiano:85, Silk:98, King:03}. 
The observed dearth in systems with simultaneously high AGN luminosity and high hydrogen column density, on the one hand, lends support to the scenario in which radiation pressure on dust clears out galactic nuclei \citep{Raimundo:10, Ricci:17, Lansbury:19}.
On the other hand, with speeds up to $\approx 0.3c$ (where $c$ is the speed of light in vacuum), the highly ionized small-scale winds known as `ultra-fast outflows' seen in $\gtrsim 40 \%$ of AGN \citep{Pounds:03, Tombesi:11} appear to pump energy into their surroundings at rates of $\gtrsim 0.01 L_{\rm AGN}$ \citep{Tombesi:12}. In a small number of systems, simultaneous detections of ultra-fast- and galaxy-wide outflows \citep{Tombesi:15, Veilleux:17, Serafinelli:19, Sirressi:19} make a compelling case that small-scale winds driven at $\ll 1 \, \rm pc$ scales do power galactic super-winds.

Virtually all galaxy evolution models based on concordance $\Lambda$CDM cosmology appeal to AGN feedback to reproduce the properties of massive galaxies as observed in the local Universe. 
In state-of-the-art models, AGN feedback accounts for (i) suppressed star formation in galaxies with stellar masses $\gtrsim 10^{11} \, \rm M_\odot$ \citep[e.g.][]{Springel:05, Bower:06, Teyssier:11, Schaye:15, Weinberger:18}, (ii) the observed galactic colour bimodality \citep{Sijacki:07, Dubois:13}, (iii) the kinematic structure and size evolution of massive galaxies \citep{Dubois:16, Peirani:17, Choi:18, vanderVlugt:19}, (iv) chemical abundance patterns such as the $\alpha$-enhancement of stellar populations \citep{Taylor:15, Segers:16}, (v) the thermodynamic and ionisation states of the circumgalactic medium \citep{vandeVoort:11, Gaspari:12, Oppenheimer:18} and (vi) the self-regulation of black hole growth \citep{DiMatteo:05, Sijacki:15, Volonteri:16}.

With a typical mass resolution of $\gtrsim 10^5 \, \rm M_\odot$ and spatial resolution of $\gtrsim 100 \, \rm pc$, black hole accretion and wind launching clearly cannot be resolved \emph{ab initio} in even high resolution `zoom-in', hydrodynamic simulations of massive galaxies.
The successes attained by all such simulations rely on phenomenological models for AGN feedback, which proceed in a simplified fashion through injection of thermal- \citep{DiMatteo:05, Springel:05,Booth:09} or kinetic energy \citep{Barai:16, Weinberger:17} around accreting black holes. The conversion efficiency $\eta$ between AGN luminosity and injected energy is often treated as a free parameter which is tuned to the value required to ensure one or multiple observables are reproduced quantitatively by the simulation. Based on galaxy merger simulations and thermal energy injection to model quasar feedback, \citet{DiMatteo:05}, for instance, find that if $\eta \, = \, 0.05$, their simulations recover the normalisation of the observed $M_{\rm BH} \-- M_{\rm \star}$ relation. In many other simulations, the feedback efficiency of quasar feedback is calibrated in the same way \citep[e.g.][]{Booth:09, Teyssier:11, Dubois:12b, Schaye:15}.
In other studies, the efficiency of AGN feedback is motivated by the properties of observed large-scale outflows \citep{Dave:19}, while others attempt to directly inject small-scale, fast winds with properties in line with broad absorption line winds in their cosmological simulations \citep[e.g.][]{Choi:12, Angles-Alcazar:17} without explicitly tuning their feedback parameters to observables.

A persisting question is the fidelity with which any of the available AGN feedback models capture the impact of physical processes such as jets, accretion disc winds and radiation pressure reliably at the $\gtrsim 100 \, \rm pc$ scales which are typically resolved. 
The extent to which the differing successes of distinct AGN feedback models should be attributed to details of the numerical implementation or to genuinely accurate modelling of the relevant physical effects is unclear.
For instance, \citet{Wurster:13} present a comparison of five popular models for black hole accretion and AGN feedback. Numerical variations in the accretion and feedback treatment introduce significant differences in the structure, temperature and density of gas in galaxy haloes and galactic nuclei, and result in order of magnitude variation in the self-regulated black hole mass.
Moreover, due to calibration, many existing simulations cannot provide insight into the origin of various observables, such as the normalisation of the scaling relations between black hole mass and host galaxy properties. Calibration also means that many available simulations cannot shed light on the physical origin of the feedback efficiencies. Existing models indeed do not exploit the wealth of information provided by analytic calculations of AGN wind solutions \citep[e.g.][]{King:03, Zubovas:12, Faucher-Giguere:12} or magneto-hydrodynamic and radiation-hydrodynamic simulations of accretion-disc winds \citep[e.g.][]{Yuan:15, Sadowski:16, Nomura:17}.  

\citet{Costa:14} showed that energy-driven bubbles can provide significant feedback in massive galaxies and established that forces $> L_{\rm AGN}/c$ are required in order for AGN to regulate star formation in massive galaxies.
This result has been confirmed more recently in radiation-hydrodynamic simulations probing the radiation pressure scenario of AGN feedback \citep[e.g.][]{Costa:18b}.
While energy-driven bubbles were understood in \citet{Costa:14} to form due to a collision between a small-scale wind and the interstellar medium, the small-scale wind was not explicitly modelled.
With increasing resolution, the growing ability to resolve smaller scales in simulations of galaxy formation, it becomes important to ensure that AGN feedback models correctly bridge the scales between large-scale outflows and the smaller-scale winds that likely power them.

The aim of this paper is thus to construct a robust, predictive and physically validated model for the generation of large-scale galactic outflows starting from small-scale AGN-driven winds.
We propose a new method to inject a small-scale winds with properties in line with those of observed ultra-fast outflows, simulations of accretion disc winds, and as typically envisaged in analytic models of large-scale outflows.
The basic model and our numerical implementation are described in detail in Section~\ref{sec:windmodel}. In Section~\ref{sec:testmodel}, we test the predictions of our model against analytic expectations and show that these can be recovered with high precision. We also quantify how the behaviour of our model is affected by degrading the resolution to the levels typically achieved in galaxy evolution simulations. In Section~\ref{sec:discs}, we test our implementation in simplified hydrodynamic simulations of galactic discs, illustrating how a more realistic gas environment leads to the emergence of complex large-scale outflows. 
We discuss the implications of our findings for the quenching of massive galaxies, how our model differs from conventional AGN feedback recipes and its various limitations in Section~\ref{sec:discussion}. We present our conclusions in Section~\ref{sec:conclusions}.

\section{AGN wind model}
\label{sec:windmodel}

In this Section, we outline our theoretical model for the nuclear AGN wind. We present its numerical implementation and conduct multiple tests. 

\subsection{Analytical background}
\label{sec_anal}
We assume that accretion onto a black hole results in the production of a quasi-spherical wind emanating from accretion-disc or dusty torus scales.
We do not consider the processes dictating how this wind is initially launched, as the associated spatial scales lie far below the resolution limits of typical galaxy formation simulations.
Instead, we posit that such a wind exists at some well-resolved radius $R$ where it has an integrated mass outflow rate $\dot{M}_{\rm w}$, speed $v_{\rm w}$ and pressure $\mathcal{P}_{\rm w}$.
In Section~\ref{sec_convergence}, we identify the spatial scales where this assumption is valid.

The mass, momentum and energy flux densities at radius $R$ are, respectively,
\begin{eqnarray}
\label{eqs_conservation_m}
\dot{m}_{\rm w}  \,  & = &  \, \rho_{\rm w} v_{\rm w} \,  =  \, \frac{\dot{M}_{\rm w}}{4 \pi b R^2 } \, ,\\
\label{eqs_conservation_p}
\dot{p}_{\rm w} \,  & = &  \, \rho_{\rm w} v_{\rm w}^2 + \mathcal{P}_{\rm w} \, ,\\
\label{eqs_conservation_e}
\dot{e}_{\rm w} \,  & = &  \, \frac{1}{2} \rho_{\rm w} v_{\rm w}^3 + \left( \frac{\gamma}{\gamma - 1} \right) \mathcal{P}_{\rm w} v_{\rm w}  \, ,
\end{eqnarray}
where $\gamma$ is the adiabatic index of the wind gas and $b \, = (4\pi)^{-1} \Omega \leq 1$ is the fractional solid angle.

The wind pressure $\mathcal{P}_{\rm w}$ evaluated at the injection radius $R$ can be determined from the adiabatic sound speed $c_{\rm w}$ at injection through the relation $\mathcal{P}_{\rm w} \, = \, \gamma^{-1} \rho_{\rm w} c_{\rm w}^2$.
For $v_{\rm w} \gg c_{\rm w}$, the momentum flux density is dominated by ram pressure ($\rho_{\rm w} v_{\rm w}^2$) and the energy flux density is dominated by the kinetic luminosity term ($\frac{1}{2} \rho_{\rm w} v_{\rm w}^3$). 
In this limit, which we shall adopt throughout this paper, the terms involving $\mathcal{P}_{\rm w}$ in Eqs.~\ref{eqs_conservation_p} and~\ref{eqs_conservation_e} are subdominant.

Integrating the ram pressure and the kinetic energy flux density, the first terms of Eqs.~\ref{eqs_conservation_p} and \ref{eqs_conservation_e} respectively, over the surface area at $R$ gives the mechanical momentum flux $\dot{P}_{\rm w}$ and the kinetic luminosity $\dot{E}_{\rm w}$ of the wind, respectively as
\begin{eqnarray}
\dot{P}_{\rm w} \, & = &\, 4 \pi b R^2 \rho_{\rm w} v_{\rm w}^2 \, = \, \dot{M}_{\rm w} v_{\rm w} \, , \\
\dot{E}_{\rm w} \, & = & \, 2 \pi b R^2 \rho_{\rm w} v_{\rm w}^3 \, = \, \frac{1}{2} \dot{M}_{\rm w} v_{\rm w}^2 \, .
\end{eqnarray}

We parametrise $\dot{P}_{\rm w}$ in terms of the momentum input rate of the AGN radiation field $L_{\rm AGN} / c$, where $L_{\rm AGN}$ is the AGN luminosity, by setting $\dot{P}_{\rm w} \, = \, \tau (L_{\rm AGN}/c)$ and the wind's velocity in terms of the speed of light $c$, by setting $v_{\rm w} \, = \, \beta c$.
This parametrisation fixes the integrated mass, momentum and kinetic energy fluxes to
\begin{eqnarray}
\label{eqs_conservation_m2}
\dot{M}_{\rm w} \, & = & \, \frac{\tau}{\beta} \frac{L_{\rm AGN}}{c^2}  \, ,\\
\label{eqs_conservation_p2}
\dot{P}_{\rm w}  \, & = & \,  \tau \frac{L_{\rm AGN}}{c} \, ,\\
\label{eqs_conservation_e2}
\dot{E}_{\rm w}  \, & = & \, \frac{\tau \beta}{2} L_{\rm AGN} \, ,
\label{eqs_conservation_set2}
\end{eqnarray}
such that the flux densities in Eqs.~\ref{eqs_conservation_m} \-- ~\ref{eqs_conservation_e} assume the final form
\begin{eqnarray}
\label{eqs_conservation_m3}
\dot{m}_{\rm w}  \,  & = &  \, \frac{1}{4 \pi c^2}  \frac{\tau}{b \beta} \frac{L_{\rm AGN}}{R^2} \, ,\\
\label{eqs_conservation_p3}
\dot{p}_{\rm w} \,  & = &  \, \frac{1}{4 \pi c}  \frac{\tau}{b} \frac{L_{\rm AGN}}{R^2} + \mathcal{P}_{\rm w} \, ,\\
\label{eqs_conservation_e3}
\dot{e}_{\rm w} \,  & = &  \,   \frac{1}{8 \pi}  \frac{\tau \beta}{b} \frac{L_{\rm AGN}}{R^2} + c \left( \frac{\gamma}{\gamma - 1} \right) \beta \mathcal{P}_{\rm w}   \, .
\end{eqnarray}

The free parameters $\tau$, $\beta$ and $b$ can be constrained from observations or from fundamental, general-relativistic magneto/radiation-hydrodynamic simulations of black hole accretion discs \citep[e.g.][]{Yuan:15}.
Plausible values for these free parameters are discussed in Section~\ref{sec_choiceparam}.

Combining Eqs.~\ref{eqs_conservation_m} and~\ref{eqs_conservation_m2} also allows us to derive the number density of the AGN wind, which is 
\begin{eqnarray}
                   \label{eq_windnh}
n_{\rm w} \, & = & \, \left( \frac{1}{4 \pi c^3 \mu m_{\rm p}} \right) \left( \frac{\tau}{\beta^2 b} \right) \left( \frac{L_{\rm AGN}}{R^2} \right) \\
                   & \approx & \, 31 \left( \frac{\tau}{b} \right) \left( \frac{\beta}{0.1} \right)^{-2}  \left( \frac{L_{\rm AGN}}{10^{45} \, \mathrm{erg \, s^{-1}}} \right)  \left( \frac{R}{\mathrm{pc}} \right)^{-2} \, \rm cm^{-3} \, ,
                   \nonumber
                   \label{eq_wind_density}
\end{eqnarray}
where $m_{\rm p}$ is the proton mass and $\mu$ is the mean particle mass, which is here assumed to correspond to the value associated with a fully ionised H and He plasma of primordial composition, i.e. $\mu \, \approx \, 0.6$.
Note the appearance of $b$ in the denominator of Eq.~\ref{eq_windnh}; in the presence of collimation, a higher wind density is required in order to keep the wind mass outflow rate $\dot{M}_{\rm w}$ constant. Note also how, at galactic halo scales $R \sim 10 \, \rm kpc$, $n_{\rm w} \lesssim 10^{-5} \, \rm cm^{-3}$ for realistic AGN luminosities. 

Eq.~\ref{eqs_conservation_set2} allows us to compute the energy efficiency of the AGN wind, which is simply
\begin{equation}
\eta \, = \, \dot{E}_{\rm w} / L_{\rm AGN}  \, = \, \frac{\tau \beta}{2} \, = \, 0.05 \tau \left( \frac{\beta}{0.1} \right).
\label{eq_efficiency}
\end{equation}

For a bright quasar, the rate at which wind kinetic energy flows with the fast wind thus corresponds a few percent of the AGN bolometric luminosity \citep[see also][]{King:03}.
If the black hole accretion rate is $\dot{M}_{\rm BH}$ and $\epsilon \dot{M}_{\rm BH} c^2$ is the total rate at which energy is generated by accretion, then, by energy conservation, $\dot{E}_{\rm w} \leq \epsilon \dot{M}_{\rm BH} c^2$. If we define the radiative efficiency as $\epsilon_{\rm r} \, = \, L_{\rm AGN} / (\dot{M}_{\rm BH} c^2)$, we can also write $\dot{E}_{\rm w} \leq \epsilon \epsilon_{\rm r}^{-1} L_{\rm AGN}$. If accretion is radiatively efficient, i.e. $\epsilon \sim \epsilon_{\rm r}$, then $\eta \leq 1$.
For radiatively inefficient accretion, the wind may be launched by e.g. hydromagnetic forces \citep[e.g.][]{Yuan:15}, such that $\epsilon_{\rm r} \ll \epsilon$. In this case, it is possible that $\eta > 1$.

\subsubsection{Outflow structure and dynamics}
\label{sec:outflowstructure}
After the wind is ejected, it moves unaccelerated into the ambient medium, causing a shock front to form ahead of the wind ejecta.
This phase, referred to as `free-expansion' in analogy with the similar supernova remnant phase, occurs on time-scales shorter than the time it takes the wind to sweep-up a mass equal to its own. If the ambient medium is homogeneous with a fixed density $\rho_{\rm 0}$, the free-expansion timescale is given by
\begin{equation}
t_{\rm free} \, = \, \left( \frac{3}{4 \pi b} \right)^{1/2} \left( \frac{\dot{M}_{\rm w}}{\rho_{\rm 0} v_{\rm w}^{3}}\right)^{1/2} \, .
\label{eq_tfree}
\end{equation}
During free-expansion, the shocked, swept-up material accumulates in a shell with an inner radius that grows as $R_{\rm sh} \propto t$ and with a velocity $\dot{R}_{\rm sh} \, \approx \, v_{\rm w}$. 
The radial distance traversed by the wind during this time is 
\begin{eqnarray}
\label{eq_rfree}
R_{\rm free} \, & \sim & \, t_{\rm free} v_{\rm w} \\
 \, & \approx & \, 10  \left( \frac{\beta}{0.1} \right) ^{-1} \left( \frac{\tau}{b} \right) ^{1/2} \left( \frac{L_{\rm AGN}}{10^{45} \, \mathrm{erg \, s^{-1}}} \right)^{1/2} \left( \frac{n_{\rm 0}}{ \mathrm{cm^{-3}}} \right)^{-1/2} \, \rm pc \, ,\nonumber
\end{eqnarray}
where Eq.~\ref{eqs_conservation_m2}, the relation $v_{\rm w} \, = \, \beta c$ and $n_{\rm 0} \, = \, \rho_{\rm 0} / (\mu m_{\rm p})$, with $\mu \, \approx \, 0.6$, are used in the second step of the equation.

When $R_{\rm sh} \sim R_{\rm free}$, the momentum of material added onto the shell is sufficient to cause it to slow down significantly.
The hydrodynamic flows that ensue have been well studied \citep{Castor:75, Weaver:77, King:03, King:05, Zubovas:12, Faucher-Giguere:12, Wagner:13, Costa:14}. 
The shell begins to push strongly against the incoming wind, causing the formation of a strong `reverse shock' across which a significant fraction of the wind kinetic energy is thermalised.
If the shock is adiabatic, about $90\%$ of the post-shock wind energy is in thermal form.
At $R \, \geq \, R_{\rm free}$, the dynamics of the outflow is controlled by the ability of the shocked wind fluid to preserve its thermal energy. If radiative losses in the \emph{shocked wind} are negligible, it expands adiabatically, doing `PdV' work on the ambient medium, driving an `energy-driven' outflow \citep{King:05}. 
There may also be outflow solutions in which the shocked wind bubble is radiative. In the limiting case that the reverse shock is isothermal, the shell of swept-up ambient gas is driven solely by the wind's ram pressure. Such wind solutions are termed `momentum-driven' \citep{King:03}.
If the shocked wind bubble does cool, though only inefficiently, the wind solution is intermediate between momentum- and energy-driven (see Section~\ref{sec:shockedwind}).

In the case of energy-driven expansion through a homogeneous medium, we expect $R_{\rm sh} \propto t^{3/5}$, $\dot{R}_{\rm sh} \propto t^{-2/5}$ and thus $\dot{R}_{\rm sh} \propto R_{\rm sh}^{-2/3}$ (see Appendix~\ref{sec:app:energy-driven}). We note that the expansion history of energy-driven shells depends on the shape of the gas density profile \citep[e.g.][]{Zubovas:12b, Faucher-Giguere:12} and on whether gravity, the thermal- and ram pressure of the ambient medium are important. As shown in e.g. \citet{Costa:14} and Section~\ref{sec:discs}, the pressure of the ambient medium and the ram pressure exerted by infalling gas cannot be safely neglected in galaxy evolution settings. However, our primary concern in this and the following Sections is to investigate a deliberately simple setup and use it to validate and test our numerical model.

Due to its role in shaping both the dynamics and thermodynamic structure of AGN-driven outflows \citep[e.g.][]{Zubovas:14, Costa:14, Costa:15, Nims:15, Richings:18, Richings:18b}, we summarise the most important radiative cooling processes in the different outflow phases in the following section.

\subsubsection{Radiative cooling}
\label{section_radiative_cooling}
As the outflowing shell decelerates to $\dot{R}_{\rm sh} \ll v_{\rm w}$, a strong reverse shock begins propagating into the free-expanding wind.
In this regime, the wind travels with a speed of approximately $| \dot{R}_{\rm sh} - v_{\rm w} | \, \approx \, v_{\rm w}$ in the frame of the shock. The post-shock temperature $T_{\rm R-shock}$ is then given by
\begin{equation}
T_{\rm R-shock} \, = \, \frac{3}{16} \frac{\mu m_{\rm p}}{k_{\rm B}} (\dot{R}_{\rm sh} - v_{\rm w})^2 \, \approx \, 1.2 \times 10^{10} \left( \frac{\beta}{0.1} \right)^2 \, \rm K \, ,
\label{eq_rshock_T}
\end{equation}
where we have assumed $\mu \, = \, 0.6$ for a fully ionised H and He plasma of primordial composition.

Similarly, a forward shock propagates into the ambient medium. If the forward shock is strong and the ambient medium is static, the Rankine-Hugoniot jump conditions give a shock propagation speed of $(\gamma + 1)/2 \dot{R}_{\rm sh} \, = \, \frac{4}{3} \dot{R}_{\rm sh}$ with respect to the ambient medium.
The corresponding post-shock temperature $T_{\rm F-shock}$ of the shocked ambient medium is then
\begin{equation}
T_{\rm F-shock} \, = \, \frac{1}{3} \frac{\mu m_{\rm p}}{k_{\rm B}} \dot{R}_{\rm sh}^2 \, \approx \, 2.4 \times 10^{7} \left( \frac{\dot{R}_{\rm sh}}{10^3 \, \mathrm{km \,s ^{-1}}} \right)^2 \, \rm K \, ,
\label{eq_fshock_T}
\end{equation}
from which the relation $T_{\rm R-shock} / T_{\rm F-shock} \sim \left( v_{\rm w} / \dot{R}_{\rm sh} \right)^2$ follows. 

\paragraph*{Shocked wind:}
Given typical temperatures $T_{\rm R-shock} \gtrsim 10^8 \, \rm K$ for $\beta > 0.01$, the main cooling routes for the shocked wind phase are thermal free-free emission and Compton scattering between AGN photons and free electrons. Free-free emission of shocked wind material occurs within a cooling radius 
\begin{equation}
R_{\rm c}^{\rm ff} \, \approx \, 4 \times 10^{-4} \left( \frac{L_{\rm AGN}}{10^{45} \, \mathrm{erg \, s^{-1}}} \right) \left( \frac{\beta}{0.1} \right)^{-3} \left( \frac{\dot{R}_{\rm sh}}{10^3 \, \mathrm{km \, s^{-1}}} \right)^{-1} \left( \frac{\tau}{b} \right) \, \rm pc \, ,
\label{eq_coolradius_ff}
\end{equation}
which is negligible even for a simultaneous choice of slower winds with $\beta \, \sim \, 10^{-2}$ and high quasar luminosities of $\sim 10^{47} \,  \rm erg \, s^{-1}$.
Combining Eqs.~\ref{eq_rfree} and ~\ref{eq_coolradius_ff} gives 
\begin{equation}
\begin{split}
\frac{R_{\rm c}^{\rm ff}}{R_{\rm free}} \, \approx \, 4\times10^{-5} & \left( \frac{L_{\rm AGN}}{10^{45} \, \mathrm{erg \, s^{-1}}} \right)^{1/2} \\  & \times \left( \frac{n_{\rm 0}}{\mathrm{cm^{-3}}} \right)^{1/2} \left( \frac{\beta}{0.1}\right)^{-2} \left( \frac{\dot{R}_{\rm sh}}{10^3 \, \mathrm{km \, s^{-1}}} \right)^{-1} \left( \frac{\tau}{b} \right)^{1/2} \, ,
\end{split}
\end{equation}
which indicates that the AGN wind should not even have thermalised within the free-free cooling radius. 

If electrons and protons reach equipartition behind the reverse shock rapidly (see Section~\ref{sec_plasma}), the most important cooling channel is inverse Compton cooling \citep{King:03}. For $\beta > 0.01$ and a fully ionised plasma of primordial composition, the cooling radius for non-relativistic Compton cooling is well approximated by 
\begin{equation}
R_{\rm c}^{\rm cpt} \, \approx \, 0.3 \left( \frac{L_{\rm AGN}}{10^{45} \, \mathrm{erg \, s^{-1}}} \right) \left( \frac{\dot{R}_{\rm sh}}{10^3 \, \mathrm{km \, s^{-1}}} \right)^{-1} \rm pc \, .
\label{eq_coolradius_cpt}
\end{equation}

The ratio between the Compton cooling radius and the free-expansion radius for a homogeneous medium, however, is
\begin{equation}
\begin{split}
\frac{R_{\rm c}^{\rm cpt}}{R_{\rm free}} \, \approx \, 0.03 & \left( \frac{L_{\rm AGN}}{10^{45} \, \mathrm{erg \, s^{-1}}} \right)^{1/2} \\ & \times \left( \frac{n_{\rm 0}}{\mathrm{cm^{-3}}} \right)^{1/2} \left( \frac{\beta}{0.1} \right) \left( \frac{\dot{R}_{\rm sh}}{10^3 \, \mathrm{km \, s^{-1}}} \right)^{-1}  \, ,
\label{eq_coolratio_R}
\end{split}
\end{equation}
indicating that non-relativistic Compton cooling can only be significant for a combination of fast winds with $\beta \sim 0.1$, high quasar luminosities $L \gtrsim 10^{47} \, \rm erg \, s^{-1}$ and high ambient medium densities \citep[see also Appendix A2 in][]{Faucher-Giguere:12}, but is otherwise unimportant \citep[cf.][]{King:03}. In Section~\ref{sec:shockedwind}, we confirm that these conclusions do not change even if we consider relativistic Compton scattering.
For some of the parameter space, we thus expect AGN-driven outflows to progress from free-expansion to an energy-driven phase without an intermediate momentum-driven phase. 

\paragraph*{Shocked ambient medium:}
After $R > R_{\rm free}$, most of the outflow mass is contained in the shocked ambient medium phase. Pressure balance across the contact discontinuity separating shocked wind and shocked ambient medium implies that $n_{0} \sim n_{\rm w} \left( v_{\rm w} / \dot{R}_{\rm sh} \right)^2 \gg n_{\rm w}$. The combination of lower post-shock temperatures (Eq.~\ref{eq_fshock_T}) and higher densities results in far shorter cooling times for the shocked ambient medium phase.
The cooling radius for free-free emission, for example, is now
\begin{equation}
R_{\rm c}^{\rm ff} \, \approx \, 3.8 \left( \frac{\dot{R}_{\rm sh}}{10^3 \, \rm km \, s^{-1}} \right)^2 \left( \frac{n_{\rm 0}}{\rm cm^{-3}} \right)^{-1} \, \rm kpc \, .
\label{eq_coolradius_ff_F}
\end{equation}

\begin{figure}
\includegraphics[width=0.485\textwidth]{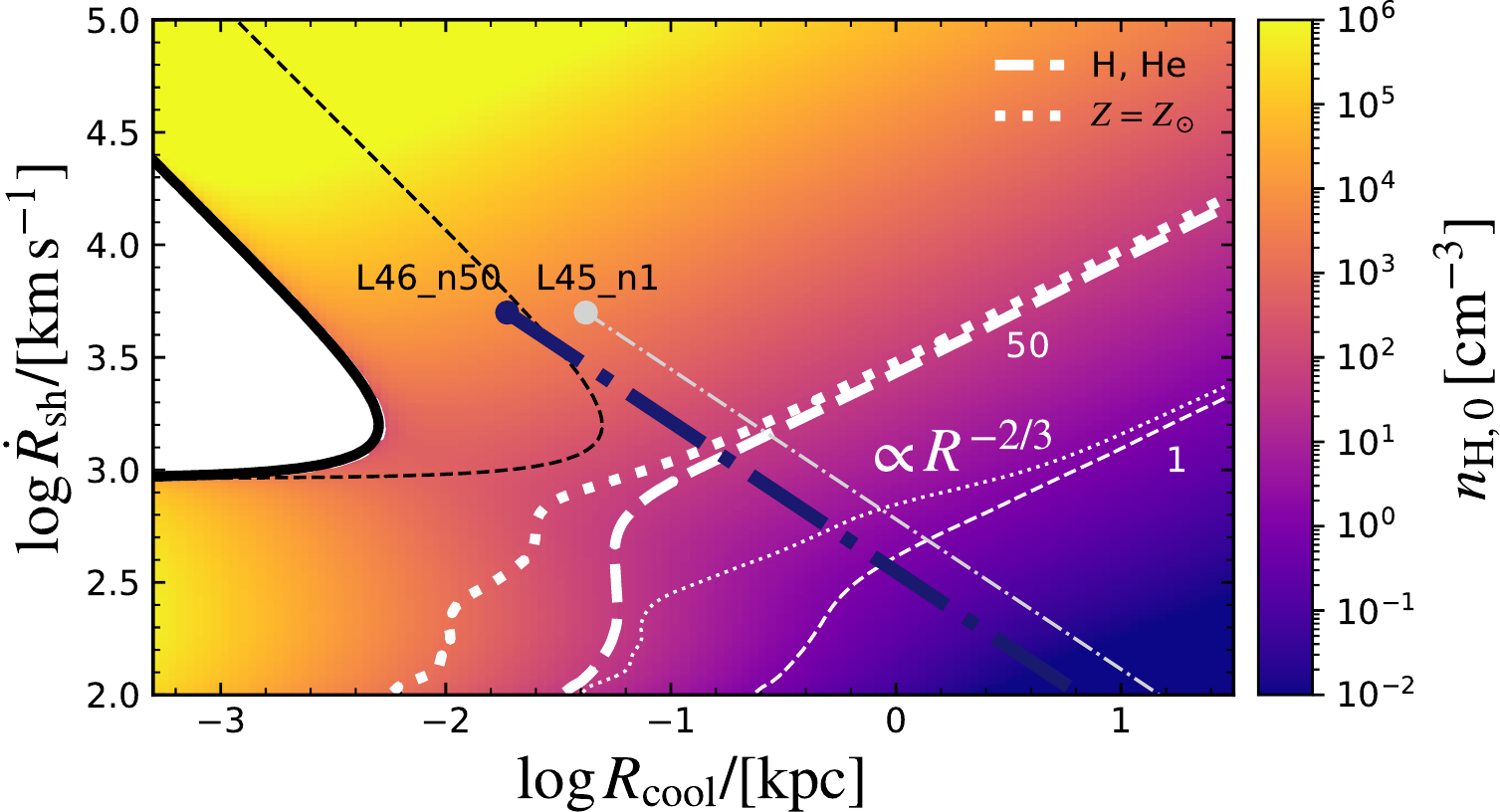}
\caption{The coloured field shows the required ambient medium hydrogen number density $n_{\rm H, 0}$ for a shell composed of shocked ambient gas with speed $\dot{R}_{\rm sh}$ to cool at any given radius $R_{\rm cool}$, assuming cooling for a collisionally ionised plasma of primordial composition. Contour levels corresponding to $n_{\rm H, 0} \, = \, 1 \, \rm cm^{-3}$ and $n_{\rm H, 0} \, = \, 50 \, \rm cm^{-3}$ are shown with white, dashed curves. The white, dotted curves show the corresponding contours assuming the ambient medium has solar metallicity. The region where Compton cooling dominates, and the cooling time is independent of density, is shaded in white for $L_{\rm AGN} \, = \, 10^{46} \, \rm erg \, s^{-1}$ and with a black dashed contour for $L_{\rm AGN} \, = \, 10^{47} \, \rm erg \, s^{-1}$. The diagonal lines show the evolutionary tracks expected for some of our simulated shells. Cooling of the shocked ambient medium phase occurs at the radii at which the tracks intersect the density contour corresponding to the ambient medium the shells propagate into.}
\label{fig_cooling}
\end{figure}

We should expect radiative cooling to become important in the shell of shocked ambient gas when its cooling timescale becomes comparable to the outflow timescale $\sim R_{\rm sh} / \dot{R}_{\rm sh}$, or when $R_{\rm sh} \sim R_{\rm c}^{\rm ff}$.
The coloured field in Fig.~\ref{fig_cooling} shows the ambient medium number density required for a shell with speed $\dot{R}_{\rm sh}$ to cool at any given radius $R_{\rm cool}$. In order to compute cooling radii, we model cooling of a H and He mixture of primordial composition, assuming collisional ionisation equilibrium, using the tabulated cooling rates of \citet{Wiersma:09}. Compton heating/cooling from AGN is modelled following \citet{Sazonov:01}, through the additional term
\begin{equation}
\Lambda_{\rm cpt} \, = \, \frac{\sigma_{\rm T}}{m_{\rm e} c^2} k_{\rm B} n_{\rm e} \frac{L_{\rm AGN}}{\pi R^2} \left(T - T_{\rm cpt} \right) \left( 1 + \frac{5}{2} \frac{k_{\rm B} T}{m_{\rm e} c^2} - 2 \pi \frac{k_{\rm B} T_{\rm cpt}}{m_{\rm e} c^2} \right)\, ,
\label{eq_NR_compton}
\end{equation}
where $\sigma_{\rm T}$ is the Thomson scattering cross-section, $m_{\rm e}$ the electron mass, $n_{\rm e}$ the electron density and $T_{\rm cpt}$ the Compton temperature. 
The Compton temperature depends on the shape of the AGN spectrum; for the average quasar, $T_{\rm cpt} \approx 2 \times 10^7 \, \rm K$ \citep{Sazonov:04}, which is the value assumed here. In Fig.~\ref{fig_cooling}, we choose $L_{\rm AGN} \, = \, 10^{46} \, \rm erg \, s^{-1}$ for illustrative purposes.

We see that, for fixed $\dot{R}_{\rm sh}$, cooling occurs at increasingly lower densities as radius increases, a result which follows directly from Eq.~\ref{eq_coolradius_ff_F}.
Raising the outflow velocity both decreases the outflow time and increases the post-shock temperature of the ambient medium, which in turn prolongs the cooling times.
Therefore, the density required for cooling increases with outflow velocity $\dot{R}_{\rm sh}$ at large radii.

At small radial distances, Compton scattering results in heating if $T_{\rm F-shock} < T_{\rm cpt}$ and in cooling if $T_{\rm F-shock} > T_{\rm cpt}$.
Thus, if $\dot{R}_{\rm sh} \lesssim 10^3 \, \rm km \, s^{-1}$, Compton heating offsets cooling losses.
With decreasing radial distance to the AGN, Compton heating becomes increasingly efficient, but can always be overcome through an increase in cooling losses, either by raising the ambient medium density or by decreasing the outflow velocity.
For $\dot{R}_{\rm sh} \gtrsim 10^3 \, \rm km \, s^{-1}$, Compton scattering leads to cooling. If Compton cooling losses dominate over other cooling processes, which is the case at small enough distances from the AGN, the cooling time becomes independent of gas density. The white region in Fig.~\ref{fig_cooling} gives the parameter combinations for which gas cooling occurs at any gas density. From below, this region is bounded by an approximately horizontal line corresponding to the outflow velocity associated with $T_{\rm F-shock} \, = \, T_{\rm cpt}$. From above, it is limited by a diagonal line that follows $\dot{R}_{\rm sh} \propto R_{\rm cool}^{-1}$ and results from the decreasing outflow times as $\dot{R}_{\rm sh}$ increases. The radii out to which this region extends increases with AGN luminosity, as shown with the dashed, black curve in Fig.~\ref{fig_cooling} for $L_{\rm AGN} \, = \, 10^{47} \, \rm erg \, s^{-1}$.

Finally, in white we also present contours for $n_{\rm H,0} \, = \, 1 \, \rm cm^{-3}$ (thin) and $n_{\rm H,0} \, = \, 50 \, \rm cm^{-3}$ (thick) for primordial cooling and Compton cooling/heating.
The dotted curves show how these contours are modified if the ambient medium has metallicity $Z \, = \, Z_\odot$.
Since metal-line cooling is efficient at $T \lesssim 5 \times 10^7 \, \rm K$, it can precipitate cooling when shell has slowed down to $\dot{R}_{\rm sh} \lesssim 10^3 \, \rm km \, s^{-1}$ \citep{Costa:14}.

Fig.~\ref{fig_cooling} can be used to work out the radius at which the shocked ambient medium phase of an AGN-driven outflow starts cooling radiatively. As an illustration, we consider a wind with  $v_{\rm w} \, =  \, 5000 \, \rm km \, s^{-1}$ ejected by an AGN with luminosity $L_{\rm AGN} \, = \, 10^{45} \, \rm erg \, s^{-1}$ into a homogeneous ambient medium of density $n_{\rm H,0}$. According to Eq.~\ref{eq_rfree}, the wind thermalises at a radial distance of $R_{\rm free} \approx 40 \, \rm pc$ from the AGN if $n_{\rm H,0} \, = \,  1 \, \rm cm^{-3}$. If $n_{\rm H,0} \, = \,  50 \, \rm cm^{-3}$ and $L_{\rm AGN} \, = \,  10^{46} \, \rm erg \, s^{-1}$, for instance, then $R_{\rm free} \approx 18\, \rm pc$.
Eqs.~\ref{eq_coolradius_cpt} and~\ref{eq_coolratio_R} indicate that the \emph{shocked wind} should not radiate its thermal energy efficiently, so it expands adiabatically into its surroundings, driving an energy-driven shell.  
Cooling of the \emph{shocked ambient medium}, however, occurs when the track described by the shell in the $(\dot{R}_{\rm sh}, R)$ plane intersects the white contour corresponding to the ambient medium it pushes into. 
The thick, blue, dash-dotted line in Fig.~\ref{fig_cooling} shows the track described by the shell if $n_{\rm H,0} \, = \, 50 \, \rm cm^{-3}$ and $L_{\rm AGN} \, = \, 10^{46} \, \rm erg \, s^{-1}$. In this case, the shocked ambient gas should cool radiatively at $R_{\rm cool} \, \approx \, 170 \, \rm pc$ when it has slowed down to a speed $\dot{R}_{\rm sh} \, \approx \, 1000 \, \rm km \, s^{-1}$.
However, if $n_{\rm H,0} \, = \, 1 \, \rm cm^{-3}$ and $L_{\rm AGN} \, = \, 10^{45} \, \rm erg \, s^{-1}$ (thin, dash-dotted, gray line), cooling would occur only at $R_{\rm cool} \, \approx \, 1.4 \, \rm kpc$ when $\dot{R}_{\rm sh} \, \approx \, 500 \, \rm km \, s^{-1}$.

In this paper we include radiative cooling down to $10^4 \, \rm K$ and do not investigate the formation of a molecular phase. We refer the reader to \citet{Richings:18,Richings:18b} for a detailed analysis of the formation of an outflowing, molecular phase.

\subsection{Choosing parameters}
\label{sec_choiceparam}

In order to fully specify all properties of the AGN wind, we must choose a wind speed $v_{\rm w}$, the momentum transfer rate $\dot{P}_{\rm w}$ of the wind in terms of $L_{\rm AGN}/c$,  the fractional solid angle $b$ subtended by wind and the initial wind temperature $T_{\rm w}$ (see Table~\ref{table:free_parameters} for a list of the free parameters of our model).
In this Section, we briefly review observational findings and results from simulations of accretion disc winds in order to guide our choice of parameters.

\subsubsection{Observational constraints}
At scales $\gtrsim 100 \, \rm pc$, there is evidence of AGN-driven galactic outflows moving at high speeds $\gtrsim 1000 \, \rm km \, s^{-1}$ \citep[e.g.][]{Sturm:11, Maiolino:12, Foerster-Schreiber:14, Harrison:14, Fluetsch:19, Veilleux:20}.
Outflow detections are typically based on emission from hydrogen recombination lines such as H$\alpha$ and H$\beta$, ionized metal lines such as [OIII] and [CII], and molecular lines such as CO.
Mass estimates, which are notoriously uncertain \citep[e.g.][]{Husemann:16, Harrison:18}, suggest outflow masses typically in the range $10^7 \, \-- \, 10^{10} \, \rm M_\odot$.
When combined with directly measured velocities, such estimates are used to compute approximate values for the outflow kinetic luminosity, typically $\dot{E}_{\rm out}/L_{\rm AGN} \approx 10^{-4} \-- 10^{-2}$, and the outflow momentum flux, which is often $\dot{P}_{\rm out} \sim L_{\rm AGN} / c$ \citep[e.g.][]{Cicone:15,Sirressi:19,Fluetsch:19} although $\dot{P}_{\rm out} > L_{\rm AGN} / c$ for many systems \citep[e.g.][]{Cicone:14,Herrera-Camus:19}.
Such massive, large-scale outflows, however, most likely consist primarily of ambient interstellar gas which is either pushed out by radiation pressure \citep[e.g.][]{Costa:18b} or swept-up by a smaller-scale AGN-driven wind originating from the galactic nucleus \citep[e.g.][]{Zubovas:12}.
In our model, we attempt to inject the small-scale wind directly and therefore should not select our parameters based on observations of large-scale outflows.  

At the smallest scales, AGN-driven winds are detected in absorption against direct X-ray emission from AGN. The most extreme winds, the ultra-fast outflows, which are observed through blue-shifted, high-ionization Fe absorption lines, can attain mildly relativistic speeds \citep[e.g.][]{Pounds:03, Cappi:09, Tombesi:13, Nardini:15, Braito:18, Pinto:18}. 
There is, however, considerable spread in the speed of ultra-fast outflows, which can range from $\sim 0.01c$ to $0.4c$ \citep[e.g.][]{Tombesi:12}.
Using a sample of $20$ systems with blue-shifted Fe K-shell absorption, \citet{Gofford:15} find that the mass outflow rate of ultra-fast outflows scales as $\dot{M}_{\rm w} \propto L_{\rm AGN}$, such that the brighter the AGN, the more mass-loaded the small-scale wind.
They also find that the scalings between integrated momentum flux $\dot{P}_{\rm w}$, kinetic luminosity $\dot{E}_{\rm w}$ and the bolometric luminosity are consistent with linear relations and, in addition, that $\dot{P}_{\rm w} \sim L_{\rm AGN} / c$ and $\dot{E}_{\rm w}/L_{\rm AGN} \approx 10^{-3} \-- 10^{-1}$.

The location of ultra-fast outflows is difficult to estimate accurately, but can be estimated from the measured ionisation parameter and column density as well as from escape velocity arguments \citep[e.g.][]{Tombesi:12,Gofford:15}.
These winds are thus thought to be launched from scales $\sim 10 \-- 100 r_{\rm g}$, where $r_{\rm g} \, = \, GM_{\rm BH}/c^2 \approx 5 \times 10^{-6} ( M_{\rm BH} / 10^8 \mathrm{M_\odot} ) \, \rm pc$.

Given reported high detection rates $\gtrsim 40\%$ \citep{Tombesi:11}, ultra-fast outflows are thought to be quasi-spherical with a fractional solid angle $b \sim 1$.
The wind geometry has also been estimated directly in a few systems; for instance, the width of the detected P-Cygni profile in two nearby quasars suggests $b \, \approx \, 0.75$ \citep{Pounds:09, Nardini:15}. 

\subsubsection{Theoretical constraints}
The efficiency at which AGN launch small-scale winds is tied to the properties of the black hole accretion disc.
If the disc is hot, optically thin and geometrically thick, accretion is radiatively inefficient and most energy is delivered mechanically in the form of bipolar, relativistic jets \citep{Narayan:94, Yuan:14}. 
Relativistic, magneto-hydrodynamic simulations of hot accretion flows also predict the existence of a quasi-spherical wind component with speeds $0.01c \-- 0.05c$ \citep[e.g.][]{Yuan:12, Sadowski:13}.
Such winds are launched through combination of centrifugal forces and magnetic pressure gradients \citep{Yuan:15}.
Even if energetically subdominant with respect to jets, winds carry significant kinetic energy, with energy fluxes in the range $\dot{E}_{\rm w} \, \approx \, (0.001 \-- 0.05) \dot{M}_{\rm BH} c^2$.
\citet{Sadowski:13} find that the radial momentum carried by wide-angle winds varies with radial distance from the supermassive black hole. 
At radii $r \gtrsim 100 r_{\rm g}$, the radial momentum flux $\dot{P}_{\rm w}$ scales weakly with radius, asymptoting to values of $\approx (0.1 \-- 1)  \dot{M}_{\rm BH} c$ depending on the spin of the accreting black hole and on the magnetic flux threading the horizon.
In the notation of Section~\ref{sec_anal}, winds launched from thick accretion discs typically have $\beta \, \approx \, 0.01 \-- 0.05$, $\tau \, \approx \, (0.1 \-- 1)  \epsilon_{\rm r}^{-1}$ and $\eta \, \approx \, (0.001 \-- 0.05)  \epsilon_{\rm r}^{-1}$.

At high accretion rates $0.01 \lesssim \frac{\dot{M}_{\rm BH}}{\dot{M}_{\rm Edd}} \lesssim 1$, black hole accretion flows are expected to cool efficiently and settle onto geometrically thin discs \citep{Shakura:73}.
Jets are not expected to form in this accretion regime \citep[e.g.][]{Sadowski:13}.
Radiation pressure on UV lines \citep[e.g.][]{Proga:00, Risaliti:10} and hydromagnetic forces \citep[e.g.][]{Contopoulos:94}, however, are expected to drive fast winds. 
Using radiation-hydrodynamic simulations of thin discs centred on black holes with mass $M_{\rm BH} \, = \, 10^6 \-- 10^9 \, \rm M_\odot$ and Eddington luminosity ratios $0.1 \-- 0.7$, \citet{Nomura:16} and \citet{Nomura:17}, for instance, find that line radiation pressure launches winds with opening angle $\approx 80^\circ$, i.e. $b \approx 0.97$, which reach terminal values of $\beta \lesssim 0.1$, $\tau \sim 1$ and $\eta \lesssim 0.05$ at $\approx 50 r_{\rm g}$, in agreement with the properties of ultra-fast outflows estimated by \citet{Tombesi:12} and \citet{Gofford:15}.
Note that, while the winds propagate along much of the solid angle, their speed is highest along the equatorial plane of the accretion disc and drops to $\beta \sim 0.01$ at low inclinations. 

Winds may also be driven from scales larger than the accretion disc.
Radiation pressure on dust at torus scales ($0.1 \-- 30 \, \rm pc$), for instance, also appears to drive winds, though at much lower speeds than ultra-fast outflows. The radiative transfer calculations presented in \citet{Roth:12}, for instance, predict $v_{\rm w} \sim 10^3 \, \rm km \, s^{-1}$ ($\beta \approx 0.003$), $\tau \, \approx \, 1 \-- 5$ and $\eta \, \approx \, 0.009$ for $L_{\rm AGN} \, \approx \, 10^{46} \, \rm erg \, s^{-1}$, the Eddington luminosity of a black hole with $M_{\rm BH} \, = \, 10^8 \, \rm M_\odot$.
In summary, in the high accretion regime, plausible choices for our free parameters are: $\beta \, \approx \, 0.003 \-- 0.3$, $\tau \, \sim 1$, $\eta \, \approx \, 0.001 \-- 0.05$ and $b \sim 1$.

\subsection{Numerical implementation}
We perform our simulations with the moving-mesh hydrodynamic code {\sc AREPO} \citep{Springel:10}. {\sc AREPO} has recently been publicly released \citep{Weinberger:19}. Here, we first briefly review {\sc AREPO} and then proceed to describe how small-scale winds are implemented numerically, as outlined in Section~\ref{sec_anal}.

\subsubsection{Moving-mesh hydrodynamics}
In {\sc AREPO}, gas is discretised on an unstructured mesh constructed from a Voronoi tessellation of a set of mesh-generating points.
Hydrodynamic fluxes across cell interfaces are computed using a directionally unsplit, second-order Godunov scheme \citep{Pakmor:16}.
The mesh-generating points move together with the fluid, overcoming the Galilean non-invariance and advection errors in supersonic flows that bedevil fixed-grid Eulerian codes. At the same time, the moving-mesh character of {\sc AREPO} ensures superior shock capturing than e.g. smoothed-particle hydrodynamic (SPH) methods, without relying on artificial viscosity.

\begin{table}
\caption{List of the free parameters in the wind injection boundary implementation for AGN feedback (left-hand column) and a brief description (right-hand column).}
\label{table:free_parameters}
\begin{center}
\begin{tabular}{ cl|l| }
  \hline
  \multicolumn{2}{|c|}{Free parameters in the injection boundary model for AGN winds} \\
  \hline
  $L_{\rm AGN}$ & Bolometric luminosity of the central AGN. \\
  $\beta$            & Ratio between the wind speed and $c$. \\
  $\tau$              & Ratio between integrated wind momentum flux and $L_{\rm AGN}/c$. \\
  $b$                  & Fractional solid angle into which wind is injected. \\
  $T_{\rm w}$     & Initial temperature of the wind. \\
  $r_{\rm sp}$     & Radius of the wind injection boundary. \\
  $n_{\rm side}$  & Number of {\sc HealPix} cells on each wind boundary layer. \\
  \hline
\end{tabular}
\end{center}
\end{table}

The Poisson equation is solved using a tree-particle-mesh ({\sc TreePM}) algorithm in order to compute gravitational accelerations for the gas cells as well as any other matter component (e.g. stars, dark matter or black holes) followed in the simulation.

\begin{figure*}
	\includegraphics[width=0.95\textwidth]{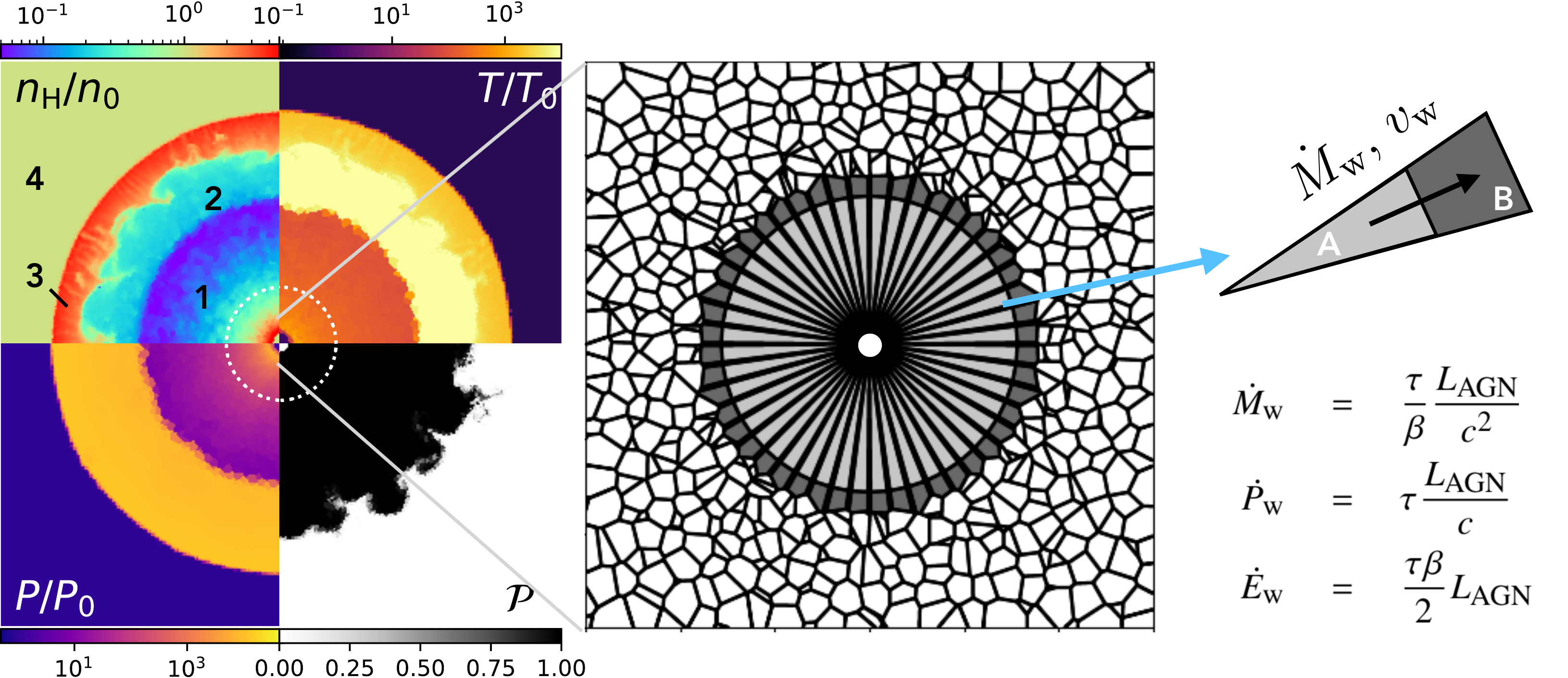}
	\caption{The AGN-driven outflow can be divided into four distinct sections: (1) the freely-expanding wind, (2) the shocked wind, (3) the shocked ambient medium and (4) the undisturbed ambient medium. The left-hand panel illustrates the density, temperature and pressure fields in units of the corresponding quantities of the assumed background medium as well as the wind tracer concentration. The wind tracer is injected together with the wind and is therefore only present in regions (1) and (2). The dotted circle gives the free-expansion radius. The central panel zooms onto the region containing the very central resolution elements, displaying the Voronoi mesh of the ambient medium and of the `wind injection boundary' across which the small-scale wind is injected. The sphere consists of two spherical layers of mesh generating points with positions determined according to a {\sc HealPix} tessellation. Mass, momentum and energy, as expected for a steady wind with fixed velocity $v_{\rm w}$, are injected at the interfaces between cells lying on each of these two layers.}\
	\label{fig_main}
\end{figure*}

In order to increase the numerical resolution, gas cells can be refined and de-refined wherever required, according to any prescribed refinement criterion.
Usually, the adopted refinement strategy ensures an approximately constant mass per Voronoi cell, such that high-density regions are resolved with more cells than low-density regions.

\subsubsection{Wind injection boundary}
We generate two concentric, spherical layers of {\sc AREPO} cells with their origin centred at the position of the black hole (see Fig.~\ref{fig_main}). 
The spatial coordinates of the mesh-generating points associated to these cells follow a {\sc HealPix} tessellation \citep{Gorski:05}, where each 2-sphere is discretised with a number of $12 n_{\rm side}^2$ pixels of equal surface area. 
The resolution $n_{\rm side}$ is a free parameter. In Appendix~\ref{sec:app:numerical}, we show that simulated outflow solutions are only weakly sensitive to its value.
This spherical structure behaves as a rigid body; the positions of its constituent cells are fixed in space relative to one another, unlike in conventional {\sc AREPO} cells.
Neither refinement nor de-refinement are allowed for the cells making up either of these two spherical layers. 
Likewise, the cells in both layers are not allowed to cool radiatively.

The two spherical layers are separated by a well-defined, spherical boundary at a radius $r_{\rm sp}$.
Wind injection is performed across the spherical boundary at the interfaces between cells belonging to the `inner layer' and the `outer layer'.
The cells pertaining to the inner layer are excluded from hydrodynamic computations and are used only to define the spherical boundary. 
For the cells belonging to the outer layer, where wind mass, momentum and energy are deposited, the hydrodynamic evolution is performed identically to all other cells in the simulation domain.
In practice, we attribute different flags to cells located in inner layer vs. those located in the outer layer (see Fig.~\ref{fig_main}) and set the fluxes between cell neighbour pairs with distinct flags to the mass, momentum and energy fluxes of Eqs.~\ref{eqs_conservation_m2} \--~\ref{eqs_conservation_e2}.

We also inject a passive, conserved scalar across the spherical boundary, along with mass, momentum and energy. This `wind tracer' advects passively with the injected wind, directly tracking its mass.
If wind material mixes with ambient gas or spreads across the simulation domain, so does the wind tracer. The tracer flux across the boundaries of this cell is thus identical to the wind mass flux across the same interfaces.
We define the wind concentration as the fraction of wind mass in any given {\sc AREPO} cell, denoting it by $\mathcal{P}$.
The wind concentration can be used to investigate the hydrodynamic evolution of the injected wind fluid, to separate shocked wind and shocked ambient medium phases and to quantify mixing with the gas the wind interacts with. 

\begin{figure}
	\centering
	\includegraphics[width=0.475\textwidth]{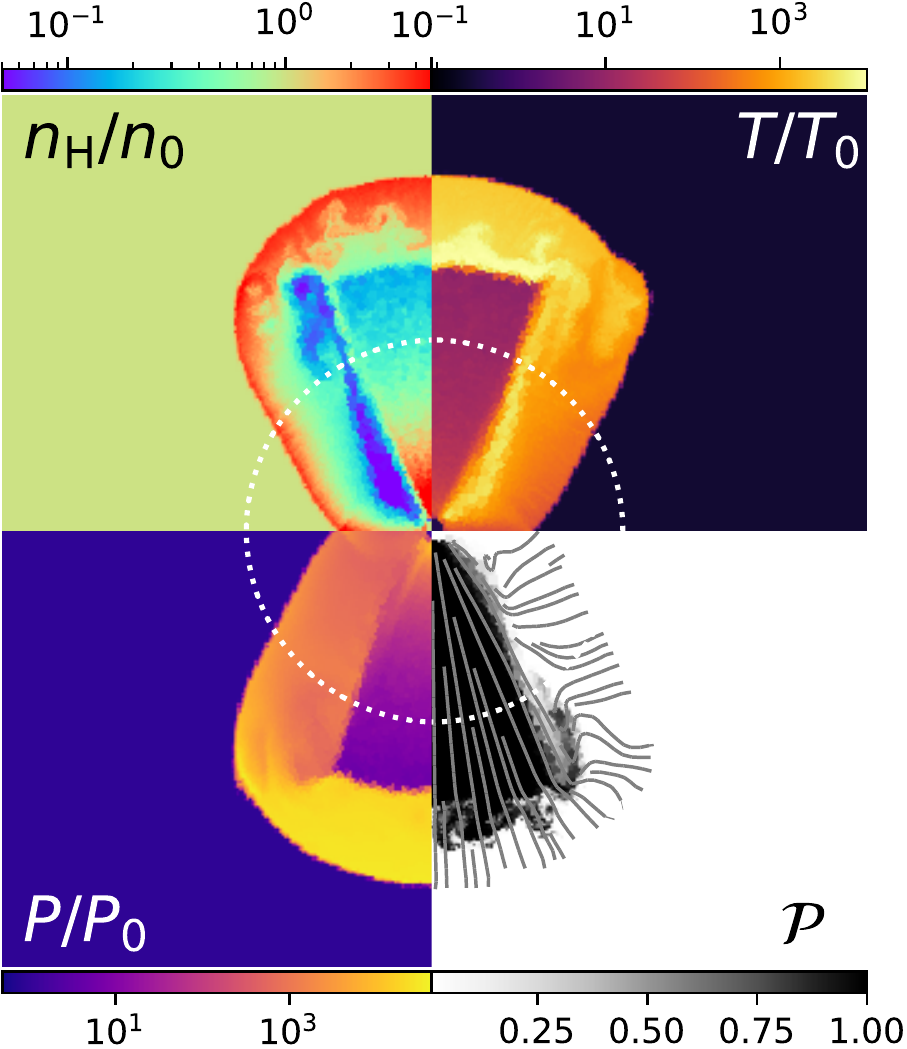}
	\caption{Density, temperature, pressure and wind tracer concentration in a simulation in which the outflow geometry is biconical with opening angle of $\theta \, = \, 30^\circ$, corresponding to $b \, = \, 0.134$. Flow streamlines are shown in the bottom right with grey curves. The dotted circle shows the free-expansion radius. The same outflow phases as in the spherical case can be discerned here: the freely-expanding wind, the shocked wind, the shocked ambient medium and the undisturbed ambient medium.}
	\label{fig_cone}
\end{figure}

While we focus on spherical winds, it is also possible to inject the wind across a fraction of the solid angle subtended by the spherical boundary.
In this case, the fluxes between `inner' and `outer' cells are set to Eqs.~\ref{eqs_conservation_m2} \--~\ref{eqs_conservation_e2}, where $b$ has then to be adjusted to the correct value, only for cells lying within the solid angle of interest.
For cell pairs lying outside of the solid angle of interest, the fluxes are set to zero.

\section{Tests to the model}
\label{sec:testmodel}
In this Section, we perform various tests to our model and demonstrate that the structure, kinematics and cooling properties of AGN-driven outflows as outlined in Section~\ref{sec:windmodel} are all reproduced in detail in our simulations. At the end of the Section we also present a number of convergence tests.

\begin{table*}
\caption{List of simulations and parameters.}
\label{table:homogeneous}
\small\addtolength{\tabcolsep}{-2pt}
\begin{center}
\begin{tabular}{ lcccccccccc| }
  \hline
  Simulation & $L_{\rm AGN}$  & $\beta$ & $n_{\rm H,0}$ & $b$ & $R_{\rm free}$ & $L$  & $m_{\rm target}$ & Compton & $r_{\rm sp}$ & $R_{\rm therm}$ \\
                    & $\rm [ erg \, s^{-1}] $ &         & $\rm [cm^{-3}]$ &   & $\rm [pc]$ & $\rm [kpc]$ & $\rm [M_\odot]$ &  cooling?   & $\rm [pc]$     & $\rm [pc]$ \\
  \hline
  \texttt{shell-L45-b0.02-n1}    &  $10^{45}$ & $0.017$ & $1$   & $1$  & $40.2$ & $1$ & $0.2$ & & $8$ & $44.8$\\
  \texttt{cone-L45-b0.02-n1}    &  $10^{45}$ & $0.017$ & $1$   & $0.134$  & $109.8$  & $1$ & $0.2$& &$8$ & $114.8$\\
  \texttt{shell-L46-b0.02-n50}  &  $10^{46}$ & $0.017$ & $50$ & $1$  & $18.0$ & $1$ & $10$ & & $8$ & $24.0$\\
  \texttt{shell-L46-b0.02-n50-HiRes} &  $10^{46}$ & $0.017$ & $50$ & $1$  & $18.0$ & $1$ & $1$ & & $8$ & $21.0$\\
  \texttt{shell-L47-b0.1-n50}  &  $10^{47}$ & $0.1$ & $50$ & $1$  & $9.5$ & $1$ & $10$ & & $2$ & $15.4$\\
  \texttt{shell-L47-b0.1-n50-Cpt}  &  $10^{47}$ & $0.1$ & $50$ & $1$  & $9.5$ & $1$ & $10$ & \checkmark & $2$ \\
  \texttt{shell-L5e47-b0.1-n1000}  &  $5 \times 10^{47}$ & $0.1$  & $10^3$  & $1$  & $4.8$ & $0.1$ & $2$ & & $0.4$ & $6.8$\\
  \texttt{shell-L5e47-b0.1-n1000-Cpt}  &  $5 \times 10^{47}$ & $0.1$  & $10^3$  & $1$  & $4.8$ & $0.1$ & $2$ & \checkmark & $0.4$\\
  \hline
\end{tabular}
\end{center}
\end{table*}

We start by making the same assumptions as in Section~\ref{sec:windmodel} and consider the propagation of small-scale AGN winds through static, homogeneous media with hydrogen number density $n_{\rm H,0}$ and temperature $T_{\rm 0} \, = \, 2 \times 10^4 \, \rm K$.
Like in Section~\ref{sec:windmodel}, we do not consider the effects of gravity, gaseous infall or density anisotropy. Our chief concern here is to test and validate our numerical model against the identical analytical setup presented in the previous Section. For an exploration of the impact of gravity, gaseous infall and anisotropy on outflow solutions, we refer the reader to \citet{King:05}, \citet{Zubovas:12}, \citet{Costa:14} and to Section~\ref{sec:discs}.
In different simulations, we probe ambient medium number densities ranging from $n_{\rm H,0} \, = \, 1 \, \rm cm^{-3}$ to $n_{\rm H,0} \, = \, 10^3 \, \rm cm^{-3}$, small-scale wind velocities of $5000 \, \rm km \, s^{-1}$ ($\beta \, = \, 0.017$) and $30000 \, \rm km \, s^{-1}$ ($\beta \, = \, 0.1$).
The ratio between wind's momentum flux and $L_{\rm AGN}/c$ is set to $\tau \, = \, 1$, the fractional solid angle to $b \, = \, 1$ and the wind temperature at injection to $T_{\rm w} \, = \, 5 \times 10^5 \, \rm K$.
Our parameters are thus close to the those of ultra-fast outflows (Section \ref{sec_choiceparam}), with our choices of $\beta$ bracketing their typical velocity range.

We name our simulations according to the AGN luminosity, wind speed and ambient medium density used; the simulation performed with e.g. $L_{\rm AGN} \, = \, 10^{47} \, \rm erg \, s^{-1}$, $\beta \, = \, 0.1$ and $n_{\rm H,0} \, = \, 50 \, \rm cm^{-3}$ is referred to as \texttt{shell-L47-b0.1-n50}.
In simulation \texttt{cone-L45-b0.02-n1}, the wind is injected along a cone with opening angle $\theta \, = \, 30^\circ$, such that $b \, = \, 2 \sin{(\theta/2)^2} \, \approx \, 0.134$.   

The simulation domain consists of a cubic box with side length $L \, = \, 1 \, \rm kpc$. 
When performed with $n_{\rm H} \, = \, 10^3 \, \rm cm^{-3}$, the box size is, instead, $L \, = \, 100 \, \rm pc$.
A list of the different simulations performed as well as the main parameters explored is provided in Table~\ref{table:homogeneous}.

The wind injection boundary is placed at the centre of the box.
Its radius is set to a value lower than the expected free-expansion radius (see Table~\ref{table:homogeneous}).
We also use $n_{\rm side} \, = \, 12$, such that the two concentric layers defining the boundary are each sampled with 1728 cells. 
In Appendix~\ref{sec:app:numerical}, we show that the wind solutions are not sensitive to this parameter, unless it becomes so small that for each boundary cell there is a very large number of conventional cell neighbours.

All simulations are performed with radiative cooling for a H and He plasma of primordial composition in photo-ionisation equilibrium with the UV background of \citet{Faucher-Giguere:09} at $z \, = \, 0$.
We do not employ a self-shielding correction and self-gravity is neglected.

\subsection{Structure of AGN-driven outflows}
When injected isotropically into an homogeneous medium, small-scale AGN winds drive outwardly-expanding shells \citep[e.g.][]{Nayakshin:10,Costa:14}.
The left-hand panel of Fig.~\ref{fig_main} shows a slice through one of our simulations (\texttt{shell-L45-b0.02-n1}), illustrating gas density, temperature, pressure and wind tracer, all normalised to the initial values of the ambient medium (except for the wind tracer, where the initial value is zero everywhere).
Four distinct flow sections can be identified: (1) the unshocked wind, whose density, temperature and pressure all fall off as it expands, (2) the low-density, hot shocked wind component, (3) the shocked ambient medium, which is in pressure equilibrium with the shocked wind, and (4) the undisturbed ambient medium.
While the shocked wind contains wind tracer, the shocked ambient medium does not.

The outflow configuration shown in Fig.~\ref{fig_main} matches the classical outflow structure usually assumed in analytic studies of small-scale wind-driven outflows \citep[e.g.][]{Weaver:77,King:03,Zubovas:12,Costa:14}.
It also agrees with the outflow structure outlined in Section~\ref{sec_anal}.
For a biconical outflow (Fig.~\ref{fig_cone}), this four-zone structure is still present, but there are additional features.
Ambient gas and wind material passing the forward- and reverse shocks near the edges of the cone are pushed aside, where the pressure is lower, leading to the formation of a laterally-expanding, hot cocoon.

\begin{figure*}
	\includegraphics[width=0.45\textwidth]{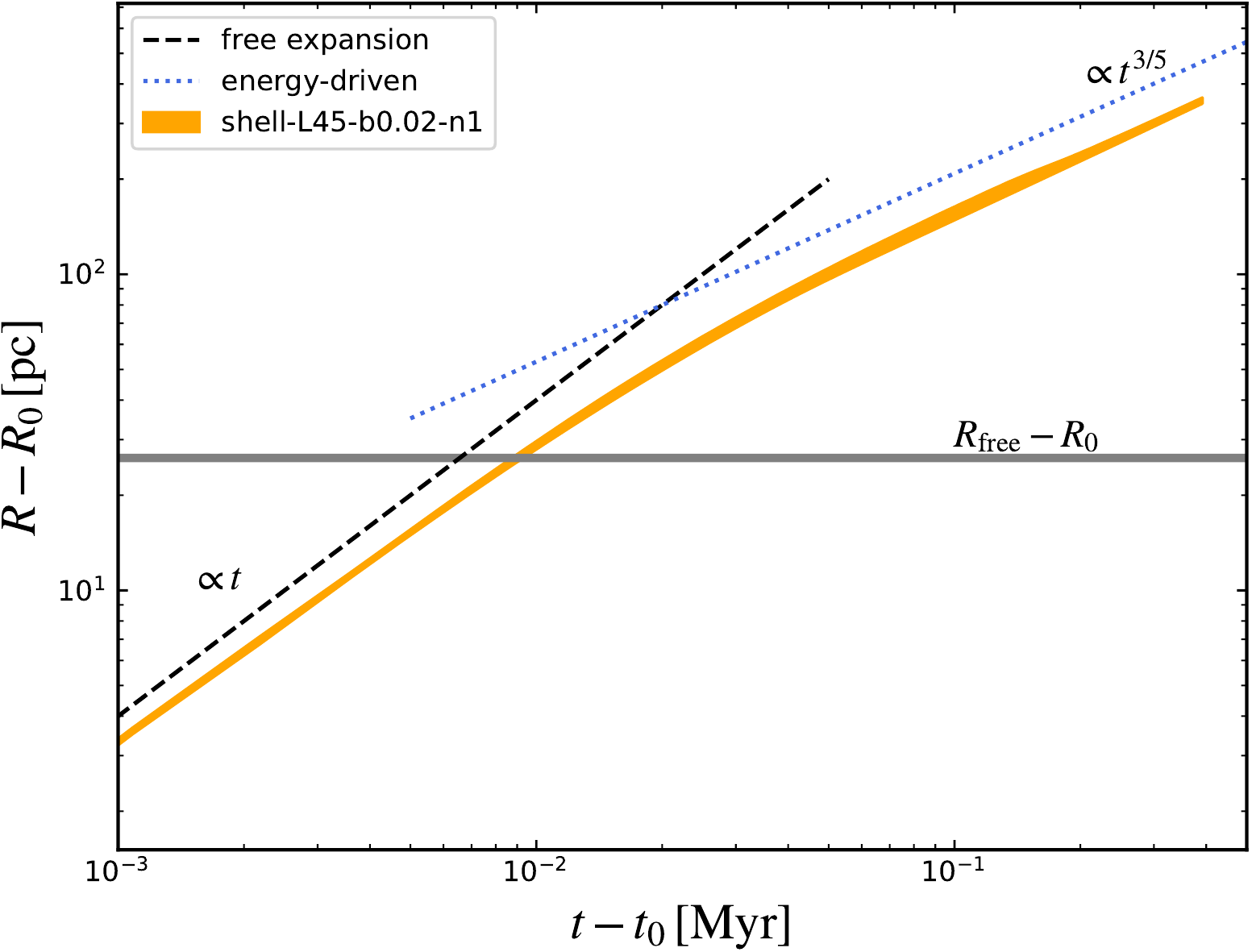}
	\includegraphics[width=0.45\textwidth]{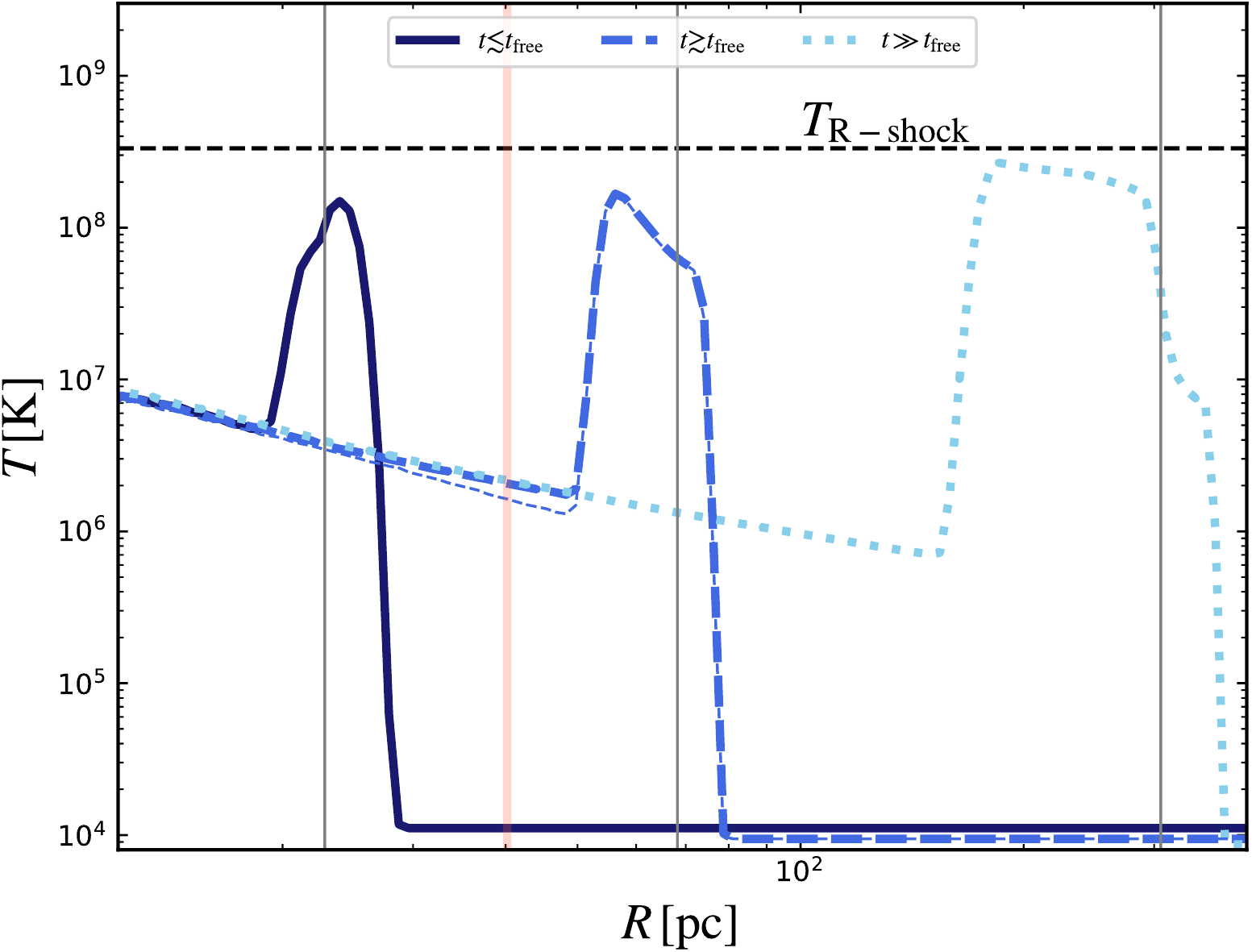}
	\caption{Left: position of the contact discontinuity as a function of time in the simulation \texttt{shell-L45-b0.02-n1} (orange curve). At early times, the shell radius grows as $R_{\rm sh} \propto t$, as expected for the free-expansion phase, while $R_{\rm sh} \propto t^{3/5}$ at later times, as expected for the energy-driven phase. The transition between both phases is gradual but it occurs close to the analytical free-expansion radius (horizontal line). Right: Temperature profile in \texttt{shell-L45-b0.02-n1} at three different times: (i) $t \lesssim t_{\rm free}$ (dark blue), (ii) $t \gtrsim t_{\rm free}$ (blue, dashed curve) and (iii) $t \gg t_{\rm free}$ (light blue, dotted curve). The horizontal line gives the post-shock temperature $T_{\rm R, shock}$ expected for a wind with $\beta \, = \, 0.017$, while the vertical, red line gives the free-expansion radius $R_{\rm free}$. Vertical, gray lines show the position of the contact discontinuity at the different times.}
	\label{fig_radtime}
\end{figure*}

\subsection{From free-expansion to energy-driving}
\label{sec:transitionrad}
In Section~\ref{sec_anal}, we have argued that the classical four-zone structure, which is reproduced in our simulations, should be valid only for $t > t_{\rm free}$.
We now verify that the simulated outflowing shells indeed experience an initial period of free-expansion, during which their radii obey $R_{\rm sh} \propto t$. 
We then test whether, after reaching the free-expansion radius $R_{\rm free}$ given by Eq.~\ref{eq_rfree}, the shells transition into an energy-driven phase, where $R_{\rm sh} \propto t^{3/5}$ (see Appendix~\ref{sec:app:energy-driven}).
Note that using $r_{\rm sp} > 0$ increases the expression obtained in Eq.~\ref{eq_rfree} by $\lesssim 5 \%$ only, because $R_{\rm free}^3 \gg r_{\rm sp}^3$ in all our simulations.

We use the wind tracer in order to locate the position of the contact discontinuity separating shocked wind and swept-up ambient phases.
We select gas cells with $\mathcal{P} > 0.5$ and compute the $95^{\rm th}$ and the $99.7^{\rm th}$ percentiles of their radial distance, corresponding to $2\sigma$ and $3\sigma$ radial distance limits, respectively.
These two values are used as lower and upper estimates for $R_{\rm sh}$.

In the left-hand panel of Fig.~\ref{fig_radtime}, we plot the time evolution of $R_{\rm sh}$ as a shaded region spanning the range between lower and upper estimates of $R_{\rm sh}$ for \texttt{shell-L45-b0.02-n1}.
We subtract the position of the shell $R_{\rm 0}$ at a very early time from that at subsequent times to clearly reveal power law behaviour when the shell is still at scales comparable to $r_{\rm sp}$.
The shell radius grows the fastest early on, when its time evolution is well approximated by $R_{\rm sh} \propto t$, as shown by comparison with the black, dashed line, as expected.
At $t \, \gtrsim \, 0.01 \, \rm Myr$, the shell begins to slow down and the time evolution of its position asymptotes towards $R_{\rm sh} \propto t^{3/5}$, marked with a dotted, blue line, as expected for the energy-driven phase. 
The transition between free-expansion and energy-driven phases is gradual, but it occurs at scales of about $R_{\rm free} \, \approx \, 40 \rm pc$ (marked with a gray, horizontal line), as anticipated (see Table~\ref{table:homogeneous}).

In order to further quantify the scale at which the wind thermalises, we search for the time and corresponding shell position $R_{\rm therm}$ at which wind material first exceeds a temperature $(1/2) T_{\rm R-shock}$.
The results are listed in Table~\ref{table:homogeneous}, where we can see that for most simulations, $R_{\rm therm}$ is indeed comparable to $R_{\rm free}$.
Note that for the simulations performed with $\beta \, = \, 0.1$, thermalisation, as we have defined it, also occurs at $R_{\rm therm} \sim R_{\rm free}$, as anticipated.

To see that the structure of the AGN-driven outflow changes after the small-scale wind thermalises, we plot radial profiles for different hydrodynamic quantities.
The right-hand panel of Fig.~\ref{fig_radtime} shows temperature radial profiles for \texttt{shell-L45-b0.02-n1} at three different times corresponding to $t \lesssim t_{\rm free}$ (dark blue, solid curve), $t \gtrsim t_{\rm free}$ (blue, dashed curve) and $t \gg t_{\rm free}$ (light blue, dotted curve). In all cases, the temperature profile takes the shape of a power law at the smallest radii, where the flow consists of the freely-expanding, isentropic wind (see Section~\ref{sec:radprof}). 
The expected free-expansion radius is shown with a vertical, red line.

\begin{figure*}
	\includegraphics[width=0.95\textwidth]{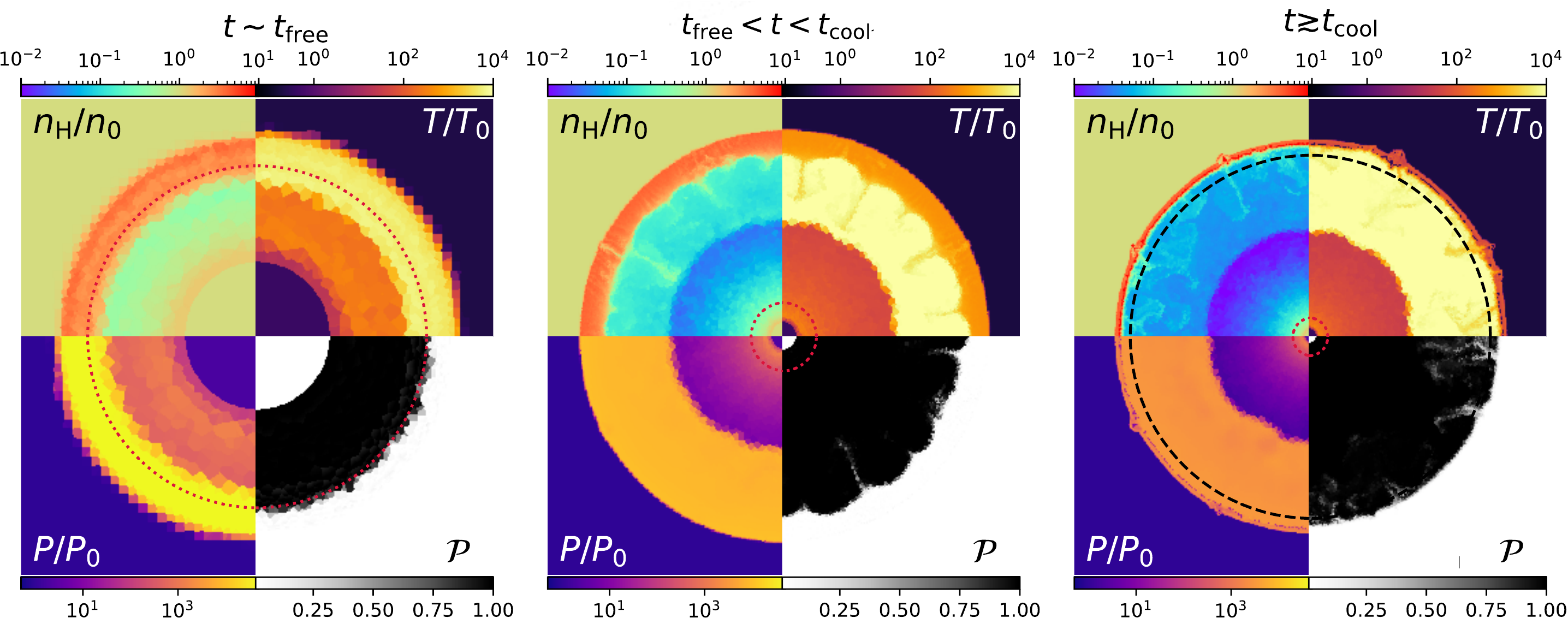}
	\caption{Time sequence showing density, temperature, pressure and wind tracer concentration in simulation \texttt{shell-L46-b0.02-n50}. The left-most panel shows the outflow configuration when the wind is passing the free-expansion radius (dotted circle). At this time, a \emph{strong} reverse shock has not yet formed and $T_{\rm R-shock} \approx T_{\rm F-shock}$. The central panel illustrates the outflow at a time $t_{\rm free} < t < t_{\rm cool}$. A strong reverse shock has generated a hot layer of shocked wind which is in pressure equilibrium with the cooler and denser shocked ambient medium layer. The right-most panel shows the outflow structure after the shocked ambient medium has passed its cooling radius (black circle). While the hot, shocked wind suffers no cooling losses, the shell of shocked ambient medium cools rapidly, collapsing into a thin, dense sheet. The ripples propagating along the cold outer envelope result from the onset of the Vishniac instability.}
\label{fig_timeseq},
\end{figure*}

At $t \lesssim t_{\rm free}$, the power law section of the flow is followed by a temperature peak. 
A closer look reveals that this in fact consists of two separate components: an outer layer with $T \gtrsim10^8 \, \rm K$ and a thinner, inner layer with $T \, \approx \, 6 \times 10^7 \, \rm K$ (see also left-hand panel in Fig.~\ref{fig_timeseq}).
The first temperature jump is associated with the wind shock, the second with the contact discontinuity (thin, gray line) and the third with the forward shock. 
At this early time, the shell of swept-up ambient gas still propagates at a speed which is only somewhat lower than $v_{\rm w}$. 
Its temperature is therefore only a factor $\approx 3$ lower than $T_{\rm F-shock} (\dot{R}_{\rm sh} = v_{\rm w}) \, \approx \, 6 \times 10^8 \, \rm K$.
Since the shell of swept-up ambient medium still has not decelerated significantly, the reverse shock is weak and its temperature is significantly lower than the expected $T_{\rm R-shock} \, \approx \, 3 \times 10^8 \, \rm K$.

At $t > t_{\rm free}$, the shape of the temperature peak reverses as the shocked wind becomes hotter than the shocked ambient medium.
A strong reverse shock becomes discernible in the right-hand panel of Fig.~\ref{fig_radtime}.
The temperature of the shocked wind behind it rises as the shell decelerates (see also Fig.~\ref{fig_timeseq}) and pushes into the wind more strongly.
As $t \gg t_{\rm free}$, we find that $T_{\rm R-shock} \, \approx \, 3 \times 10^8 \, \rm K$.
The temperature of the outer layer, on the other hand, falls off with radius, because the shell decelerates and the post-shock temperature of the gas drops (see Eq.~\ref{eq_fshock_T}).
Thus only at  $t \gg t_{\rm free}$ does the outflow structure takes one the classical energy-driven structure \citep{King:03,Faucher-Giguere:12} presented in Section~\ref{sec_anal}, with its inner hot, over-pressurised bubble and the outer, slower, higher-density shell.

In Fig.~\ref{fig_radtime}, the thin dashed line for $t \gtrsim t_{\rm free}$ shows the temperature profile in a simulation identical to \texttt{shell-L45-b0.02-n1}, but in which entropy conservation is enforced in the freely-expanding wind. In the supersonic section of the flow, the energy is dominated by the kinetic energy component. Due to explicit energy conservation in the default setup of our {\sc AREPO} simulations, even small errors in the kinetic energy estimation can lead to spurious entropy production \citep[see][]{Springel:10} and flatter temperature profiles than expected.
As can be seen in Fig.~\ref{fig_radtime}, this effect does not change the position of the shell and does not affect its dynamics.

Finally, we recall that $T_{\rm w} \, = \, 5 \times 10^5 \, \rm K$. The temperature of the freely-expanding wind seen in Fig.~\ref{fig_radtime}, however, is higher than this value. 
This overestimate is caused by small discretisation errors in the evaluation of kinetic energy. Since energy is explicitly conserved in {\sc AREPO}, these small errors show up as additional thermal energy.
In Appendix~\ref{sec:app:numerical}, we show that this issue is resolved with sufficient resolution and that as long as the actual wind temperature is $T \ll T_{\rm R-shock}$, the outflow dynamics is not affected.

\subsection{Radial profiles}
\label{sec:radprof}
\begin{figure}
	\includegraphics[width=0.495\textwidth]{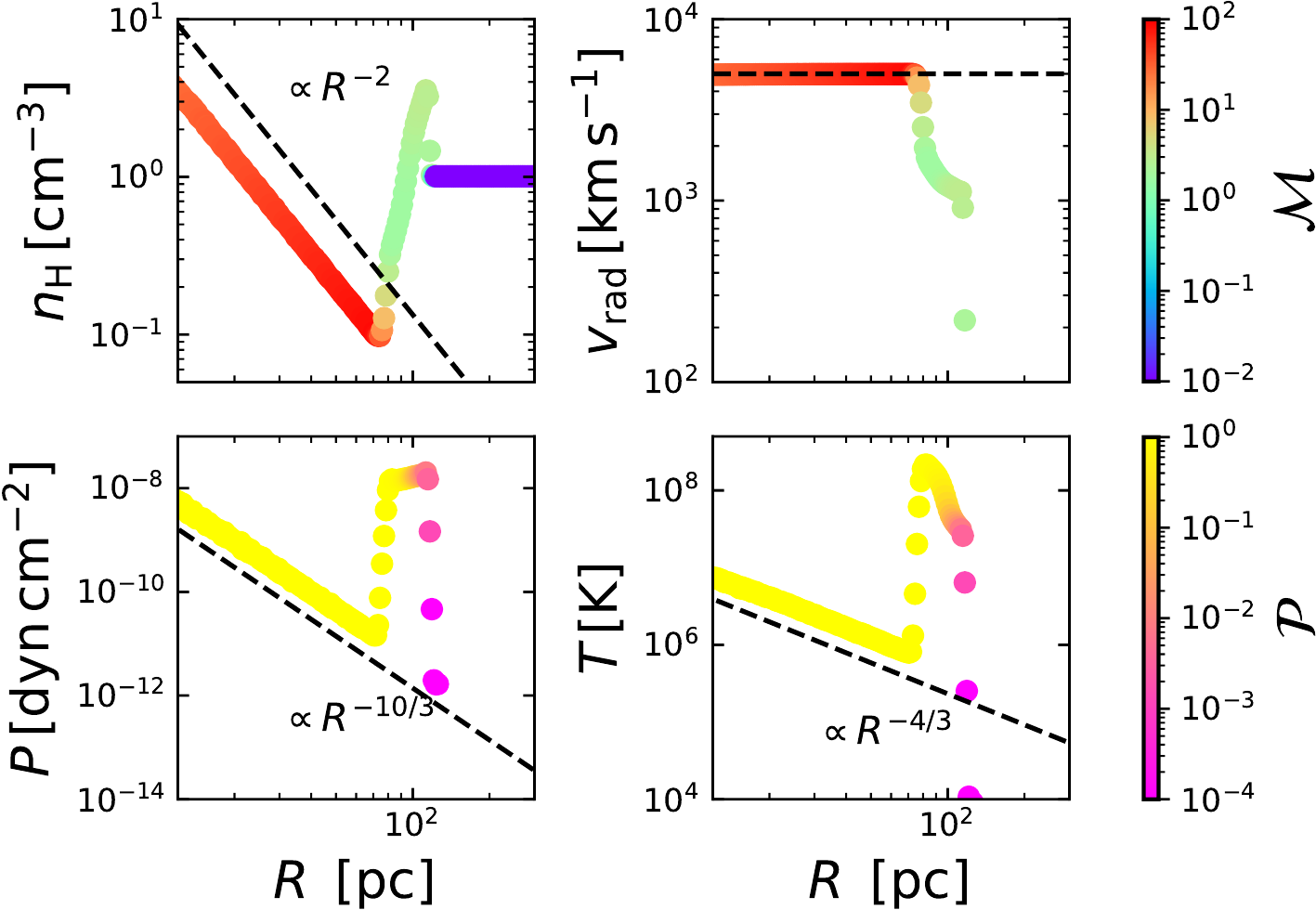}
	\caption{Radial profiles of hydrogen number density (top left), radial velocity (top right), pressure (bottom left) and temperature (bottom right) in \texttt{shell-L45-b0.02-n1} at $t \, \approx \, 0.048 \, \rm Myr \, = \, 6 t_{\rm free}$. The plot symbols are colour-coded according to local Mach number (top row) and to the local wind tracer concentration (bottom row). The scalings of the various hydrodynamic quantities with radius for the isentropic wind, the post-shock temperature of the different outflow components and the overall structure of the outflows all agree with analytical expectations.}\
	\label{fig_profiles}
\end{figure}

We present radial profiles of gas density, radial velocity, pressure and temperature in Fig.~\ref{fig_profiles} at $t \, \approx \, 0.048 \, \rm Myr \, = \, 6 t_{\rm free}$ for \texttt{shell-L45-b0.02-n1}.
In the top row, the plot symbols are colour-coded according to the local Mach number, while in the bottom row, they are colour-coded according to the local wind tracer concentration. 

At $t \, = \, 6 t_{\rm free}$, a strong reverse shock into the wind has already formed and the outflow has settled into the classical configuration assumed in \citet{King:03}, \citet{Faucher-Giguere:12} and \citet{Costa:14}.
The four expected flow regions are easily recognisable. At the smallest radii, we find the isentropic, freely-expanding wind.
The density indeed follows $\rho \propto r^{-2}$ and the radial velocity is spatially constant at $v_{\rm rad} \, \approx \, \beta c$, as expected. 
Also the pressure and temperature follow clear power laws, dropping as $P \propto r^{-2 \gamma}$ and $T \propto r^{-2 (\gamma - 1)}$, respectively, as expected for an adiabatic flow.

All hydrodynamic quantities then jump sharply at $R \, \approx \, 80 \, \rm pc$, the position of the reverse shock that slows down the AGN wind.
The Mach number changes abruptly from high values $\mathcal{M} \gg 1$ to $\mathcal{M} \lesssim 1$ behind the strong reverse shock, as expected.
At $R \, \approx \, 100 \, \rm pc$, there is a kink in the density and temperature profiles at which the wind tracer concentration drops to negligible values, marking a transition from shocked wind to shocked ambient medium phases.
The density and temperature jumps associated with the contact discontinuity are smoothed out due to significant mixing between wind and ambient medium fluids.
The last discontinuity in the radial profiles corresponds to the forward shock that propagates into the undisturbed ambient medium.
The gas density jumps by a factor $(\gamma + 1) / (\gamma - 1) \, = \, 4$, as expected for a strong, adiabatic shock, while the post-shock temperature of $T_{\rm F-shock} \, \approx \, 3 \times 10^7 \, \rm K$ is consistent with the expectation for a shock velocity $\approx 1100 \, \rm km \, s^{-1}$ (see Eq.~\ref{eq_fshock_T}).

In Fig.~\ref{fig_cooling} we presented evolutionary tracks for $\dot{R}_{\rm sh}$ as a function of radius for two different wind solutions.
The grey track concerns a wind with $v_{\rm w} \, = \, 5000 \, \rm km \, s^{-1}$ powered by an AGN with $L_{\rm AGN} \, = \, 10^{45} \, \rm erg \, s^{-1}$ propagating into a homogeneous medium with $n_{\rm H,0} \, = \, 1 \, \rm cm^{-3}$, the same parameters as  \texttt{shell-L46-b0.02-n1}.
The velocity at the position of the discontinuity is $\approx 1000 \, \rm km \, s^{-1}$, close, but somewhat lower than na{\"i}vely expected in Fig.~\ref{fig_cooling}, where $\dot{R}_{\rm sh} \approx 1800 \, \rm km \, s^{-1}$ at the same radius. This small inconsistency is likely caused by the idealisation that the shell moves at speed $\dot{R}_{\rm sh} \, = \, v_{\rm w}$ when $R < R_{\rm free}$ and that it instantly enters the energy-driven phase when $R = R_{\rm free}$. In reality, this transition is more gradual, as shown in Fig.~\ref{fig_radtime}.

Thus far, we demonstrated that the structure of the outflow and its dynamics, as captured in our model, are in agreement with the basic expectations outlined in Section~\ref{sec_anal}.
While we have focussed on \texttt{shell-L45-b0.02-n1} so far, we obtain similar results for our other simulations.
In Fig.~\ref{fig_timeseq}, we show slices at three different critical times during the outflow evolution for \texttt{shell-L46-b0.02-n50}.
As before, there is a short, initial free-expansion phase in which the wind, and the outflow as a whole, moves outwards at roughly the small-scale wind speed. 
The reverse shock is still weak and, therefore, there isn't a significant hot, shocked wind component (left-hand panel). 
Beyond the free-expansion radius (red, dotted circle), a strong shock forms, slowing down the wind, which thermalises and starts driving out the ambient medium through its pressure in an energy-driven outflow (middle panel).

As the shell crosses the cooling radius (black, dashed circle) for the shocked ambient medium (third panel), it collapses as its pressure drops due to effective radiative cooling.
Cooling is indeed expected in \texttt{shell-L46-b0.02-n50} (see Fig.~\ref{fig_cooling}).
The protuberances that can be seen at this point have significant azimuthal velocity components. They therefore ripple through the outflow's thin outer sheet, in a likely instance of the Vishniac instability \citep{Vishniac:83,Nayakshin:12}.

Instabilities form along the contact discontinuity in all our simulations at $t \lesssim t_{\rm free}$ (e.g. Fig.~\ref{fig_main}) at early times.
At $R \lesssim R_{\rm free}$, the shocked wind is colder and denser than the gas it encounters.
Since the shell of shocked ambient gas always decelerates, the flow is Rayleigh-Taylor unstable at $R \lesssim R_{\rm free}$ \citep[see e.g.][for a similar insight]{Gull:73}.
In our specific configuration, however, this instability is short-lived, because the shocked wind becomes less dense than the shocked ambient medium at $R \sim R_{\rm free}$.
 
\subsection{Radiative cooling in energy-driven shells}
\label{sec:cooling}

\begin{figure}
	\includegraphics[width=0.475\textwidth]{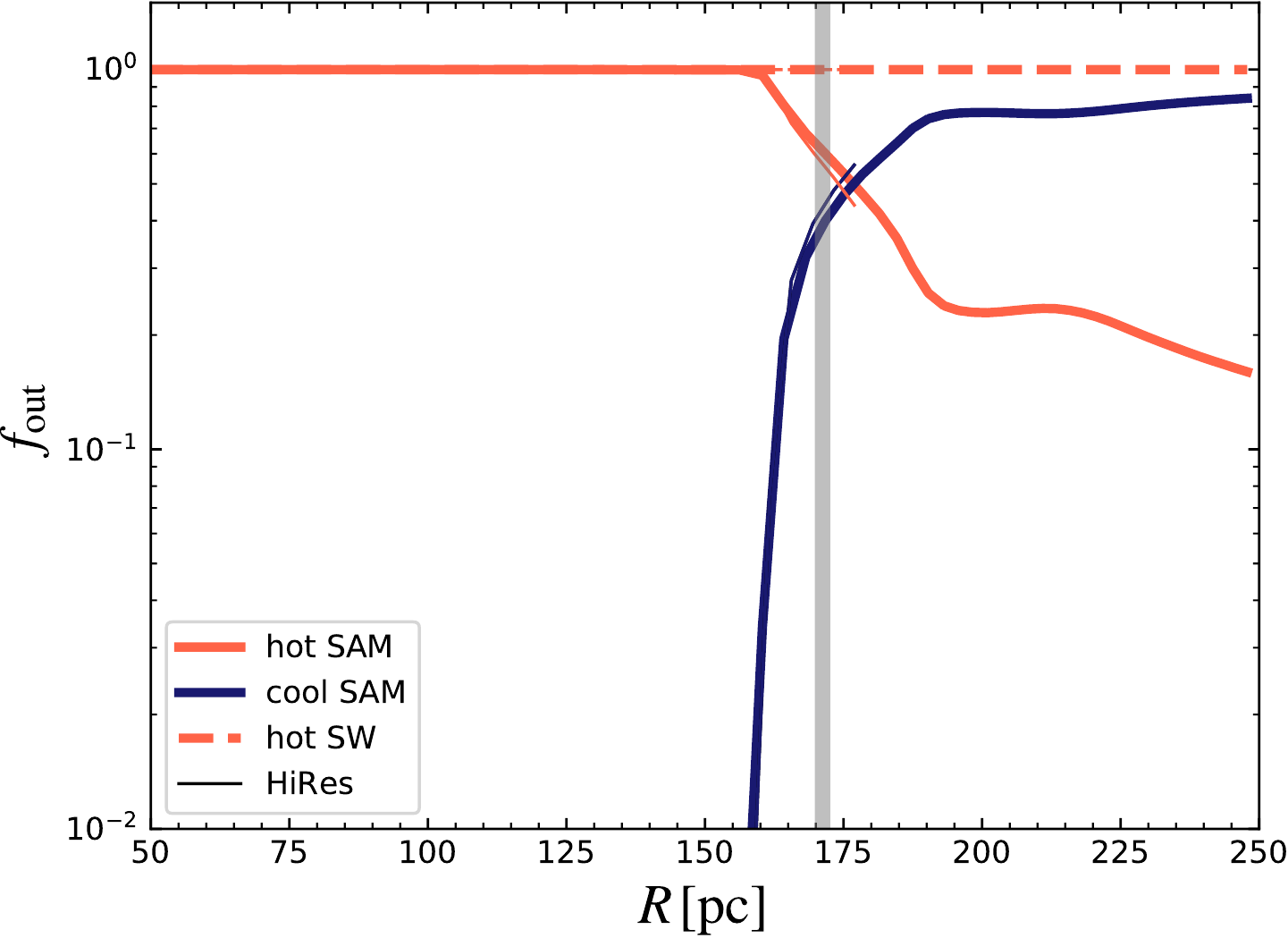}
	\caption{Mass fraction of the outflowing shocked ambient phase with temperature $T > 2 \times 10^4 \, \rm K$ (orange curves) and with $T \leq 2 \times 10^4 \, \rm K$ (blue curves) in \texttt{shell-L46-b0.02-n50}. Thick, solid curves give results for the simulation with $m_{\rm target} \, = \,10 \, \rm M_\odot$, while the thin curves give the results for $m_{\rm target} \, = \,1 \, \rm M_\odot$. The radius at which the shocked ambient medium cools down is numerically converged and agrees with the cooling radius derived analytically (vertical, gray line). The shocked wind medium (dashed, orange line) does not cool radiatively.}\
\label{fig_hotwarm}
\end{figure}

Based on Section~\ref{section_radiative_cooling}, we expect the shocked ambient medium to have cooled down in \texttt{shell-L46-b0.02-n50} by the time it has crossed a radial distance of $\approx 170 \, \rm pc$. In all simulations, we expect the reverse shock to be adiabatic and the outflows to be energy-driven. 

\subsubsection{Shocked ambient medium}
In order to isolate the shocked ambient medium, we select gas cells with $\mathcal{P} < 0.5$ and $v_{\rm r} > 10 \, \rm km \, s^{-1}$.
This selection filters out the un-shocked and shocked wind fluids as well as the undisturbed ambient medium. We then separate the shocked ambient medium component into a hot phase, which we define as having a temperature $> 2 \times 10^4 \, \rm K$, and a cool phase with $\leq 2 \times 10^4 \, \rm K$. In Fig.~\ref{fig_hotwarm}, we plot the mass fraction of both components as a function of the shell position in \texttt{shell-L46-b0.02-n50}.
The mass fraction of outflowing hot gas is shown with red, solid curves while the mass fraction in cool gas is shown in dark blue, solid curves.

We see that the shocked ambient medium outflow phase is entirely composed of hot gas in the innermost 150 pc. 
However, the proportion of hot gas eventually drops, while the proportion of cool gas rises. 
At $R \, \approx \, 170 \, \rm pc \, \approx \, R_{\rm cool}$, about half of the mass in the shell is in the cool phase. As the simulation progresses, the fraction of cold material continues to rise and, by $R \, = \, 250 \rm pc$, $90\%$ of the shocked ambient medium mass is cold.
Comparison with \texttt{shell-L46-b0.02-n50-HiRes} shows that the time at which radiative cooling begins to affect the structure is well converged.

The dashed curve in Fig.~\ref{fig_hotwarm} shows the evolution of hot gas mass for the shocked wind phase, which we identify as that with $\mathcal{P} > 0.5$.
As expected, this phase is always hot and does not cool effectively.
In \texttt{shell-L45-b0.02-n1}, however, the shell should not cool within 1 kpc and should remain hot in the whole simulated domain.
We verified that both the shocked wind and the shocked ambient medium remain hot throughout the entire simulation, for which $R_{\rm sh} \leq 0.7 \, \rm kpc$ due to the box size.

\subsubsection{Shocked wind and inefficient Compton cooling}
\label{sec:shockedwind}
The only plausible cooling channel for the shocked wind is inverse Compton scattering \citep{King:03}. 
In order to maximise the effects of Compton cooling, we use the relativistic expression for the cooling rate. The associated cooling rate per unit volume $V$ is given by
\begin{equation} 
\Lambda_{\rm cpt} / V \, = \, 4 n_{\rm e} \sigma_{\rm T} \frac{L_{\rm AGN}}{\pi R^2} \left( \frac{k_{\rm B} T}{m_{\rm e} c^2} \right)^2 \, ,
\label{eq_rel_compton}
\end{equation}  
and exceeds the non-relativistic expression by a factor $4 k_{\rm B} T / (m_{\rm e} c^2)$. We use Eq.~\ref{eq_rel_compton} only if gas has $k_{\rm B} T \geq m_{\rm e} c^2$ and Eq.~\ref{eq_NR_compton} otherwise.

\begin{figure}
	\includegraphics[width=0.475\textwidth]{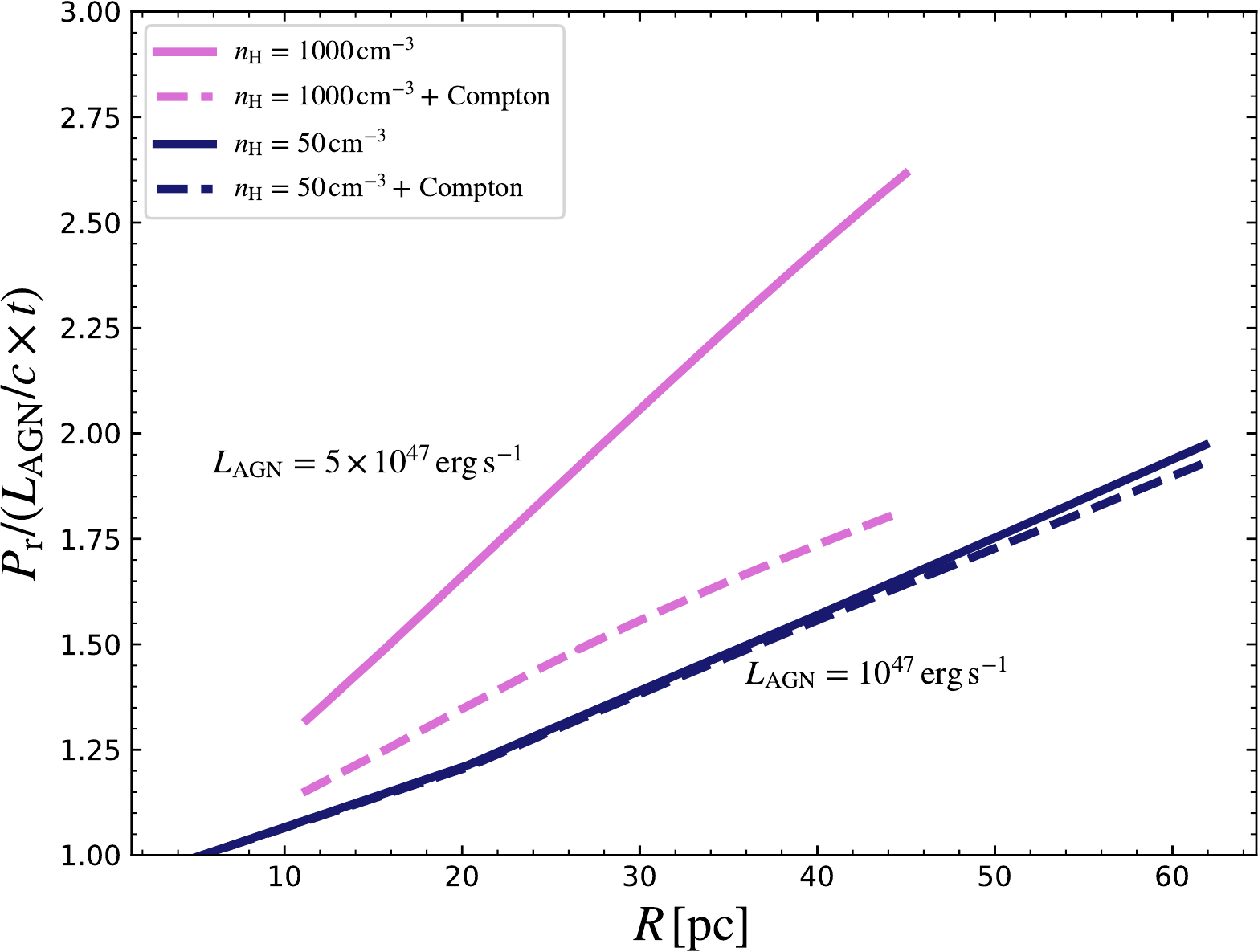}
	\caption{Total radial momentum input rate normalised by the total momentum injected by AGN radiation as a function of shell position in \texttt{shell-b0.1-L47-n50} (dark blue curves) and in \texttt{shell-b0.1-L5e47-n1000} (pink curves). Solid curves give the radial momentum evolution for simulations for which we do not follow Compton cooling, while dashed curves illustrate the effect of switching on Compton cooling. Switching on relativistic Compton cooling does not strongly affect the outflow solution even for $L_{\rm AGN} \, = \, 10^{47} \, \rm erg \, s^{-1}$ and $n_{\rm H} \, = \, 50 \, \rm cm^{-3}$ \-- the outflow is energy- rather than momentum-driven. Compton cooling only has an impact if the density of the ambient medium through which the wind propagates is high enough to ensure the wind thermalises rapidly, as seen for $L_{\rm AGN} \, = \, 5\times10^{47} \, \rm erg \, s^{-1}$ and $n_{\rm H} \, = \, 1000 \, \rm cm^{-3}$. Even in this case we see that $\dot{P}_{\rm r} > L_{\rm AGN}/c$ and the solution is not purely momentum-driven.}\
	\label{fig_compton}
\end{figure}

Eq.~\ref{eq_coolratio_R} shows that winds with high $\beta$ are the most likely to thermalise within the Compton cooling radius.
We therefore focus on our simulations with $\beta \, = \, 0.1$, recalling that for \texttt{shell-L47-b0.1-n50-Cpt}, $R_{\rm free} \, \approx \, 9.5 \, \rm pc$ and for \texttt{shell-L5e47-b0.1-n1000}, $R_{\rm free} \, \approx \, 4.8 \, \rm pc$.
From Eq.~\ref{eq_coolradius_cpt}, we can compute the Compton cooling radii as $R_{\rm c}^{\rm cpt} \, = \, 30 (\dot{R}_{\rm sh}/\mathrm{10^3 \, km \, s^{-1}})^{-1}\, \rm pc$ for \texttt{shell-L47-b0.1-n50-Cpt} and $R_{\rm c}^{\rm cpt} \, = \, 150 (\dot{R}_{\rm sh}/\mathrm{10^3 \, km \, s^{-1}})^{-1}\, \rm pc$ for \texttt{shell-L5e47-b0.1-n1000}.
Using the relativistic Compton cooling rate increases these cooling radii by a factor $9/16 (m_{\rm p} / m_{\rm e}) \beta^2 \approx 10 \left( \beta / 0.1\right)^2$.

Using the expected scaling for the free-expansion phase, $\dot{R}_{\rm sh} \approx v_{\rm w}$, and for the energy-driven phase, $\dot{R}_{\rm sh} \propto R_{\rm sh}^{-2/3}$, we obtain a mean velocity $\langle \dot{R}_{\rm sh} \rangle \, \approx \, 10^4 \, \rm km \, s^{-1}$ within $R_{\rm c}^{\rm cpt}$ in \texttt{shell-L47-b0.1-n50-Cpt} and $\langle \dot{R}_{\rm sh} \rangle \, \approx \, 2000 \, \rm km \, s^{-1}$ at $R_{\rm c}^{\rm cpt}$ in \texttt{shell-L5e47-b0.1-n1000-Cpt}.
Using Eq.~\ref{eq_coolradius_cpt}, gives $R_{\rm c}^{\rm cpt}/R_{\rm free} \approx 3$ and $R_{\rm c}^{\rm cpt}/R_{\rm free} \approx 160$, respectively.
Thus, in \texttt{shell-L47-b0.1-n50-Cpt}, Compton cooling should just about have an effect on the solution, while it should affect outflow dynamics significantly in \texttt{shell-L5e47-b0.1-n1000-Cpt}.

\begin{figure*}
	\includegraphics[width=0.325\textwidth]{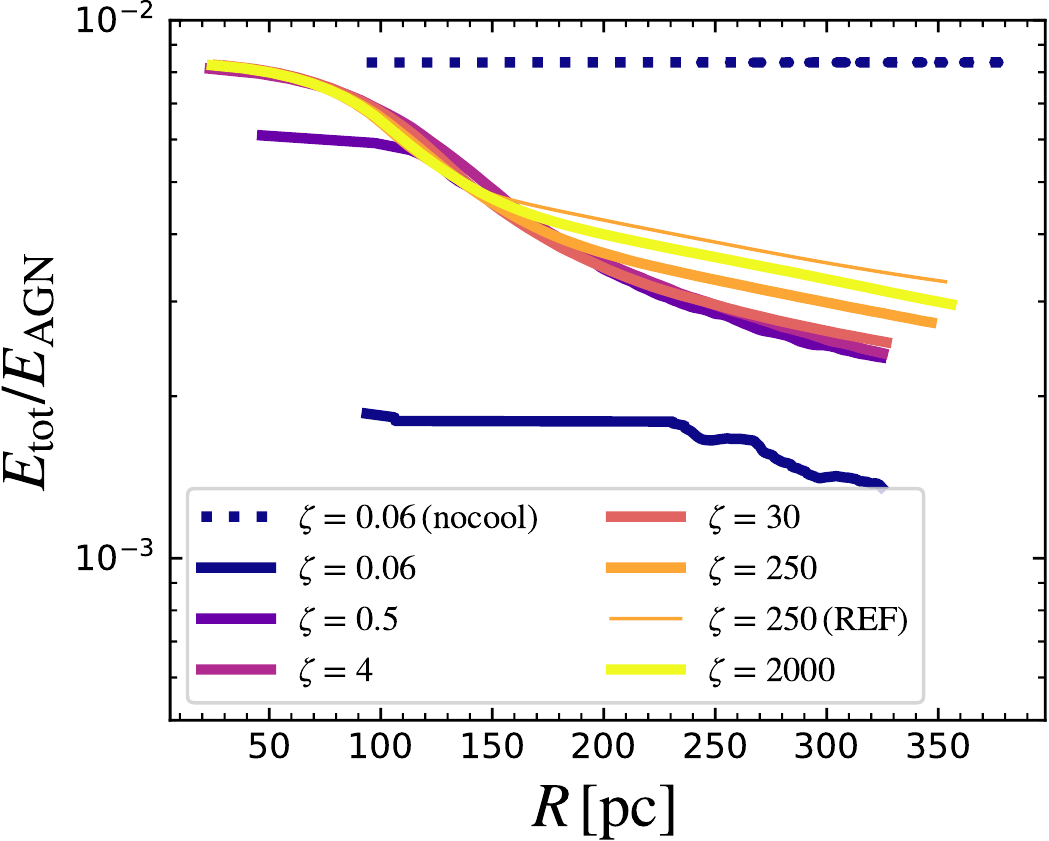}
	\includegraphics[width=0.325\textwidth]{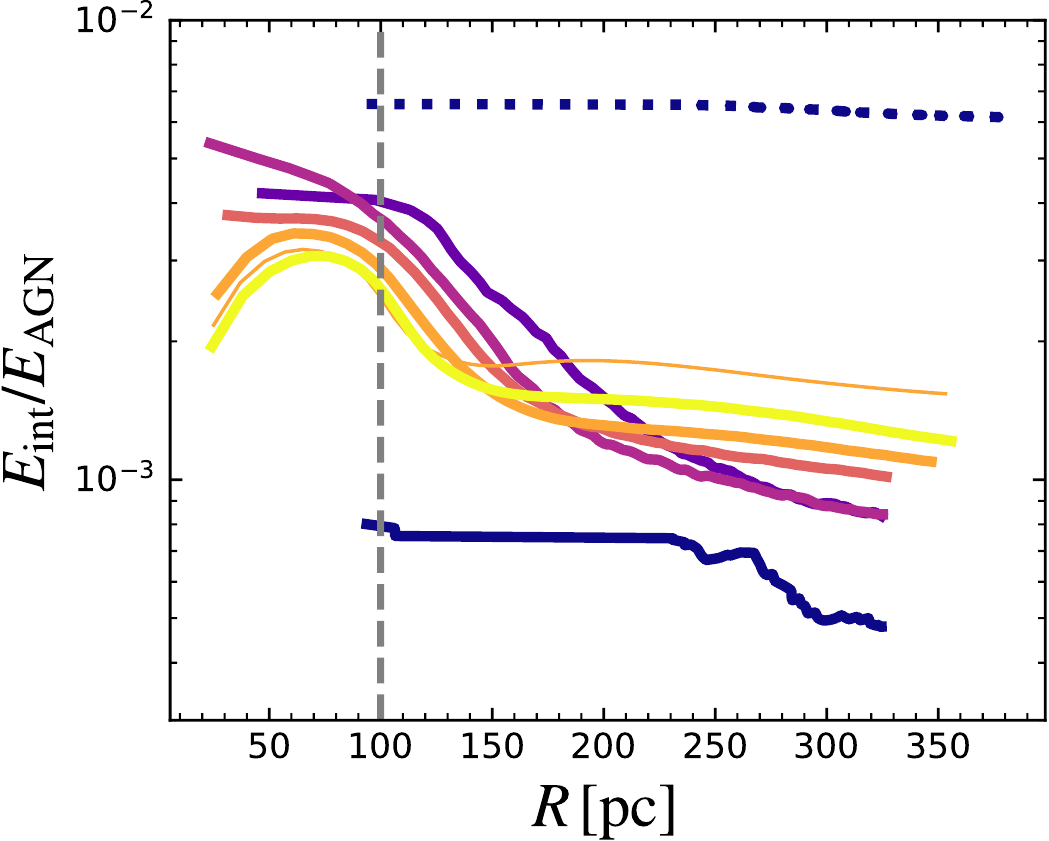}
	\includegraphics[width=0.325\textwidth]{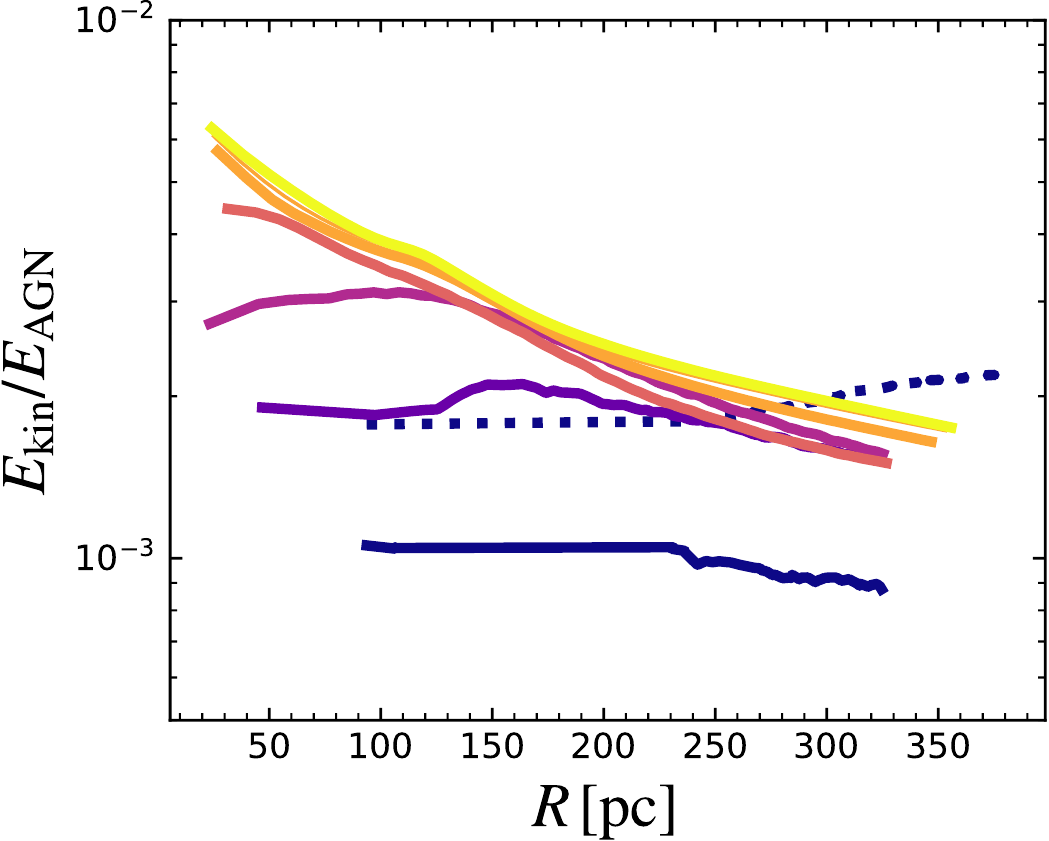}
	\caption{The evolution of the total (left-hand panel), thermal (middle panel) and kinetic (right-hand panel) energy in outflowing material, normalised by the net energy liberated by the AGN, as a function of shell position. Different curves illustrate the evolution for different values of $\zeta$, the number of cells with which the injected wind is resolved at thermalisation. The vertical, dashed line in the central panel shows the location of $R_{\rm cool}$. As long as $\zeta > 1$, the wind solution is well converged at small radii, but phase mixing between shocked wind and shocked ambient medium can result in spurious cooling (see text). Failing to resolve the wind shock radius, however, can lead to overcooling. In the example shown here (for $\zeta \, = \, 0.06$), poor resolution results in a factor $2 \-- 3$ lower total energy in the outflow.}
	\label{fig_energy_conv}
\end{figure*}

In Fig.~\ref{fig_compton}, we plot the total radial momentum input rate  $\dot{P}_{\rm r}$ normalised by total momentum injected by AGN radiation as a function of shell position in \texttt{shell-L47-b0.1-n50} (dark blue curves) and in \texttt{shell-L5e47-b0.1-n1000} (violet curves).
As expected, switching on relativistic Compton cooling (dashed lines) does not affect the outflow solution even for $L_{\rm AGN} \, = \, 10^{47} \, \rm erg \, s^{-1}$ and $n_{\rm H} \, = \, 50 \, \rm cm^{-3}$.
Compton cooling only has an impact if the density of the ambient medium through which the wind propagates is high enough to ensure the wind thermalises rapidly, as seen for $L_{\rm AGN} \, = \, 5\times10^{47} \, \rm erg \, s^{-1}$ and $n_{\rm H} \, = \, 1000 \, \rm cm^{-3}$. Even in this case we see that $\dot{P}_{\rm r} > L_{\rm AGN}/c$ and the solution is not purely momentum-driven.

Our results do not mean that momentum-driven solutions are impossible. 
Instead, they show that AGN-driven outflows progress from free-expansion to their energy-driven phase directly for a wide range of the parameter space.
Even if Compton cooling affects the solution, this may be only partially momentum-driven and still have $\dot{P}_{\rm r} > L_{\rm AGN}/c$.

\subsection{Convergence properties}
\label{sec_convergence}

The total wind mass injected at thermalisation is given by $M_{\rm free} \, \sim \, \dot{M}_{\rm w} t_{\rm free}$ and can be estimated using Eqs.~\ref{eq_wind_density} and~\ref{eq_tfree} as
\begin{equation}
M_{\rm free} \, = \, 57.5  \left( \frac{\beta}{0.1} \right) ^{-3} \left( \frac{\tau^3}{b} \right) ^{1/2} \left( \frac{L_{\rm AGN}}{10^{45} \, \mathrm{erg \, s^{-1}}} \right)^{3/2} \left( \frac{n_{\rm 0}}{ \mathrm{cm^{-3}}} \right)^{-1/2} \, \rm M_\odot \, .
\label{eq_mfree}
\end{equation} 
If the cell target mass $m_{\rm target} \gtrsim M_{\rm free}$, the free-expansion phase of the outflow cannot be resolved. Instead, as it moves across the wind injection boundary, the wind thermalises instantly.
The failure to separately resolve the shocked wind component, which is unlikely to cool (Section~\ref{section_radiative_cooling}), and the shocked ambient component, which can cool, may lead to numerical overcooling if $R_{\rm cool} \sim R_{\rm free}$.

We define $\zeta \, = \, M_{\rm free} / m_{\rm target}$ as the number of {\sc AREPO} cells with which the AGN-driven wind is resolved at thermalisation.
The expectation is that the wind solution converges as $\zeta \gg 1$, but that it diverges as $\zeta \lesssim 1$.
We again simulate the propagation of a spherical wind through a homogeneous medium using $n_{\rm H,0} \, = \, 300 \, \rm cm^{-3}$, $L_{\rm AGN} \, = \, 10^{47} \, \rm erg \, s^{-1}$ and $\beta \, = \, 0.017$ .
For these parameters, we obtain $M_{\rm free} \approx 5 \times 10^5 \, \rm M_\odot$, $R_{\rm cool} \, \approx \, 100 \, \rm pc$ and $R_{\rm free} \approx 24 \, \rm pc$.
We then vary the mass resolution in our various simulations.
For every factor of $8$ increase in the cell target mass $m_{\rm target}$, we increase $r_{\rm sp}$ by a factor of $2$, such that $r_{\rm sp}$ always matches the mean intercell distance at the beginning of the simulation.
For our highest resolution simulation, where $m_{\rm target} \, = \, 250 \, \rm M_\odot$, i.e. $\zeta \, = \, 2000$, we use $r_{\rm sp} \, = \, 2 \, \rm pc$. For the lowest resolution simulation, where $m_{\rm target} \, = \, 8.2 \times 10^6 \, \rm M_\odot$, i.e. $\zeta \, = \, 0.06$, we set $r_{\rm sp} \, = \, 64 \, \rm pc$.

Fig.~\ref{fig_energy_conv} shows the evolution of the total energy (left-hand panel), thermal energy (middle panel) and kinetic energy (right-hand panel) normalised by $E_{\rm AGN} \, = \, L_{\rm AGN} t$, as a function of the shell position.
Different curves illustrate how the energy evolution varies with $\zeta$.
For $\zeta > 1$, $E_{\rm tot} / E_{\rm AGN}$ starts at $\approx 0.008 E_{\rm AGN} \, \approx \, \eta E_{\rm AGN}$ and decays slowly with radius. 
As the wind thermalises at the smallest radii, the fractional internal energy $E_{\rm int} / E_{\rm AGN}$ rises, whereas the fractional kinetic energy $E_{\rm kin} / E_{\rm AGN}$ drops.
As $\zeta$ decreases, we find that $E_{\rm tot} / E_{\rm AGN}$ remains unchanged at small radii, though the fractional thermal energy becomes higher and the kinetic energy proportionally lower.
The reason for this behaviour is that, as $\zeta$ is reduced, the free-expansion phase of the wind is resolved more poorly. 
The cells comprising the wind interact with more massive cells and shock-heat just after injection.

\begin{figure}
	\includegraphics[width=0.475\textwidth]{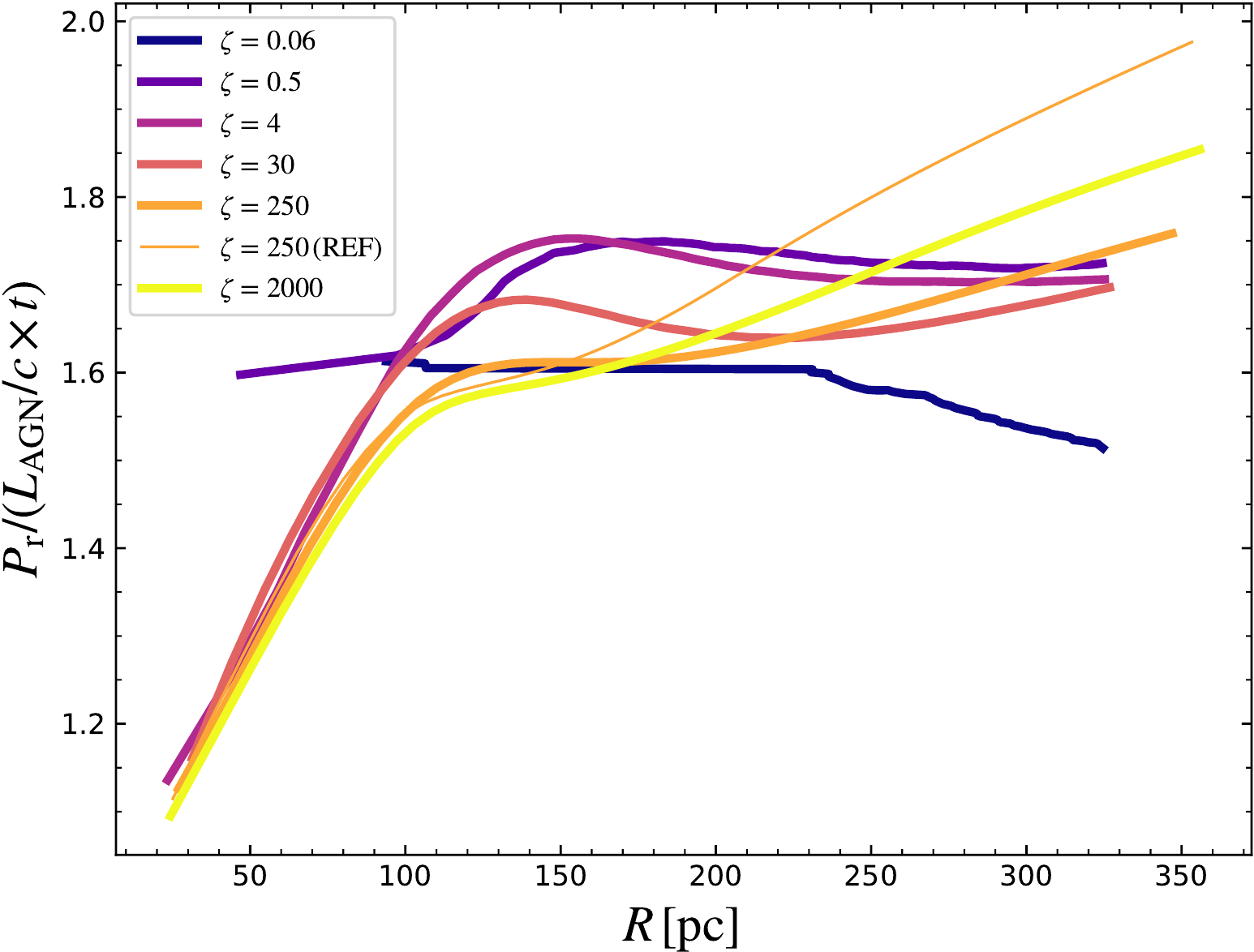}
	\caption{Total radial momentum normalised by $L_{\rm AGN} / c \times t$ as a function of the shell position for simulations with varying values for $\zeta$. Two numerical effects can suppress the net radial momentum of the outflow; (i) blending of shocked wind and shocked ambient medium components (see curve for $\zeta \, = \, 0.06$) and (ii) numerical mixing between shocked wind and shocked ambient medium phases (at large radii), both leading to artificial overcooling. The former can be resolved by increasing the mass resolution, while the latter can be mitigated by sharpening the resolution in regions with steep density gradients (thin curve).}
	\label{fig_ptot_conv}
\end{figure}

At $R_{\rm sh} \, \sim \, R_{\rm cool} \, \approx \, 100 \, \rm pc$, the shell of swept-up ambient medium cools radiatively and, accordingly, the fractional internal and total energies of the outflow drop.
At $R > R_{\rm cool}$, $E_{\rm int} / E_{\rm AGN}$ settles at an approximately constant value, as the internal energy of the shocked ambient medium continues to be radiated away rapidly, but preserved in the shocked wind phase.
Nevertheless, at $R > R_{\rm cool}$, the total, thermal and kinetic energies, as well as the total radial momentum (Fig.~\ref{fig_ptot_conv}) agree within a factor $2$ between simulations with $\zeta \, = \, 1$ and $\zeta \, = \, 2000$.

Another clear trend is the systematic underestimate in total, thermal and kinetic energy components at $R \gtrsim R_{\rm cool}$ as $\zeta$ decreases. This effect is most pronounced in the thermal energy curve.
After the outer shell cools radiatively, phase mixing between the shocked wind and shocked ambient medium components enhances radiative cooling in the shocked wind phase.
This effect can be mitigated either by increasing the resolution, and hence $\zeta$, or by adopting a more aggressive refinement strategy. 
The thin curves in Fig.~\ref{fig_energy_conv} show the energy evolution in a simulation with $\zeta \, = \, 800$, but where additional refinement is introduced to ensure that the volume ratios between adjacent cells do not vary by more than a factor $4$. 
This strategy leads to sharper resolution in regions with steep density gradients, such as across the contact discontinuity separating shocked wind and shocked ambient medium phases, and reduces the efficiency of mixing and hence shocked wind cooling.

As anticipated, we find poorer convergence if $\zeta < 1$ (dark blue curves in Figs.~\ref{fig_energy_conv} and~\ref{fig_ptot_conv}).
In this case, the hot, shocked wind bubble cannot be resolved at early times and is blended with outflowing material pertaining to the shocked ambient phase.
Since the latter cools efficiently, the effect of dropping $\zeta$ below unity is to suppress the energy-driven phase, causing the radial momentum to drop by a factor $\approx 1.4$ at the end of the simulation. 
Within the simulated domain, we thus see changes in energy and momentum of a factor $\lesssim 2 \--  3$ by varying the mass resolution by almost $4$ orders of magnitude.
Our model therefore yields wind solutions which are robust to changes in the resolution.

We also performed a similar resolution study for wind solutions with $\beta \, = \, 0.1$, where it is typically harder to resolve free-expansion (Eq. ~\ref{eq_mfree}).
Perhaps surprisingly, we find similarly good convergence as for solutions with $\beta \, = \, 0.017$. 
The higher energy injection rate in wind solutions with $\beta \, = \, 0.1$, however, offsets cooling losses more effectively and compensates for the lower $M_{\rm free}$ values.

\section{Outflows from disc galaxies}
\label{sec:discs}

In this Section, we investigate how small-scale winds power galactic super-winds in systems comprising a gaseous disc embedded in a homogeneous, spherically symmetric galactic halo.
Our aims are (i) to illustrate how the structure of large-scale outflows is modified by this more complex density field, (ii) to evaluate how properties of large-scale outflows vary with respect to the properties of the small-scale winds, (iii) to identify the channels via which AGN winds modify star formation, and (iv) to further test whether our new AGN wind model can indeed be reliably applied at the more typical resolution reached in galaxy formation simulations.   

\subsection{Numerical Setup}

\subsubsection{NFW halo and disc}
The dark matter component is modelled as a static Navarro-Frenk-White (NFW) potential \citep{Navarro:97} with halo concentration $C \, = \, 7.2$ and mass $(1 - f_{\rm gas}) M_{\rm 200}$, where $f_{\rm gas} \, = \, 0.17$ and $M_{\rm 200} \, = \, 10^{12} \, \rm M_\odot$.
A rotating spherical gas cloud with dimensionless spin parameter $\lambda \, = \, 0.05$ and mass $f_{\rm gas} M_{\rm 200}$ is placed at the centre of the dark matter potential.
The cloud's density profile follows the same NFW profile as the dark matter component.

Radiative cooling, which is modelled simply for a primordial mixture of H and He in photoionisation equilibrium with the \citet{Faucher-Giguere:09} UV background at $z \, = \, 0$, triggers a cooling flow at the start of the simulation.
As a result, the spinning gas cloud collapses towards the centre of the halo, where it settles into a disc.
Star formation is treated following \citet{Springel:03}, where the effects of unresolved physical processes operating within the interstellar medium (ISM), including thermal instability, evaporative heating of cold gas clouds and heating due to supernova explosions, are captured by an effective equation of state that is applied to all gas with $n_{\rm H} > n_{\rm th}$.
This effective equation of state is stiffer than that of isothermal gas, because it accounts for additional pressure provided by supernova explosions within the ISM.

Stellar particles are spawned stochastically from gas with $n_{\rm H} > n_{\rm th}$ at a rate
\begin{equation}
\frac{d \rho_{\rm \star}}{dt} \, = \, (1 - \beta) \frac{\rho_{\rm c}}{t_{\rm \star}} \, ,
\end{equation}
where $\beta$ is the mass fraction of massive stars assumed to instantly explode as supernovae, $\rho_{\rm c}$ is the density of cold clouds \citep[see][for details]{Springel:03} and $t_{\rm \star} \, = \, t_{\rm \star}^{\rm 0} \left(n / n_{\rm th} \right)^{-1/2}$ is the star formation timescale.
In our simulations, we adopt $\beta \, = \, 0.1$, $n_{\rm th} \, = \, 0.28 \, \rm cm^{-3}$ and $t_{\rm \star}^{\rm 0} \, = \, 1.5 \, \rm Gyr$.

We do not model supernova-driven winds. Our simulations should be regarded as idealised experiments aimed at illustrating how small-scale AGN winds drive large-scale outflows from systems comprising a galactic disc. They are designed to enable an accurate evaluation of the outflow energy and momentum contents and their effect on star formation, without `contamination' or non-linear coupling with supernova-driven winds \citep{Costa:15, Biernacki:18}.

While we do probe the effects of AGN winds in our simulations (see below), we do not model the gravitational field generated by the supermassive black holes. Since we neglect explicit supernova feedback, the central density, velocity dispersion and gravitational potential gradient are likely higher than they realistically should. In this study, however, we focus on  the differential effect of varying AGN luminosity and duty cycle in otherwise identical simulations, such that we quantify the importance of these parameters in shaping outflow dynamics. Rigorous, quantitatively-accurate predictions for e.g. the properties of outflows will require accurately modelling supernova feedback, including the gravitational potential of the black hole as well as black hole accretion (see Section~\ref{sec_plasma}).

\subsubsection{AGN wind injection boundary}
In order to model small-scale AGN winds, a spherical wind injection boundary, as described in Section~\ref{sec:windmodel}, is introduced at the location of the dark matter potential minimum in the initial conditions. 
We set $n_{\rm side} \, = \, 12$, such that each of the two layers used to define the wind injection boundary is sampled with $1728$ Voronoi cells.
The radius of the inner spherical layer is set to $7.5 \, \rm pc$ and of the outer layer to $12.5 \, \rm pc$, such wind injection occurs at radius $R_{\rm inj} \, = \, 10 \, \rm pc$. 
In our fiducial simulations, the mean radius of gas cells within $100 \, \rm pc$ of the potential minimum is $\approx 8 \, \rm pc$ and thus wind injection occurs at a scale commensurate with the size of the smallest Voronoi cells in our simulations. 

In cases where large masses are ejected from the nucleus, the global potential minimum may deviate from the dark matter potential minimum. 
In order to ensure that the wind injection boundary remains centred at the global potential minimum, we impart a small drift velocity to the cells comprising the two wind injection layers.
The drift velocity is directed towards the global potential minimum, which is computed once every global time-step, and is chosen to have a magnitude of $v_{\rm drift} \, \lesssim \, 20 \, \rm km \, s^{-1}$, comparable to the lowest sound speeds found in the galactic disc. 
We find that the wind injection boundary does not drift more than $\approx 50 \, \rm pc$ from the dark matter potential minimum in our simulations.
 
\subsubsection{Simulations}
In one of our simulations, we do not inject an AGN wind across the spherical boundary. This simulation, referred to as \texttt{disc-noAGN}, illustrates the evolution of the disc galaxy when no AGN feedback is present. It is used as reference e.g. when we test the ability of small-scale winds to drive galactic outflows (Section~\ref{sec:outflowprops}) or when we quantify the impact of outflows on the host galaxy (Section~\ref{sec:impact}). 

\begin{figure*}
	\includegraphics[width=0.985\textwidth]{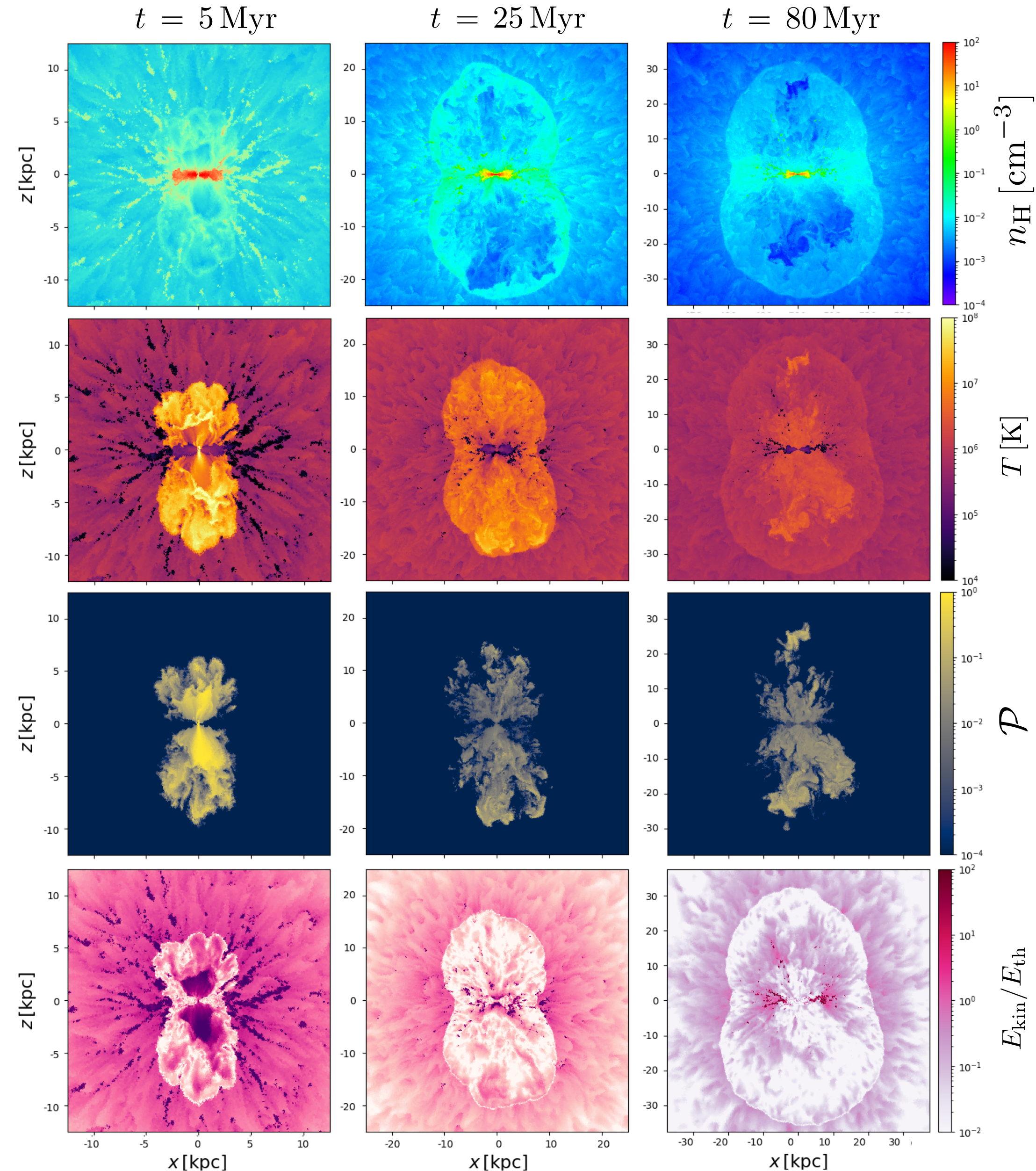}
	\caption{Time-sequence showing density (first row), temperature (second row), wind tracer (third row) and $E_{\rm kin}/E_{\rm th}$ (fourth row) slices through the simulation with $v_{\rm w} \, = \, 5000 \, \rm km \, s^{-1}$ and $L_{\rm AGN} \, = \, 10^{46} \, \rm erg \, s^{-1}$. Spherical small-scale winds power energy-driven bubbles which break-out of the galactic disc, expanding through the halo in large-scale bipolar outflows. As the outflow propagates to large radii, it isotropises coupling to most of the gaseous halo and interrupting gas inflow.}
	\label{fig_outflowevolution}
\end{figure*}

In all other simulations, we include an AGN wind. Wind injection is chosen to start at $t \, = \, 150 \, \rm Myr$, shortly after the star formation history rate reaches its peak.
We assume that an AGN wind injection episode lasts $t_{\rm AGN} \, = \, 5 \, \rm Myr$ and that episodes are cyclical.
We define the duty cycle $\eta_{\rm duty}$ as the time fraction of each cycle during which wind injection occurs.
Each cycle then has a period $t_{\rm cycle} \, = \, t_{\rm AGN} / \eta_{\rm duty}$.

We sample duty cycles in the range $\eta_{\rm duty} \, = \, (0.001, 0.5)$. In our simulation with $\eta_{\rm duty} \, = \, 0.5$, we adopt $L_{\rm AGN} \, = \, 10^{45} \, \rm erg \, s^{-1}$, such that by $t \, = \, 500 \, \rm Myr$, an energy $E_{\rm AGN} \approx 10^{61} \, \rm erg$ would have been liberated in the form of radiation by the AGN. 
For every duty cycle choice, we rescale the AGN luminosity, such this $E_{\rm AGN}$ remains fixed. 
Thus, for $\eta_{\rm duty} \,  = \, 0.1$, for example, we employ $L_{\rm AGN} \, = \, 5 \times 10^{45} \, \rm erg \, s^{-1}$, whereas for $\eta_{\rm duty} \,  = \, 0.005$, we use $L_{\rm AGN} \, = \, 10^{47} \, \rm erg \, s^{-1}$.

We explore two different wind speeds.
In most simulations, we adopt $v_{\rm w} \, = \, 5000 \, \rm km \, s^{-1}$, i.e. $\beta \, = \, 0.017$, while in others we set $v_{\rm w} \, = \, 30000 \, \rm km \, s^{-1}$, i.e. $\beta \, = \, 0.1$. 
In all cases, the winds are injected isotropically, i.e. $b \, = \, 1$, with $\tau \, = \, 1$ and $T_{\rm w} \, = \, 10^6 \, \rm K$.

The target mass resolution is $m_{\rm target} \, = \, 1.6 \times 10^5 \, \rm M_\odot$ in our fiducial simulations. 
We also perform higher-resolution simulations with $m_{\rm target} \, = \, 1.6 \times 10^4 \, \rm M_\odot$ and a few low resolution simulations with $m_{\rm target} \, = \, 1.6 \times 10^6 \, \rm M_\odot$.
Since the density of the ambient medium around the injection boundary at $t \, = \, 150 \, \rm Myr$ is $n_{\rm H, 0} \sim 10^3 \, \rm cm^{-3}$, the ratio between the wind mass at thermalisation and the mass resolution is $\zeta < 1$ in many for our simulations.
In some simulations, $\zeta > 10$, such that free-expansion can be resolved.
The simulation with $L_{\rm AGN} \, = \, 10^{46} \, \rm erg \, s^{-1}$, in particular, is performed at three different resolutions, where $\zeta \, \approx \, 1$ is achieved in the highest-resolution simulation, which allows to quantify the effect of not directly resolving free-expansion.
The most important parameters of our simulations are summarised in Table~\ref{table:disc}.

\begin{table}
\small\addtolength{\tabcolsep}{-3pt}
\caption{List of simulations and parameters. The first column lists the simulation names, the second shows the AGN luminosity, the third gives the small-scale wind velocity in units of c, the fourth gives the assumed AGN duty cycle, the fifth shows the number of cells with which the wind is resolved at thermalisation and the last column lists the cell target mass.}
\begin{center}
\begin{tabular}{ lccccccc| }
  \hline
  Simulation & $L_{\rm AGN}$     & $\beta$ & $\eta_{\rm duty}$ & $\zeta$ & $m_{\rm target}$ \\
                    & $\rm [ erg \, s^{-1}] $ & & & & $\rm [M_\odot]$ \\
  \hline
  \texttt{disc-noAGN}        &  \-- & \-- & \-- & \-- & $1.6 \times 10^5$ \\
  \texttt{disc-L45-b0.02}    &  $10^{45}$ & $0.017$ & $0.5$   & $0.002$  & $1.6 \times 10^5$ \\
  \texttt{disc-L5e45-b0.02}    &  $5 \times 10^{45}$ & $0.017$ & $0.1$ & $0.03$  & $1.6 \times 10^5$ \\
  \texttt{disc-L46-b0.02}    &  $10^{46}$ & $0.017$ & $0.05$ & $0.1$   & $1.6 \times 10^5$ \\
  \texttt{disc-L47-b0.02}    &  $10^{47}$ & $0.017$ & $0.005$ & $2$   & $1.6 \times 10^5 $ \\
  \texttt{disc-L5e47-b0.02}    &  $5 \times 10^{47}$ & $0.017$ & $0.001$ & $30$   & $1.6 \times 10^5$ \\
  \texttt{disc-L45-b0.1}      &  $10^{45}$ & $0.1$     & $0.5$   & $10^{-5}$   & $1.6 \times 10^5$ \\
  \texttt{disc-L46-b0.1}      &  $10^{46}$ & $0.1$  & $0.05$  & $4 \times 10^{-4}$   & $1.6 \times 10^5$ \\
  \texttt{disc-L45-b0.02-H}      &  $10^{45}$ & $0.017$  & $0.5$  & $0.05$   & $1.6 \times 10^4$ \\
  \texttt{disc-L46-b0.02-H}      &  $10^{46}$ & $0.017$  & $0.05$  & $1$   & $1.6 \times 10^4$ \\
  \texttt{disc-L46-b0.02-L}      &  $10^{46}$ & $0.017$  & $0.05$  & $0.01$   & $1.6 \times 10^6$ \\
  \hline
\end{tabular}
\end{center}
\label{table:disc}
\end{table}

\subsection{Large-scale outflows from small-scale winds}
Fig.~\ref{fig_outflowevolution} illustrates the time-evolution of different hydrodynamic quantities after the onset of AGN feedback in simulation \texttt{disc-L46-b0.02-H}. 
We show gas density (first row), temperature (second row), wind tracer (third row) and the ratio between radial kinetic energy and thermal energy $E_{\rm kin} / E_{\rm th}$ (fourth row) slices.
The properties of the small-scale wind at injection are, to a large extent, `forgotten' by the large-scale outflow \citep[see also][]{Costa:14, Costa:15, Nelson:19}. 
The wind, which is injected isotropically and at constant speed, powers an anisotropic, large-scale outflow with a more complex velocity structure.
The approximate axisymmetry of the gas disc, which is seen edge-on, collimates the outflow towards the galactic poles, which becomes bipolar as a result \citep[see also][]{Zubovas:12c, Gabor:14, Costa:15, Curtis:16, Hopkins:16, Hartwig:18}.
At scales much larger than the disc radius, however, the outflow isotropises, as the forward shocks on the outer rim of both bubbles overlap, forming a lateral shock that propagates along the disc plane. 

As in the spherical case (Section~\ref{sec:testmodel}), the outflow can be separated into different flow zones. 
At the smallest scales and while $t < t_{\rm AGN}$, we find a small biconical region enclosing the freely-expanding (unshocked) AGN wind (phase 1 in Fig.~\ref{fig_main}), appearing as a smooth flow with high $E_{\rm kin} / E_{\rm th}$ at early times. 
The AGN wind then passes the reverse shock, where the temperature jumps to $T \approx 3 \times 10^8 \, \rm K$ and $E_{\rm kin} / E_{\rm th}$ decreases to values $< 1$. 
Behind the reverse shock, the shocked wind (phase 2 in Fig.~\ref{fig_main}), contained in the hot, turbulent plumes seen above and below the disc, rises through the galactic halo, expanding adiabatically. 
Since they both consist of wind fluid, both these phases of the outflow are well populated by the wind tracer, shown in the third row.
The outer, denser and slightly less warm layer, which does not contain wind tracer fluid, instead consists of swept-up ambient gas (phase 3 in Fig.~\ref{fig_main}).
This phase is clearest in the third column of Fig.~\ref{fig_outflowevolution}.
It is bounded on the outside by the forward shock, the thin, sharp layer across which both temperature and density jump sharply and kinetic energy is dissipated.

\begin{figure*}
	\includegraphics[width=0.995\textwidth]{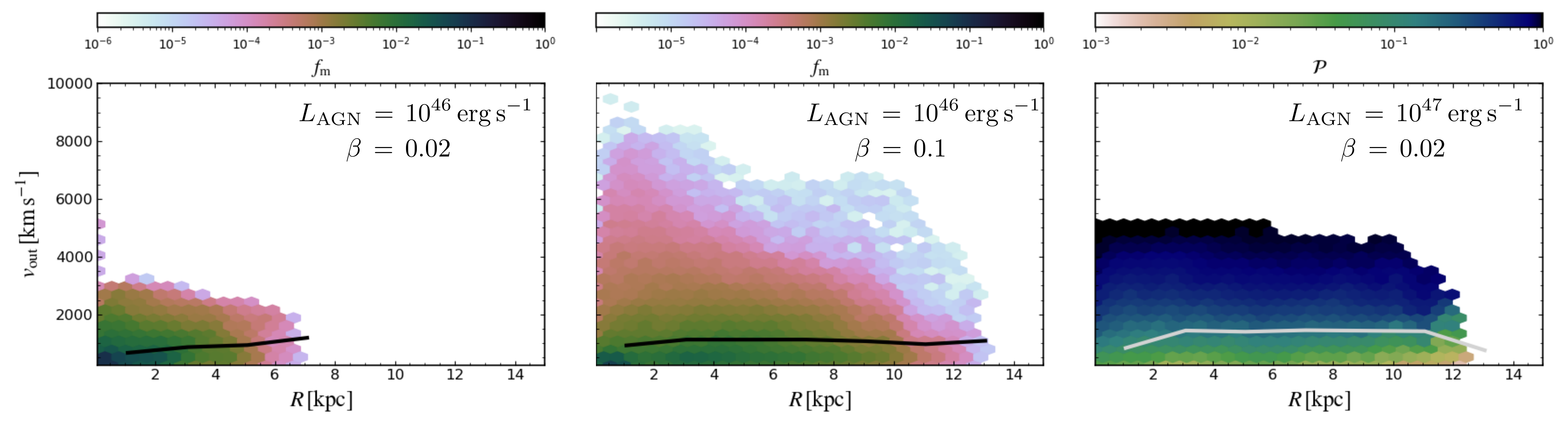}
	\caption{The distribution of gas radial velocity as a function of radial distance from the AGN in \texttt{disc-L46-b0.02} (left-hand panel), \texttt{disc-L46-b0.1} (central panel) and \texttt{disc-L47-b0.02} (right-hand panel) during the time period $t_{\rm AGN}$ over which an AGN wind is being injected. In colour, we show the relative mass contribution to each bin in the left-hand and central panels and the mean wind tracer concentration in the right-hand panel. The thick lines show the mass-weighted mean velocity at every radius. The velocity structure of the resulting large-scale outflows is complex, but various simple trends emerge: the brighter the AGN or the faster the small-scale wind, the faster and more spatially-extended the large-scale outflow becomes.}
	\label{fig_outflowrad}
\end{figure*}

We next take a closer look at outflow properties (velocity structure, momentum and kinetic energy contents) as well as the conditions that need to be fulfilled in order for the small-scale wind to power a galactic outflow. 

\subsection{Outflow properties and conditions for launch}
\label{sec:outflowprops}
The properties of the small-scale wind are set by the fluxes enforced at the spherical boundary.
While guided by observational and theoretical constraints (Section~\ref{sec_choiceparam}), the mass, momentum and energy fluxes associated with the small-scale winds are `put-in by hand'.
The properties of the large-scale outflows that develop as the small-scale wind interacts with surrounding gas, on the other hand, constitute simulation predictions.

In Fig.~\ref{fig_outflowrad}, we plot the distribution of gas radial velocities as a function of radial distance from the AGN in three representative simulations of our set. 
We stack the distributions from all the snapshots in the time interval $\left[ t_{\rm 0}, t_{\rm 0} + t_{\rm AGN} \right]$, corresponding to the first outflow episode.
The left-hand panel concerns the simulation with $L_{\rm AGN} \, = \, 10^{46} \, \rm erg \, s^{-1}$ and $v_{\rm w} \, = \, 5000 \, \rm km \, s^{-1}$, the central panel shows results for the simulation with $L_{\rm AGN} \, = \, 10^{46} \, \rm erg \, s^{-1}$, but with a faster wind speed of $v_{\rm w} \, = \, 30000 \, \rm km \, s^{-1}$, while the right-hand panel corresponds to the simulation with a brighter AGN with $L_{\rm AGN} \, = \, 10^{47} \, \rm erg \, s^{-1}$ and $v_{\rm w} \, = \, 5000 \, \rm km \, s^{-1}$. 

In all simulations and at all scales, the spread in outflow velocity at fixed radius is very large, in stark contrast with the homogeneous medium scenario \citep[see also][]{Costa:14}.
The bulk of the outflow has typical speeds $v_{\rm out} \, = \, 1000 \-- 2000 \, \rm km \, s^{-1}$, with a sparsely populated tail approaching $v_{\rm out} \approx v_{\rm w}$.
Significant masses moving at speeds exceeding $1000 \, \rm km \, s^{-1}$ are present all the way out to radii $5 \-- 15 \, \rm kpc$ for all cases shown.

Fig.~\ref{fig_outflowrad} allows us to identify various clear trends:
\begin{itemize}
\item At fixed small-scale wind speed, the maximum, and to a smaller extent the mean (solid curve), velocities of the large-scale outflows increase with AGN luminosity, as seen by comparing the leftmost and rightmost panels.
\item At fixed AGN luminosity, the maximum, and to a smaller extent the mean (solid curve), velocities of the large-scale outflows increase with the speed of the small-scale wind, as can be concluded from comparison between the leftmost and central panels.
\item The brighter the AGN, the more spatially extended the large-scale outflow. Similarly, the faster the small-scale wind, the more spatially extended the large-scale outflow.
\item While some gas remains ultra-fast, with $v_{\rm out} > 3000 \, \rm km \, s^{-1}$ out to $\sim \, \rm kpc$ scales, the maximum outflow velocity tends to fall with radius.
\item The highest velocity component is associated with pure wind fluid with $\mathcal{P} \approx 1$ and with a hybrid phase consisting of a mixture of ambient and wind gas with $\mathcal{P} \gtrsim 0.1$. Gas with lower wind concentration is slower, typically with  $v_{\rm out} \lesssim 2000 \, \rm km \, s^{-1}$.
\end{itemize} 

All these trends are consistent with the expectation that outflows become faster and more mass-loaded if the AGN injects more energy.
This can be achieved either by raising the speed of the small-scale wind or the AGN luminosity.

\begin{figure*}
	\includegraphics[width=0.32\textwidth]{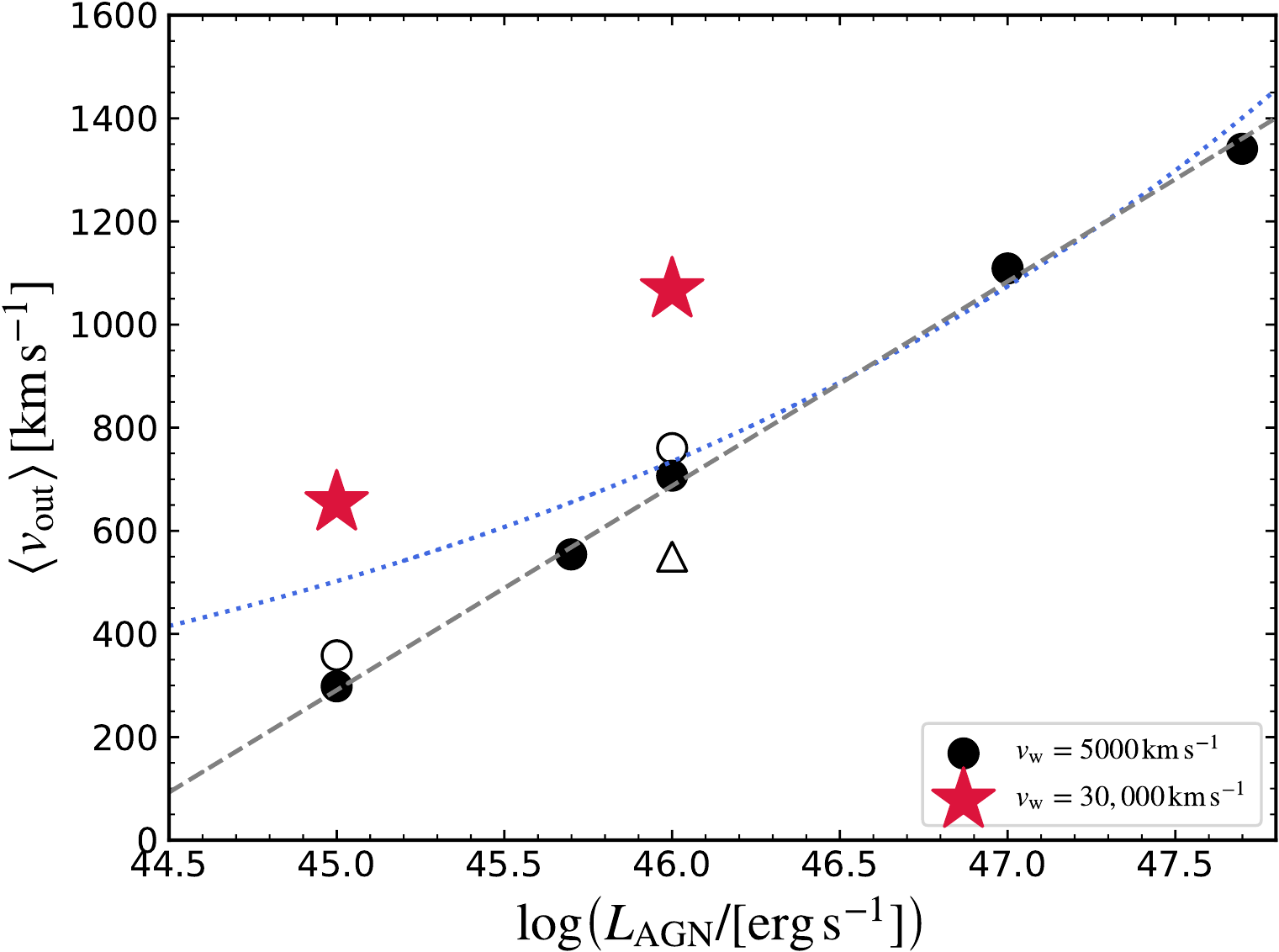}
	\includegraphics[width=0.32\textwidth]{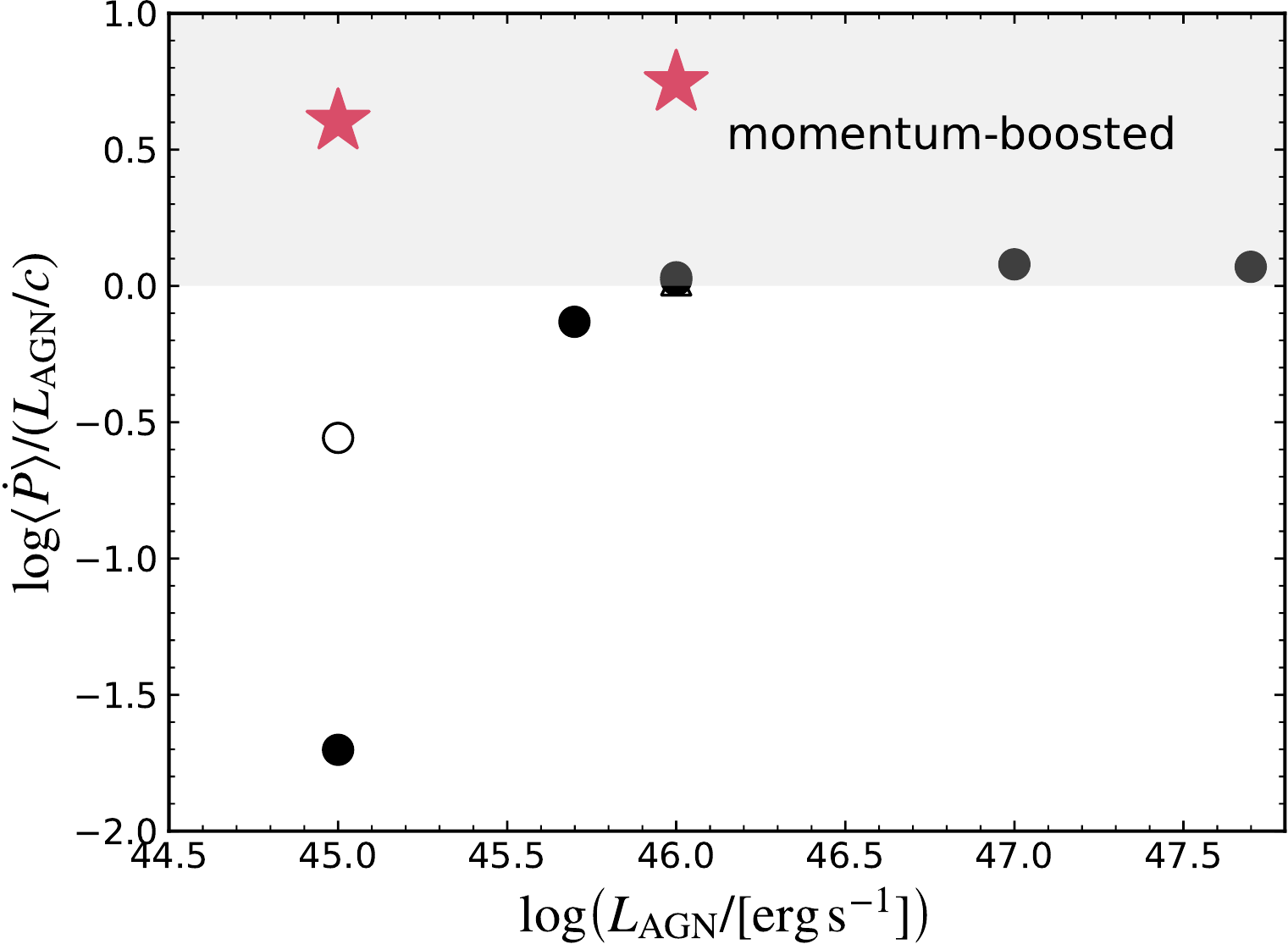}
	\includegraphics[width=0.32\textwidth]{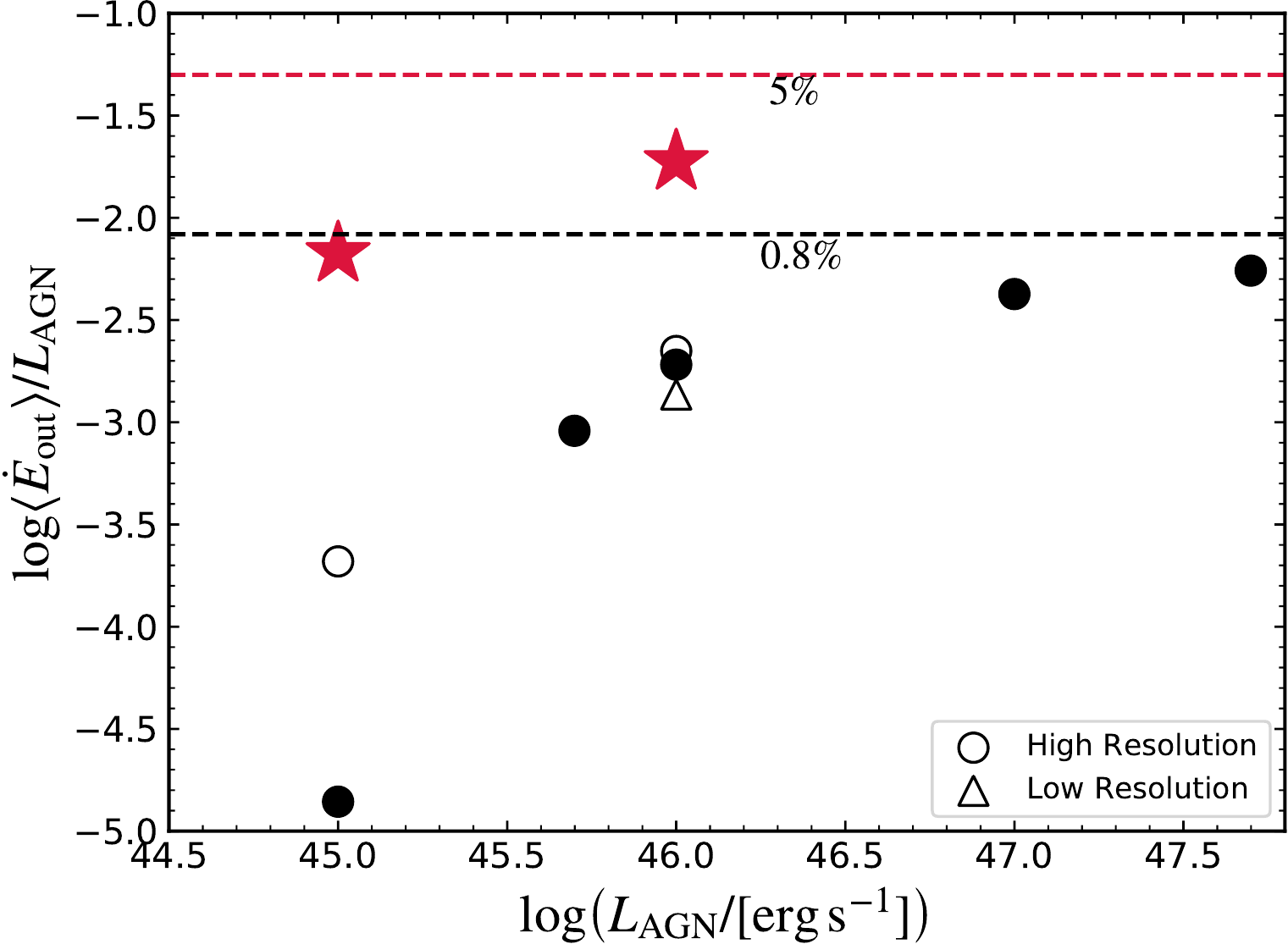}
	\caption{The speed (left-hand panel), radial momentum flux (central panel) and kinetic luminosity loading factors (right-hand panel) of the large-scale outflow as a function of AGN luminosity for a small-scale wind with $v_{\rm w} \, = \, 5000 \, \rm km \,s^{-1}$ (black symbols) and $v_{\rm w} \, = \, 30000 \, \rm km \,s^{-1}$ (red symbols). At fixed small-scale wind speed, the speed of the large-scale outflow scales with AGN luminosity. The dashed, black line on the left-hand panel shows a logarithmic scaling $v_{\rm out} \propto \log_{\rm 10}{L_{\rm AGN}}$ and the dotted, blue curve shows the best-fit power law applied to the fiducial data (black circles), excluding the lowest data point (see text for details). For $v_{\rm w} \, = \, 5000 \, \rm km \,s^{-1}$, the large-scale momentum fluxes are moderate $\lesssim L_{\rm AGN} / c$ because the small-scale winds carry only a small fraction of the total liberated AGN energy, while for $v_{\rm w} \, = \, 30000 \, \rm km \,s^{-1}$, we find that $\langle \dot{P}_{\rm out} \rangle > L_{\rm AGN} / c$. At the highest luminosities, the large-scale outflow asymptotes toward a value of order the kinetic luminosity of the small-scale wind. However, at lower AGN luminosities, the outflowing bubble has to do significant work to climb up the gravitational potential and to counter the ambient pressure.}
	\label{fig_outflowprops}
\end{figure*}

\subsubsection{Moderate small-scale winds}
We now take a closer look at how the properties of the large-scale outflow vary with AGN luminosity at fixed (small-scale) wind speed $v_{\rm w}$.
To start with, we investigate the \texttt{disc-LXX-b0.02} simulation set, in which, we recall, the wind speed is $v_{\rm w} \, = \, 5000 \, \rm km \, s^{-1}$.

We compute mass-weighted mean radial velocities $\langle v_{\rm out} \rangle$ by averaging over all cells in the simulation with radial velocities $v_{\rm rad} > v_{\rm noAGN} \,  = \, 250 \, \rm km \, s^{-1}$.
The threshold velocity $v_{\rm noAGN}$ corresponds to the maximum radial velocity reached in \texttt{disc-noAGN}. The resulting mass-weighted mean radial velocities are finally averaged over all simulation snapshots in the time interval $[t_{\rm 0}, \, t_{\rm 0} + t_{\rm AGN}]$, the period associated with the first outflow event.  
Similarly, we add up the radial momentum $p_{i} \, = \, m_{i} v_{\mathrm{rad}, \, i}$ and kinetic energy $E_{\mathrm{kin}, i} \, = \, \frac{1}{2} m_{i} v_{\mathrm{rad}, \, i}^2$ of each cell with $v_{\rm rad} > v_{\rm noAGN}$ to obtain expressions for the time-averaged momentum flux and kinetic luminosity $\langle \dot{P}_{\rm out} \rangle \, = \, \left( \sum_{i}{ p_i} \right) / t_{\rm on}$ and $\langle \dot{E}_{\rm kin} \rangle \, = \, \left( \sum_{i}{ E_{\mathrm{kin}, i}} \right) / t_{\rm on}$, respectively, where $t_{\rm on} \, = \, t - t_{\rm 0}$.

Fig.~\ref{fig_outflowprops} shows mass-weighted velocity (left-hand panel), radial momentum flux- (central panel) and kinetic luminosity loading factors (right-hand panel) as a function of AGN luminosity.
The black symbols, which illustrate the results for simulations performed with $v_{\rm w} \, = \, 5000 \, \rm km \, s^{-1}$, show a correlation between the speed of the large-scale outflow and the AGN luminosity. For $L_{\rm AGN} \, = \, 10^{45} \, \rm erg \, s^{-1}$, the mean speed of the outflow is $v_{\rm out} \approx 320 \, \rm km \, s^{-1}$, barely above $v_{\rm noAGN}$, while for $L_{\rm AGN} \, = \, 10^{47} \, \rm erg \, s^{-1}$, it exceeds $v_{\rm out} \approx 1100 \, \rm km \, s^{-1}$. 
In our fiducial simulations, the data shown on the left-hand panel of Fig.~\ref{fig_outflowprops} are well fit by a $v_{\rm out} \propto \log_{\rm 10}{L_{\rm AGN}}$ scaling (dashed, black line), because there is an accelerated drop in the outflow velocity as the AGN luminosity falls. At higher luminosities, however, the data exhibit power-law behaviour. If we neglect our simulation with  $L_{\rm AGN} \, = \, 10^{45} \, \rm erg \, s^{-1}$, we find $v_{\rm out} \, \approx \, 500 \left( \frac{L_{\rm AGN}} {10^{45} \mathrm{[erg \, s^{-1} ]}}\right)^{0.165} \, \rm km \, s^{-1}$ (dotted, blue curve). This relation only applies at luminosities significantly higher than the threshold luminosity for outflows (see Section~\ref{sec:condition}).

The outflow radial momentum flux $\dot{P}_{\rm out}$ also scales with AGN luminosity, though, for all simulations with $v_{\rm w} \, = \, 5000 \, \rm km \, s^{-1}$, we find $\dot{P}_{\rm out} \lesssim L_{\rm AGN}/c$. Note that the central wind has $\dot{P}_{\rm w} \, = \, L_{\rm AGN}/c$ and thus $\dot{P}_{\rm out} \lesssim \dot{P}_{\rm w}$. Similarly, while the small-scale wind in these simulations has a kinetic luminosity $\dot{E}_{\rm w} \, \approx\, 0.008 L_{\rm AGN}$, the kinetic luminosity of the large-scale outflows is always considerably lower with $\dot{E}_{\rm out} \, \approx \, 0.004 L_{\rm AGN} \approx 0.5 \dot{E}_{\rm w}$ for $L_{\rm AGN} \, = \, 10^{47} \, \rm erg \, s^{-1}$ and $\dot{E}_{\rm out} \, < \, 10^{-4} L_{\rm AGN} \ll \dot{E}_{\rm w}$ for $L_{\rm AGN} \, = \, 10^{45} \, \rm erg \, s^{-1}$.

\subsubsection{Fast small-scale winds}
At fixed AGN luminosity, more energetic small-scale winds result in faster outflows. 
The properties of the large-scale outflows in the two \texttt{disc-LXX-b0.1} simulations, in which $v_{\rm w} \, = \, 30000 \, \rm km \, s^{-1}$, are shown with red stars in Fig.~\ref{fig_outflowprops}.
There is a significant large-scale outflow even at $L_{\rm AGN} \, = \, 10^{45} \, \rm erg \, s^{-1}$, with a speed $v_{\rm out} \, \approx \, 700 \, \rm km \, s^{-1}$, i.e. higher by a factor $2$ than in the corresponding simulation with $v_{\rm w} \, = \, 5000 \, \rm km \, s^{-1}$.
As also seen in the previous section, the mean outflow speed also rises with AGN luminosity, exceeding $v_{\rm out} \, \approx \, 1000 \, \rm km \, s^{-1}$ at $L_{\rm AGN} \, = \, 10^{46} \, \rm erg \, s^{-1}$.
The slope of $\langle v_{\rm out} \rangle$ with AGN luminosity is consistent with the linear fit found for the simulations with moderate small-scale winds, but with a normalisation that is higher by about $400 \, \rm km \, s^{-1}$.

We now find $\dot{P}_{\rm out} > L_{\rm AGN}/c$, i.e. $\dot{P}_{\rm out} > \dot{P}_{\rm w}$; for \texttt{disc-L45-b0.1} and \texttt{disc-L46-b0.1}, we obtain $\dot{P}_{\rm out} \, \approx \, 3L_{\rm AGN}/c$ and $\dot{P}_{\rm out} \, \approx \, 5L_{\rm AGN}/c$, respectively.
Similarly, we find kinetic luminosities ranging from $\approx 0.008 L_{\rm AGN} = 0.16 \dot{E}_{\rm out}$ to $\approx 0.02 L_{\rm AGN} = 0.4 \dot{E}_{\rm out}$, once again finding that $\dot{E}_{\rm out} \ll \dot{E}_{\rm w}$ at lower AGN luminosities.

\subsubsection{Condition for launching outflows}
\label{sec:condition}
The fact that $\dot{P}_{\rm out} \lesssim L_{\rm AGN}/c$ in all simulations with $v_{\rm w} \, = \, 5000 \, \rm km \, s^{-1}$ may seem surprising. Energy-driving is typically expected to generate outflows with $\dot{P}_{\rm out} > L_{\rm AGN}/c$ and observed outflows with $\dot{P}_{\rm out} \lesssim L_{\rm AGN}/c$ are often interpreted as momentum-driven or radiation pressure-driven.

It is possible for $\dot{P}_{\rm out} \lesssim L_{\rm AGN}/c$ even in the presence of expanding, shocked wind bubbles.
For an energy-driven outflow, the momentum transfer rate is, at most, equal to $\dot{p}_{\rm out}^{\rm max} \, = \, 2 f \dot{E}_{\rm w}/v_{\rm out} \, = \, \beta f L_{\rm AGN} / v_{\rm out} \, = \, f \left( v_{\rm w} / v_{\rm out} \right) L_{\rm AGN} / c$, where $f$ is the fraction of the wind's energy that ends up in kinetic energy of the large-scale outflow. For outflow velocities $v_{\rm out} \, \approx \, 1000 \, \rm km \, s^{-1}$ and $v_{\rm w} \, = \, 5000 \, \rm km \, s^{-1}$, we expect $\dot{p}_{\rm out}^{\rm max} \, = \, 5 f L_{\rm AGN} / c$. If, for instance, about one quarter or one half of the wind's original energy is retained in kinetic form ($f \, = \, 0.25 \-- 0.5$) in the large-scale outflow \citep[e.g.][]{Faucher-Giguere:12}, we should find $\dot{p}_{\rm out}^{\rm max}  \, \lesssim \, 1 \-- 2 L_{\rm AGN} / c$, as seen for the higher luminosity simulations in Fig.~\ref{fig_outflowprops}.

 \begin{figure}
	\includegraphics[width=0.45\textwidth]{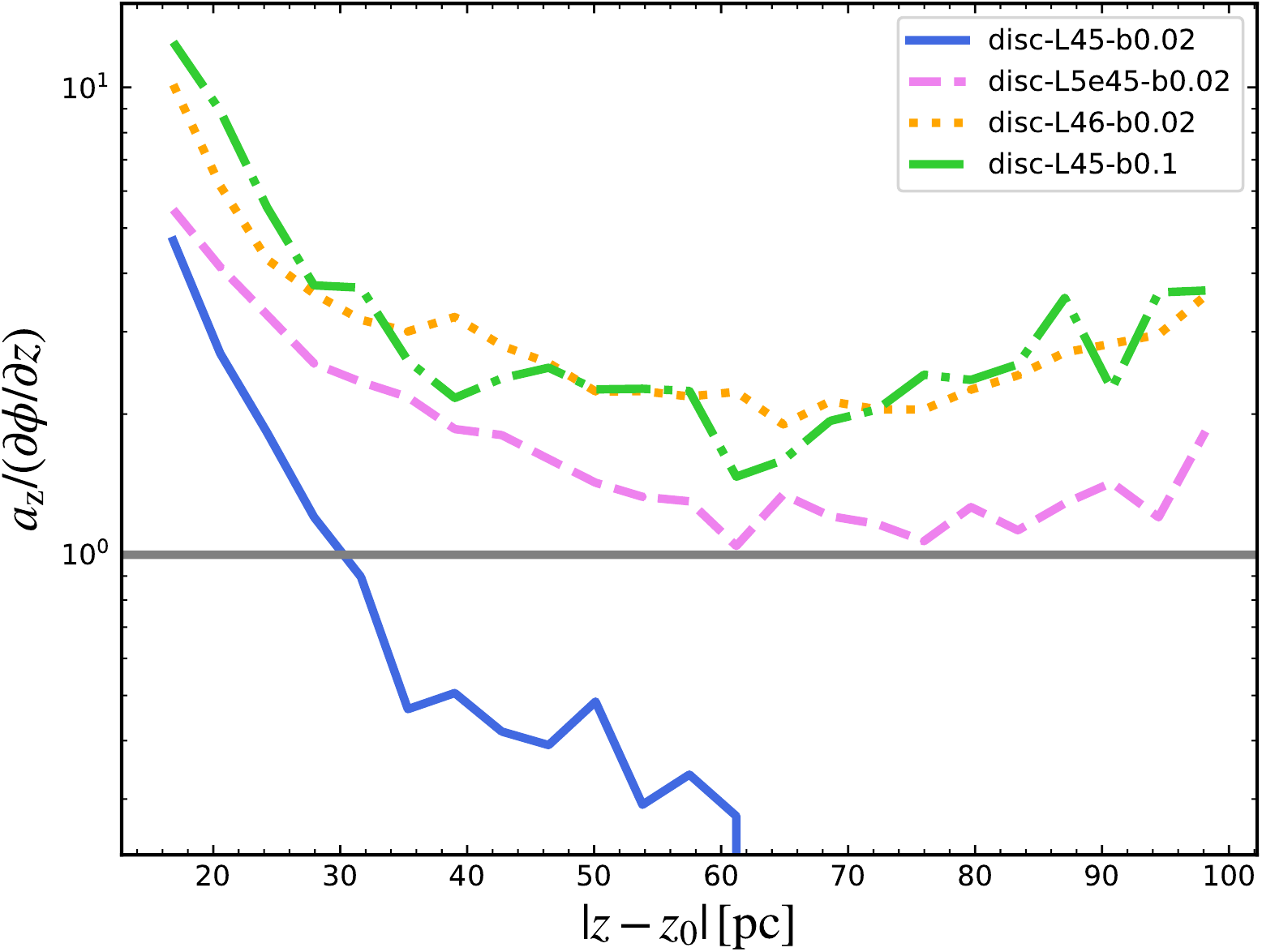}
	\caption{The net acceleration along the disc rotation axis for outflowing gas as a function of absolute vertical distance from the AGN for the simulations with a small-scale wind with $v_{\rm w} \, = \, 5000 \, \rm km \, s^{-1}$ and $L_{\rm AGN} \, = \, 10^{45} \, \rm erg \, s^{-1}$ (solid, blue curve), $L_{\rm AGN} \, = \, 5 \times 10^{45} \, \rm erg \, s^{-1}$ (dashed, violet curve) and $L_{\rm AGN} \, = \, 10^{46} \, \rm erg \, s^{-1}$ (dotted, orange curve). The green, dash-dotted curve shows the acceleration profile for the simulation with $v_{\rm w} \, = \, 30000 \, \rm km \, s^{-1}$ and $L_{\rm AGN} \, = \, 10^{45} \, \rm erg \, s^{-1}$. At lower AGN luminosities and lower small-scale wind speeds, the outflow stalls within a few $10 \, \rm pc$ from the AGN. At higher luminosities or small-scale wind speeds, there is acceleration all the way out to the disc height, after which outflows break-out and begin propagating into the halo.}
	\label{fig_force}
\end{figure}

At low luminosities, where large-scale outflows are typically slower, we, however, find that $\dot{p}_{\rm out} / (L_{\rm AGN} / c)$ drops, while, na\"{i}vely, it would have been expected to rise as $\left( v_{\rm w} / v_{\rm out} \right)$ increases.
Fig.~\ref{fig_outflowprops}, in fact, suggests that $\dot{p}_{\rm out} \rightarrow 0$ and $\dot{E}_{\rm out} \rightarrow 0$ as $L_{\rm AGN}$ approaches a threshold value of $L_{\rm crit} \sim 10^{45} \, \rm erg \, s^{-1}$.

\begin{figure*}
	\includegraphics[width=0.475\textwidth]{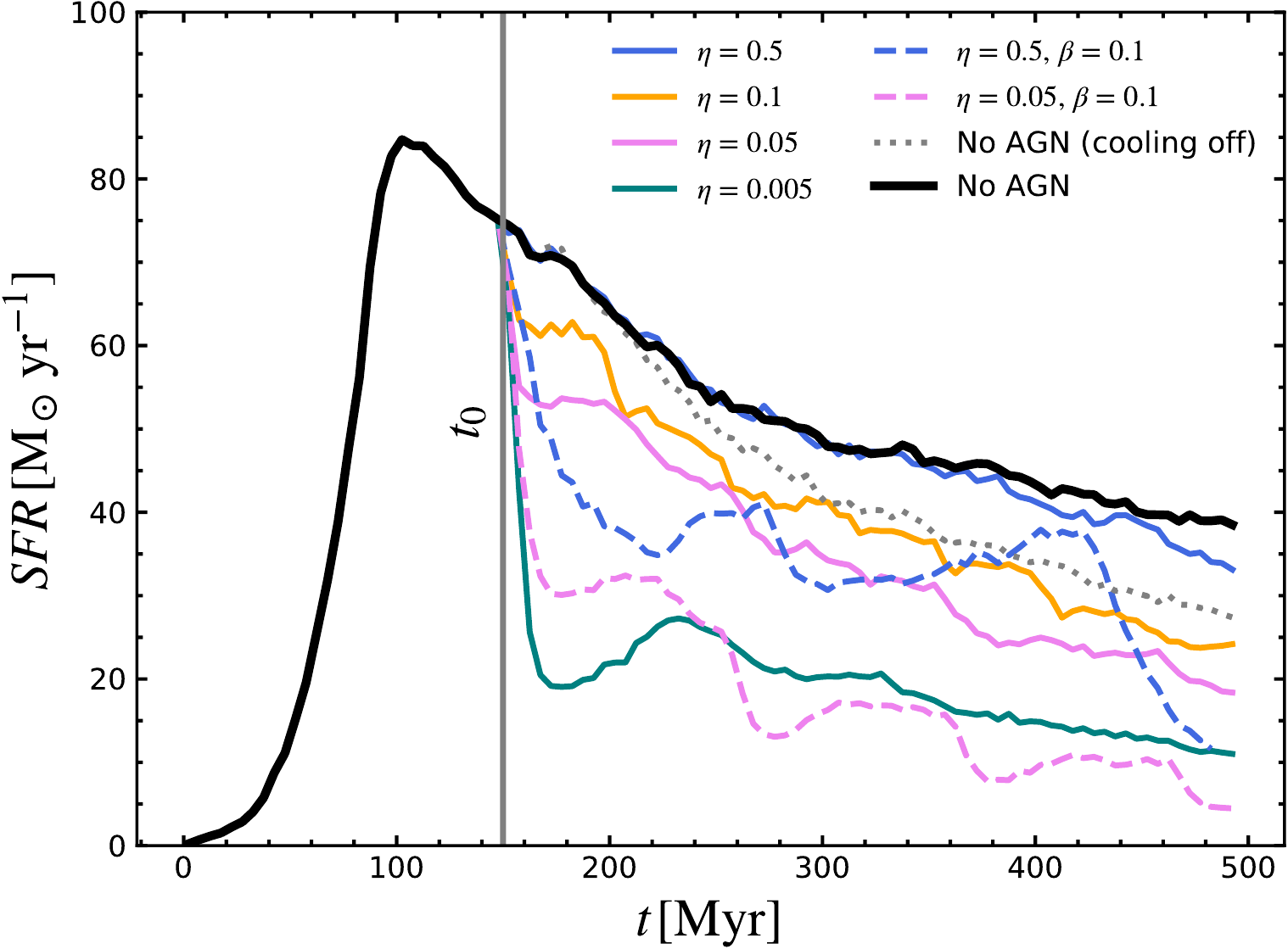}
	\includegraphics[width=0.475\textwidth]{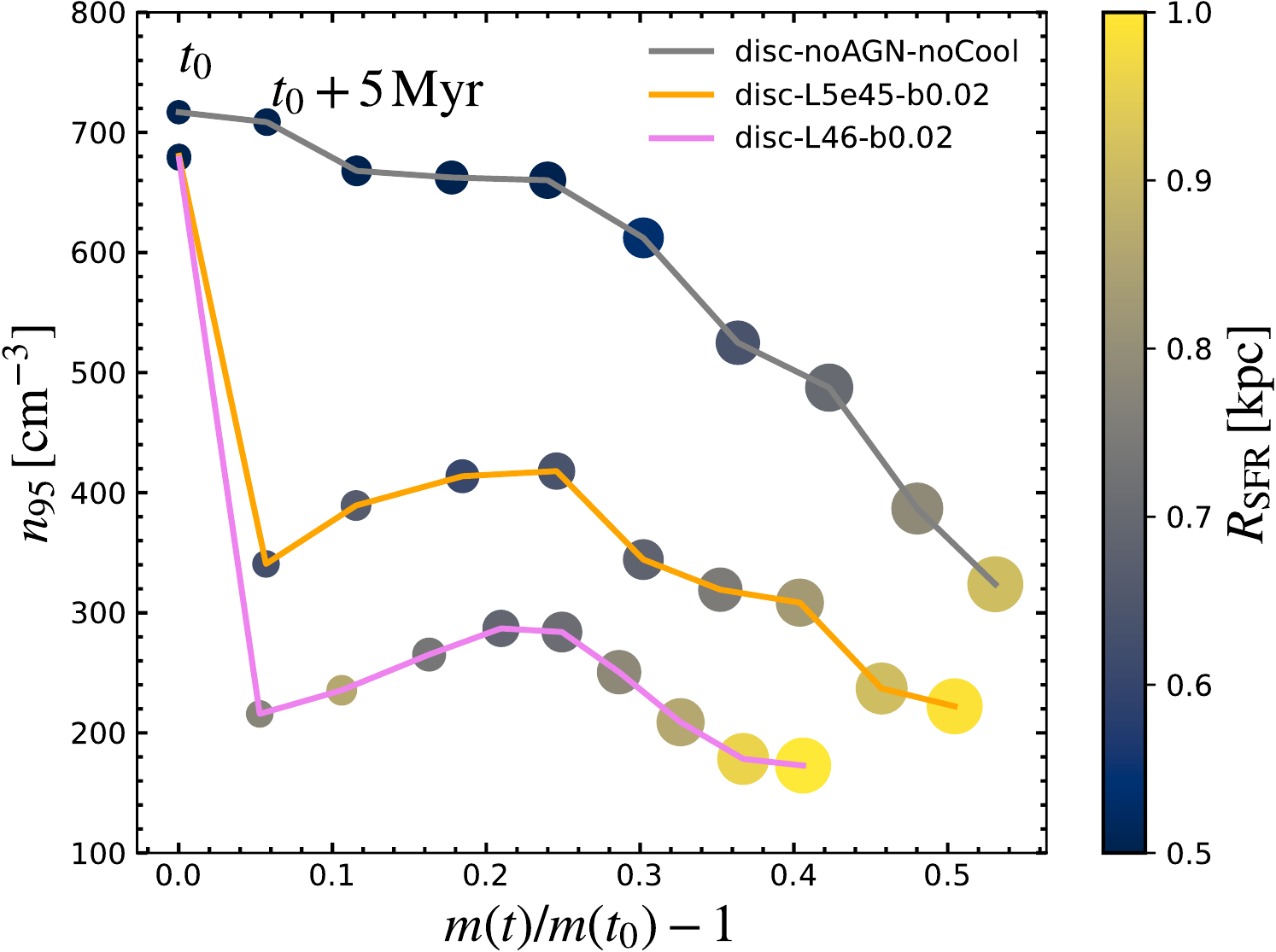}
	\caption{Left: Star formation history in the simulation without AGN feedback (black curve) and in our various simulations that include AGN winds. The star formation rate initially increases, reaching a peak at $t \, \approx \, 100 \, \rm Myr$. As the high density tail of the gas phase is converted into stars and gas accretion drops, the star formation rate falls off. AGN winds clearly suppress star formation, though only by a factor $<4$. For the simulated halo, the reduction in star formation exceeds that expected from preventing halo gas cooling (gray, dotted curve). Right: Gas density modulation (y-axis) vs. variation in the star-forming gas mass (x-axis) as a function of time (which is represented with size of the plot symbols). The mass in star-forming gas \emph{increases} with time even in the presence of AGN winds. The drop in the star formation rate is instead predominantly caused by preferential elimination of high density gas through its interaction with AGN-driven winds. In the absence of AGN winds, star formation shifts to larger radii (the colour of the symbol gives the radius enclosing half of the total star formation rate) as the dense nuclear reservoir is depleted, but AGN winds accelerate this shift.}
	\label{fig_sfr}
\end{figure*}

In order to understand the origin of this critical luminosity, we compute the acceleration along the disc rotation axis, which is also the outflow propagation axis, by evaluating the hydrodynamic momentum equation 
\begin{equation}
a_{z} \, = \, \frac{\partial v^z_i}{\partial t} \, = \, - \frac{1}{\rho_i} \frac{\partial \mathcal{P}_i}{\partial z} - \left(v^x_i \frac{\partial v^z_i}{\partial x} + v^y_i \frac{\partial v^z_i}{\partial y} + v^z_i \frac{\partial v^z_i}{\partial z} \right) - \frac{\partial \phi_i}{\partial z} \, ,
\label{eq_acc}
\end{equation}
for every gas cell $i$ with $\mathcal{P} \geq 10^{-3}$, such that all wind material (shocked and un-shocked) as well as any swept-up ambient medium component experiencing significant mixing with the wind fluid are taken into account.

In Fig.~\ref{fig_force}, we plot the $90^{\rm th}$ percentile of the $a_{\rm z}$ distribution as a function of $|z - z_{\rm 0}|$, the absolute value of the height above or below the disc plane. 
The net acceleration $a_{\rm z}$ is normalised to $\partial \phi / \partial z$.
For \texttt{disc-L45-b0.02} (solid, blue curve), we find net acceleration away from the disc at $|z - z_{\rm 0}| \lesssim 60 \, \rm pc$. Already at $|z - z_{\rm 0}|  \gtrsim 30 \, \rm pc$, the net outward acceleration is comparable, in magnitude, to the inward, gravitational acceleration. At $|z - z_{\rm 0}| \gtrsim 60 \, \rm pc$, the net acceleration is negative, i.e. towards the disc. Thus, while there is some acceleration within the central $\approx 60 \, \rm pc$, the pressure gradient associated with the shocked wind bubble is not sufficient for it to break out of the galaxy and launch a large-scale outflow.
For \texttt{disc-L5e45-b0.02} (dashed, violet curve), we instead find net outward acceleration at all scales, though most strongly in the central $\approx 60 \, \rm pc$. At larger scales we find that $|a_{\rm z}| \approx \partial \phi / \partial z $. 
In this simulation, even though the outflow breaks out of the galaxy, its pressure is still comparable to the pressure of the ambient medium.
Following the same trend, in \texttt{disc-L1e46-b0.02} (dotted, orange curve), we find yet higher acceleration at all scales with a magnitude that significantly exceeds that induced by the gravitational potential.
The acceleration in \texttt{disc-L46-b0.02} is similar to that of \texttt{disc-L45-b0.1}, shown with a dash-dotted, green curve in Fig.~\ref{fig_force}. The wind energy injection rate is $8.3 \times 10^{43} \, \rm erg \, s^{-1}$ in the former and $5 \times 10^{43} \, \rm erg \, s^{-1}$ in the latter, so the resulting acceleration should be expected to be similar.

In the absence of any AGN wind, the ambient medium pressure roughly balances gravity in the z-direction.
Introducing a fast, nuclear wind at a relatively low AGN luminosity only weakly disturbs this equilibrium, resulting in a small shocked bubble that stalls as its pressure drops.
The significant amount of work done to overcome the pressure of the surrounding gas means that comparatively little energy remains in kinetic form, resulting in the dramatically low large-scale outflow kinetic luminosities seen for the low-luminosity end in Fig.~\ref{fig_outflowprops}.
In the limit of high AGN luminosities, however, the pressure gradient is set entirely by the shocked wind bubble, which then expands insensitively to either counter-pressure from the ambient gas or gravity. 
As a large proportion of injected energy is retained in kinetic form in the large-scale outflow, its kinetic luminosity begins to asymptote towards the maximum possible value.

If $\beta \, = \, 0.1$ and given the same AGN luminosity, The total energy injection rate increases by a factor of $6$. Since the energy is split evenly between shocked wind and shocked ambient medium components \citep[see e.g.][]{Weaver:77,Faucher-Giguere:12}, the (predominantly thermal) energy and, hence, pressure associated with the shocked wind bubble should be higher by a factor $\approx 3$.
In contrast with \texttt{disc-L45-b0.02} (solid, blue curve), there is now significant acceleration at all scales.

\subsubsection{Resolution considerations}
The open circles and triangles in Fig.~\ref{fig_outflowprops} show results from the high-resolution and low-resolution simulations, respectively.
At $L_{\rm AGN} \, = \, 10^{46} \, \rm erg \, s^{-1}$, there is clear convergence in our results, as the difference between the fiducial and high-resolution data points is significantly smaller than that between the fiducial and low-resolution data points. For instance, the mean outflow speed grows only by $\lesssim 5\%$ between fiducial and high-resolution simulations, while it drops by $\approx 25 \%$ between fiducial and low-resolution simulations.  

At  $L_{\rm AGN} \, = \, 10^{45} \, \rm erg \, s^{-1}$, the outflow velocity also appears to change only little with resolution.
The discrepancy between fiducial and high resolution simulations is, however, much larger for the momentum flux and kinetic luminosity.
In this regime, where there is a steep gradient in $\dot{P}_{\rm out} / (L_{\rm AGN}/c)$ and $\dot{E}_{\rm out} / L_{\rm AGN}$, convergence is naturally harder to achieve.
It is therefore likely that, with increasing resolution, the precise value of the threshold luminosity at which large-scale outflows break-out of the disc is somewhat lower than seen in our simulations.

\subsection{Impact on the host galaxy}
\label{sec:impact}

The large-scale outflows launched in our simulations can affect star formation through three channels:
\begin{enumerate}
\item Prevention of accretion of new material from the gaseous halo, via heating and ejection of halo gas.
\item Depletion of the star-forming gas reservoir through net ejection from the host galaxy (galaxy `blow-out').
\item Modulation of gas density, i.e. the prolongation of the gas depletion timescale, through preferential removal or destruction of high-density gas. 
\end{enumerate}

Gas ejection plays a role in all channels.
In (i), halo gas is ejected, in (ii) the mass in star-forming gas drops and in (iii) AGN causes a drop in the gas density, while the mass in star-forming gas could remain unchanged.

In order to narrow in on the dominant star formation channel, we perform an additional simulation, which we call \texttt{disc-noAGN-noCooling}, where, starting at $t \, = \, t_{\rm 0}$, we disable radiative cooling for gas with $n_{\rm H} < 0.028 \, \rm cm^{-3}$. This density threshold corresponds to $0.1 n_{\rm th}$, the density at which the gas temperature jumps from $T \, \approx \, 10^4 \, \rm K$ to $T > 10^6 \, \rm K$. In this simulation, gas which is already cold can continue to collapse and form stars, but hot halo gas can no longer cool and contribute to star formation.
If the prevention of gas accretion is the main star formation suppression channel, we should expect the star formation history of simulations performed with AGN winds to resemble the star formation history of \texttt{disc-noAGN-noCooling}.

Star formation histories for \texttt{disc-noAGN} (thick, black curve), \texttt{disc-noAGN-noCooling} (grey curve) and the simulations performed with AGN winds are shown in Fig.~\ref{fig_sfr}. Comparing \texttt{disc-noAGN} and \texttt{disc-noAGN-noCooling}, we see that it takes several $10 \, \rm Myr$ for the suppression of halo gas accretion to cause noticeable differences in the star formation history. 
The star formation rate begins to fall off more steeply in \texttt{disc-noAGN-noCooling} than in \texttt{disc-noAGN} only at $t \, \gtrsim \, 210 \, \rm Myr$.
In contrast, star formation is suppressed immediately at $t \, \approx \, 150 \, \rm Myr$ in all simulations with AGN winds. 
Clearly, channel (i) does not operate on its own.

We also see that, as the AGN luminosity increases, and as $\eta_{\rm duty}$ decreases, the magnitude of immediate star formation suppression increases.
Suppression also becomes more efficient if the speed of the small-scale wind increases, as seen by comparing, for instance, the two violet curves.
These results, however, are consistent with both suppression channels (ii) and (iii).

In order to establish which of channels (ii) and (iii) is dominant, it is useful to define the quantity $m(t) \, = \, M_{\rm ISM} + M_{\rm \star}$. If $m(t)$ increases, replenishment from inflowing gas more than compensates for losses due to gas ejection, if $m(t)$ decreases, destruction or ejection through AGN winds dominate. Finally, if $m(t)$ does not change, either gas destruction by AGN winds and replenishment due to inflow compensate exactly or the galactic gas is simply converted into stars.
On the right-hand panel of Fig.~\ref{fig_sfr}, we plot the evolution of the $95^{\rm th}$ percentile of the density of star-forming gas, i.e. gas with density greater than $n_{\rm th}$, against $m(t)$.
While the star formation rate drops in all simulations, including those without AGN winds, we find that $m(t)$ \emph{increases} in every case, growing by about $40 \-- 50\%$ in the $50 \, \rm Myr$ following the first AGN outburst. 
This result unambiguously shows that the drop in star formation seen on the left-hand panel of Fig.~\ref{fig_sfr} is not driven by a net loss of the star-forming gas reservoir.
Instead, the drop in star formation is caused by a sudden decline in the abundance of high-density gas. 

With time, the abundance of high-density gas thins out in all simulations. Star formation activity shifts to lower density gas residing in the extended disc well outside of the galactic nucleus (see coloured symbols on the right-hand panel of Fig.~\ref{fig_sfr}). For instance, half of star formation occurs within the innermost $\approx 400 \, \rm pc$ at $t \, = \, 150 \, \rm Myr$, but at  $\approx 1.3 \, \rm kpc$ at $t \, = \, 250 \, \rm Myr$ in the simulations without AGN feedback. 
In contrast, the AGN-driven outflow only propagates out to a radius of $\lesssim 500 \, \rm pc$ along the disc plane.  
Since, at later times, much of the star-forming gas is shielded from the wind, which is funnelled towards the galactic poles (Fig.~\ref{fig_outflowevolution}), it becomes increasingly unlikely for the AGN to have an appreciable direct effect on the star formation history. 
Accordingly, the amplitude of the star formation leaps associated with AGN outbursts becomes smaller with time (Fig.~\ref{fig_sfr}).

 \begin{figure}
	\includegraphics[width=0.45\textwidth]{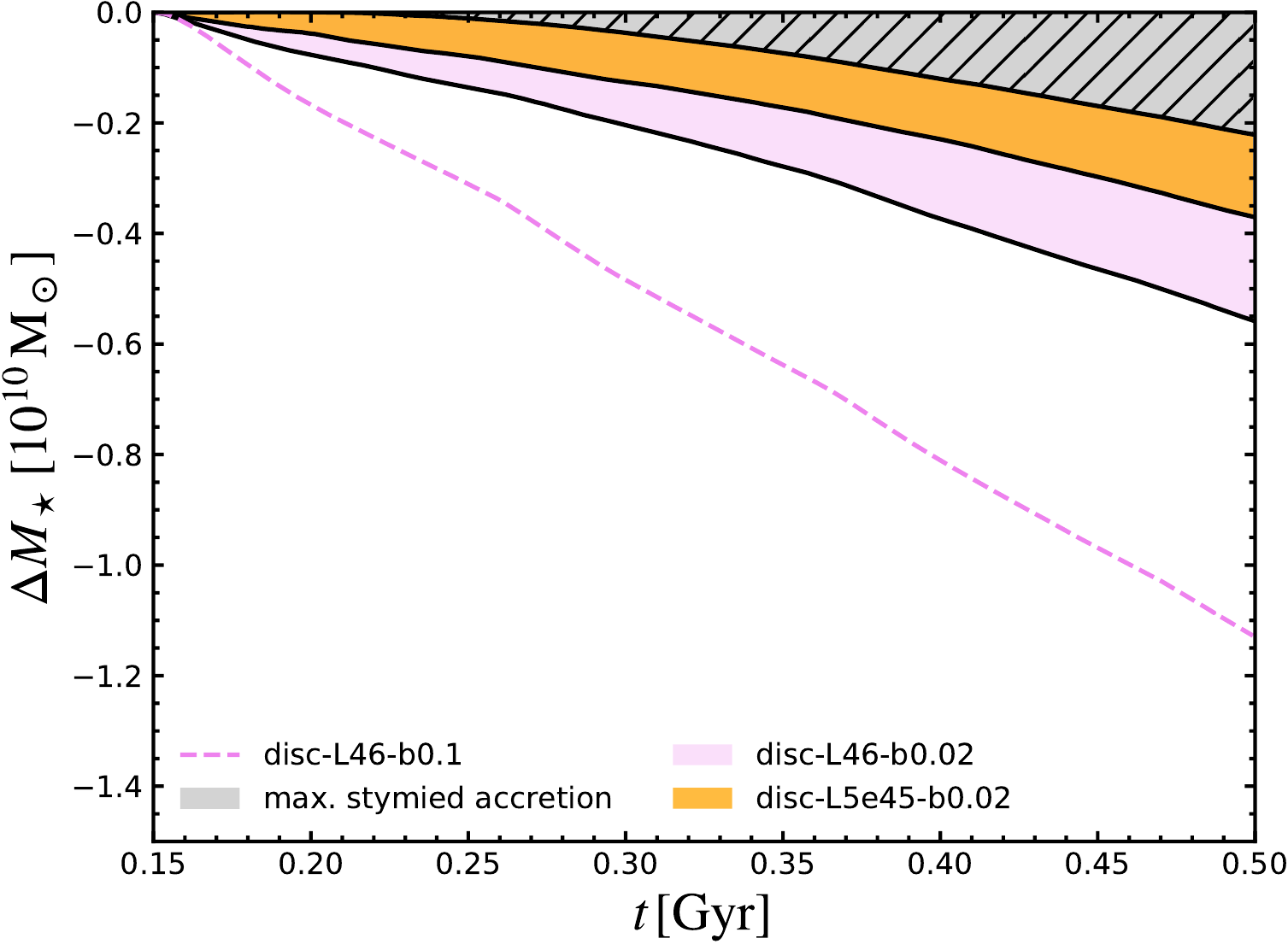}
	\caption{Decrement in cumulative stellar mass since $t \, = \, 150 \, \rm Myr$ as a function of time. At early times, stellar mass growth is impeded mostly via removal of dense gas by the outflows. The higher the AGN luminosity or the faster the small-scale wind, the more significant the suppression. At later times, interruption of halo gas accretion emerges as a more dominant channel for star formation suppression in most simulations.}
	\label{fig_ejecVSacc}
\end{figure}

Instead, AGN winds begin to operate primarily through their effect on halo gas. The left-hand panel of Fig.~\ref{fig_sfr} shows that the star formation histories of all simulations performed with AGN winds indeed share the same shape as \texttt{disc-noAGN-noCooling} after $t \, \approx \, 250 \, \rm Myr$. 
We verify that, by this time, gas accretion is countered in all our simulations with AGN winds except \texttt{disc-L1e45-b0.02}. 
In \texttt{disc-L5e45-b0.02}, we measure an inflow rate of $\approx 40 \, \rm M_\odot \, \rm yr^{-1}$ at $R \, = \, 10 \, \rm kpc$ at  $t \, \approx \, 300 \, \rm Myr$ and $\approx 25 \, \rm M_\odot \, \rm yr^{-1}$ at  $t \, \approx \, 450 \, \rm Myr$.
In \texttt{disc-noAGN}, the corresponding values are $\approx 80 \, \rm M_\odot \, \rm yr^{-1}$ and $\approx 45 \, \rm M_\odot \, \rm yr^{-1}$, and in \texttt{disc-noAGN-noCool}, they are $\approx 30 \, \rm M_\odot \, \rm yr^{-1}$ and $\approx 20 \, \rm M_\odot \, \rm yr^{-1}$ at the respective times.

The difference in stellar mass between simulations \texttt{disc-noAGN-noCooling} and \texttt{disc-noAGN} can be used as an estimate of the maximum decrement in stellar mass that can be caused by suppressing halo gas accretion. 
Remaining differences between our simulations with winds with respect to \texttt{disc-noAGN-noCooling} thus can be attributed entirely to ejection and destruction of star-forming gas.
In Fig.~\ref{fig_ejecVSacc}, we plot the difference in the cumulative stellar mass between our simulations with respect to \texttt{disc-noAGN}. The hatched region gives the decrement caused by stymied halo gas accretion, or channel (i) in the list presented at the beginning of this section. The coloured regions instead display the decrement caused by the removal of dense gas from the host galaxy, or channel (iii), in our different simulations.
At the onset of AGN wind injection, virtually all star formation suppression is caused by the removal of dense gas by the first outflow episode. The magnitude of the stellar mass decrement grows with AGN luminosity and wind speed, since the resulting outflows are more energetic.
Suppression of halo gas accretion gains importance in most simulations at $t \, \approx \, \rm 450 \, Myr$, a few $100 \, \rm Myr$ after the first outburst.

While the results presented in Fig.~\ref{fig_ejecVSacc} can be used to quantify how variations in AGN luminosity or wind speed translate into changes in the integrated stellar mass, we should emphasise that the magnitude of these changes depends on the time at which the AGN switches-on (which we do not attempt to model self-consistently in this study), the gas density in the nucleus, which is sensitive to other forms of feedback neglected here, and on the AGN lifetime. For instance, we experimented with reducing the AGN lifetime to $t_{\rm AGN} \, = \, 1 \, \rm Myr$ in a simulation otherwise identical to \texttt{disc-L46-b0.02}. The lower AGN lifetime makes it harder to expel cold, dense gas from the nucleus and results in a factor $1.1 \-- 1.2$ higher star formation rates than in \texttt{disc-L46-b0.02}.

\section{Discussion}
\label{sec:discussion}

\subsection{Numerically modelling AGN winds}
In the smoothed-particle hydrodynamic (SPH) simulations of \citet{Springel:05} and \citet{DiMatteo:05}, black holes are treated as sink particles and black hole accretion is modelled as a Bondi-Hoyle flow.
The density, speed of sound and gas velocity around the black hole sinks, which are required to estimate the accretion rate, are estimated by performing an SPH average over a number of neighbours which is typically $N_{\rm ngb} \, = \, 16-512$.
AGN feedback, in turn, is modelled by injecting an amount of thermal energy $E_{\rm AGN}$ into the SPH neighbours at a rate that is directly proportional to the AGN bolometric luminosity. 

While some modifications have been introduced to reduce numerical overcooling \citep{Booth:09}, black hole accretion and quasar feedback follow this basic model in most state-of-the-art cosmological simulations \citep{Vogelsberger:13, Schaye:15, Weinberger:17}.  
In other models, however, energy is injected in kinetic form into a number of cell/particle neighbours \citep[e.g. the `low accretion mode' in][]{Weinberger:17}, ostensibly producing stronger feedback \citep[e.g.][]{Choi:12, Barai:16}.

In this section, we review the motivation behind modelling AGN feedback via thermal- and kinetic energy injection and highlight the differences with respect to our new model.

\subsubsection{The cases for and against pure thermal energy injection}
Depositing a thermal energy $E_{\rm AGN}$ into a region with mass $M_{\rm ngb}$ increases its temperature by, at most, $\alpha E_{\rm AGN}/M_{\rm ngb}$, where $\alpha \, = \, \left( \gamma - 1 \right) \mu m_{\rm p} k_{\rm B}^{-1} $.  
Depending on the shape of the weighting kernel determining how much energy each resolution element receives, the temperature rise may amount to $\ll \alpha E_{\rm AGN}/M_{\rm ngb}$ for many cells/particles.
While the value of the pressure exerted by the injection region and its radial profile thus depend on numerical parameters such as $M_{\rm ngb}$, resolution and the shape of the weighting kernel, thermal energy injection always operates by generating hot, over-pressurised bubbles that expand through `PdV' work on surrounding gas \citep[e.g.][]{Costa:14}.

The ability of this model to physically capture wind-powered AGN feedback may be questioned, because explicit mass and momentum deposition are neglected  \citep[e.g.][]{Ostriker:10}.
During free-expansion, when the wind's ram pressure dominates over its thermal pressure (see Section~\ref{sec:testmodel}), thermal energy injection indeed provides a poor approximation. 
However, free-expansion takes place at scales of $\approx 0.1 \-- 100 \, \rm pc$ (Eq.~\ref{eq_rfree}).
This scale is either unresolved or only marginally resolved in typical galaxy evolution simulations, as we have verified with our own simulated disc galaxies (Section~\ref{sec:discs}). 
Moreover, as we have shown in Section~\ref{sec:testmodel}, outflows driven by a small-scale wind become energy-driven after the wind crosses the free-expansion radius.
At this point, the bulk of the wind's kinetic energy is converted into thermal form; for a strong shock, $\approx 89 \%$ of the wind energy just behind the shock is thermal.
If the free-expansion radius is not resolved, the outflow enters the energy-driven phase at the resolution scale directly (see Section~\ref{sec_convergence}). In this case, its energy content will be predominantly thermal immediately after injection (Fig.~\ref{fig_energy_conv}), not unlike in the hot bubbles generated via thermal energy injection.
Consequently, and as shown directly in \citet{Costa:14}, the dynamics of energy-driven outflows can, in principle, be reproduced accurately with standard thermal energy injection models.

Nevertheless, there is consensus that, on its own, continuous deposition of thermal energy into gas at scales of $\sim 100 \, \-- 1000 \, \rm pc$, the typical resolution afforded by high-resolution cosmological `zoom-in' simulations, does not regulate star formation in massive galaxies effectively \citep[e.g.][]{vanderVlugt:19}.
As pointed out in \citet{Booth:09}, this inefficiency, however, is largely a consequence of \emph{numerical} cooling losses caused by insufficient resolution and also, as shown in \citet{Weinberger:18}, by the enforcement of an effective equation of state to model dense star-forming gas. Deviation from purely energy-driven dynamics and weakened feedback can thus arise due to violation of adiabaticity caused by insufficient resolution.

We recall that the shocked wind bubble should typically not undergo significant cooling losses (see Sections~\ref{section_radiative_cooling} and~\ref{sec:shockedwind}) and severe cooling losses are therefore at odds with the wind-based feedback scenario explored in this study.
One way to prevent numerical overcooling and reconcile the qualitative dynamics of outflows, as generated via thermal energy injection, with physical, energy-driven outflows is to increase the numerical resolution and reduce $M_{\rm ngb}$ \citep[e.g.][]{Curtis:15}.
If higher resolution cannot be achieved, numerical corrections that circumvent overcooling, such as storing up thermal energy and injecting it only once it can offset radiative cooling \citep{Booth:09}, likely lead to better qualitative agreement with the energy-driven outflows generated in our simulations.
Quantitative differences in the thermodynamic properties of the different outflow zones should nevertheless remain. 
For instance, while in our model, the post-shock temperature, pressure and density of the wind is determined self-consistently based on $v_{\rm w}$, $\tau$ and $L_{\rm AGN}$, the properties of the over-pressurised bubble, as generated via thermal energy injection, are set by the number of cell/particle neighbours, the resolution and the shape of the weighting kernel.

Unlike in models based on thermal energy injection, the model presented in this paper also accurately reproduces the free-expansion phase, provided this can be resolved.
It thus correctly predicts when the solution becomes energy-driven or if it should be momentum-driven (see Section~\ref{sec:shockedwind}).
It can therefore be meaningfully applied to higher-resolution, smaller-scale simulations, predicting the correct outflow dynamics in e.g. studies targeting the interaction between AGN winds and the interstellar medium or accretion flows within galactic nuclei.
The energy-driven models used in \citet{Costa:14} or \citet{Curtis:15} and models based on thermal energy injection in general instead implicitly assume that thermalisation occurs instantaneously, which may be invalid at $\lesssim 100 \rm \, pc$ scales.
Thus, while we expect agreement at galactic halo scales, differences between the effects of outflows driven in thermal energy injection models and our new model are likely to be most pronounced at the scale of the host galaxy and, in particular, the galactic nucleus.

\subsubsection{The cases for and against pure kinetic energy injection}
In various models, energy is injected in kinetic form, typically producing stronger feedback than achieved by thermal energy injection. 
In one variant, no wind mass is explicitly added into the simulation domain \cite[e.g.][]{Weinberger:17}. Instead a wind is generated by depositing an energy $E_{\rm AGN}$ into a region of fixed mass $M_{\rm ngb}$, such that the velocity of each gas cell should be incremented, at most, by   
\begin{equation}
v_{\rm kick} \, = \, \left( \frac{2 E_{\rm AGN}}{M_{\rm ngb}} \right)^{1/2} \, .
\end{equation}
Since the velocity imparted to local gas depends on the AGN luminosity through $E_{\rm AGN}$ and on the choice of $M_{\rm ngb}$, we refer to this model the variable-speed kinetic energy (VSK) injection model.
In order to understand how a wind launched using VSK evolves, we recall the basic expectations outlined in Section~\ref{sec:windmodel} and confirmed in Section~\ref{sec:testmodel} with simulations.
If $v_{\rm kick}$ significantly exceeds the sound speed of the medium surrounding the injection region, the kinetic energy-dominated wind initially moves outwards ballistically. Initially, its mass is large compared to the mass it has encountered. If it propagates through a homogeneous medium of density $\rho_{\rm 0}$, the wind, assumed to propagate into a solid angle $\Omega \, = \, 4 \pi b$, will have swept-up a mass equal to its own and thermalised at a radius
\begin{eqnarray}
\label{eq:Rfreekin}
R_{\rm free}^{\rm kin} \, & = & \, \left( \frac{3}{\Omega} \frac{M_{\rm ngb}}{\rho_{\rm 0}} \right)^{1/3} \\
                    \, & \approx & \, 540 \, b^{-1/3} \left( \frac{M_{\rm ngb}}{10^8 \, \rm M_\odot} \right)^{1/3} \left( \frac{n_{\rm 0}}{10 \, \rm cm^{-3}} \right)^{-1/3}\, \rm pc \, , \nonumber 
\end{eqnarray}
assuming $\mu \, = \, 0.6$.
At $R_{\rm free}$, there will be two separate shocked phases: a hot, internal shocked wind phase and a cooler outer shell of swept-up gas \citep{Costa:14, Costa:15}, as also seen in Fig. 1 of \citet{Weinberger:17} or Fig. 2 of \citet{Nelson:19}.
Eq.~\ref{eq:Rfreekin} can be compared with the free-expansion radius (Eq.~\ref{eq_rfree}) of a small-scale AGN wind as modelled in this paper, assuming this propagates through the same medium towards the same solid angle:
\begin{eqnarray}
\label{eq:Rfreeratio}
\frac{R_{\rm free}^{\rm kin}}{R_{\rm free}} \, & \approx & \, 170  \left( \frac{M_{\rm ngb}}{10^8 \, \rm M_\odot} \right)^{1/3} \left( \frac{n_{\rm 0}}{10 \, \rm cm^{-3}} \right)^{1/6} \nonumber \\  & & \times \left( \frac{\beta}{0.1} \right) \left( \frac{L_{\rm AGN}}{10^{45} \, \mathrm{erg \, s^{-1}}} \right)^{-1/2}  b^{1/6}  \tau^{-1/2} \, .
\end{eqnarray}
For the typical resolution reached in large cosmological simulations, i.e. $m_{\rm target} \sim 10^6 \, \rm M_\odot$ and a number of neighbours $\sim 100$, winds driven in VSK typically thermalise one to two orders of magnitude larger scales than predicted by our model. 
Since kinetic energy cannot be radiated away, it is no surprise that VSK results in stronger feedback than achieved via continuous thermal energy injection at low resolution.
Eq.~\ref{eq:Rfreeratio} also indicates that bringing VSK and our model into quantitative agreement is not straightforward and requires fine-tuning of six different variables.

While $v_{\rm kick}$ is variable at injection in VSK in time due to its dependence on $E_{\rm AGN}$ and spatially due to kernel-weighting, the wind speed $v_{\rm w}$ is a constant in our model. In our model, the choice of $v_{\rm w}$ is based on fundamental radiation-hydrodynamic simulations of disc-driven winds \citep[e.g.][]{Nomura:16}, which predict that small-scale winds attain well-defined terminal speeds at scales $R \sim 100 r_{\rm g}$. The assumption of a fixed wind speed at injection allows us to connect and test our numerical model on robust, analytical descriptions of the interaction between accretion disc winds and the surrounding medium \citep{King:03, Faucher-Giguere:12}.
The speed of the wind determines the temperature of the shocked wind bubble (see Eq.~\ref{eq_rshock_T}) and, hence, the strength of the energy-driven phase. Since $v_{\rm w}$ is fixed in our model, it predicts a much narrower range of post-shock temperatures than in VSK, and therefore impacts the abundance and thermodynamic properties of the hottest gas phase. 

We highlight also that, since they typically thermalise at larger scales, the winds launched in VSK can `disguise' as cold large-scale outflows, because, by construction, thermalisation occurs much later and potentially at kpc scales. If $M_{\rm ngb} \sim 10^8 \, \rm M_\odot$, a significant fraction of the outflow mass thus may be cold, consisting of freely-expanding ejecta and not e.g. of swept-up ambient medium which has cooled down \citep[as in][]{Costa:15} or of clouds which survive after passing a forward shock.

Another significant difference is the mass and density associated with the shocked wind component. In our model the wind density is remarkably low (Eq.~\ref{eq_windnh}) and even for $L_{\rm AGN} \, = \, 10^{47} \, \rm erg \, s^{-1}$, $n_{\rm w} \sim 10^{-5} \, \rm cm^{-3}$ at $R \, = \, 10 \, \rm kpc$. Instead, launching $\sim 10^8 \, \rm M_\odot$ at high-speed leads to densities $\sim 10^{-3}  \, \rm cm^{-3}$ at the same scale irrespective of the AGN luminosity.
The much higher densities in VSK, which cause the thermalisation to occur at larger radii, have a number of repercussions: 1) there are likely to be significant differences in predictions for e.g. X-ray emissivity or the Sunyaev-Z'eldovich signal associated with hot gas and 2) it is now virtually impossible for solutions to become momentum-driven if AGN Compton cooling was to be included. 

In contrary to VSK, in our model, changes in AGN luminosity translate into variations in the mass flux across the spherical boundary (see e.g. Eq.~\ref{eqs_conservation_m3}).
It is also possible to devise a variable-mass kinetic energy (VMK) injection model, where a fixed wind speed $v_{\rm kick}$ is assumed \citep[e.g.][]{Choi:12, Barai:16, Angles-Alcazar:17} and, given an energy $E_{\rm AGN}$, inject it into a variable mass $M_{\rm kick}$ given by
\begin{equation}
M_{\rm kick} \, = \, \frac{2 E_{\rm AGN}}{v_{\rm kick}^2} \, = \, \frac{2 \eta}{v_{\rm kick}^2} L_{\rm AGN} \Delta t \, ,
\end{equation}
where $\Delta t$ is the duration of a timestep and $\eta$ is a free parameter giving the fraction of the instantaneous AGN bolometric luminosity which is converted into kinetic energy.
The ejected mass $M_{\rm kick}$ cannot be smaller than the mass resolution $m_{\rm target}$. The latter sets a characteristic energy $E_{\rm min}$ that needs to be accumulated before $N_{\rm kick}$ cells/particles can be ejected:
\begin{equation}
E_{\rm min} \, = \, \eta \sum{L_{\rm AGN} \Delta t} \, = \, \eta \langle L_{\rm AGN} \rangle t_{\rm AGN} \, = \, \frac{1}{2} N_{\rm kick} m_{\rm target} v_{\rm kick}^2 \, .
\end{equation}
This expression can be rearranged into
\begin{eqnarray}
N_{\rm kick} \, & \approx & \, 16 \left( \frac{\eta}{0.05}\right) \left(\frac{\langle L_{\rm AGN} \rangle}{10^{46} \, \rm erg \, s^{-1}} \right) \left( \frac{t_{\rm AGN}}{\rm Myr} \right) \nonumber 
\\ && \times \left( \frac{v_{\rm kick}}{10^4 \, \rm km \, s^{-1}} \right)^{-2} \left( \frac{m_{\rm target}}{10^6 \, \rm M_\odot} \right)^{-1} \, .
\end{eqnarray}
We see that relatively long timescales of $t_{\rm AGN} \sim \, \rm Myr$ are required even for high time-averaged AGN luminosities of $\langle L_{\rm AGN} \rangle \sim 10^{46} \, \rm erg \, s^{-1}$ before $16$ resolution elements are ejected if $m_{\rm target} \sim 10^6 \, \rm M_\odot$. Raising the wind speed to $v_{\rm kick} \, = \, 30000 \, \rm km \, s^{-1}$ or setting $\eta \, = \, 0.005$ extends the required mean luminosity or the required timescale by yet another order of magnitude. 

Low resolution effectively decouples the ejected mass from the instantaneous AGN luminosity, in contradiction with e.g. Eq.~\ref{eqs_conservation_m3}. Limiting the number of ejected resolution elements to a number even smaller than $16$ would reduce $E_{\rm min}$. However, it would also mean that the wind solid angle is severely under-sampled. At low $N_{\rm kick}$, the wind is discretised into a small number of `bullets' that are ejected in a few directions, significantly reducing $b$ in Eq.~\ref{eq_rfree}.
If the solid angle is under-sampled, the free-expansion radius can be extended, and the transition into the energy-driven phase may occur at larger radii than expected. As in VSK, a delay in thermalisation renders VMK more efficient than continuous thermal energy injection.

In our model, the AGN wind is injected by updating the fluxes across a boundary and not by ejecting gas cells explicitly.
It therefore ensures that the wind mass flux remains coupled to the instantaneous AGN luminosity. Since injection occurs across the desired solid angle by construction, our model also does not risk overestimating the thermalisation radius.

\subsubsection{Injection into nearest neighbours in Lagrangian codes}
Injection into nearest neighbours is susceptible to various other numerical problems.
In Lagrangian codes, the resolution around the accreting black hole is lost when the central resolution elements are driven outwards through AGN outflows. 
Since energy is injected into a fixed number of neighbours (and not into a fixed volume), the spatial scale at which energy is deposited tends to increase, artificially compensating for adiabatic cooling losses. In extreme cases, injection may occur directly at several $\rm kpc$ scales, sometimes resulting in conspicuous holes in the centre of simulated galaxies. The injection procedure proposed in this study alleviates these problems in two ways: i) injection occurs at a fixed spatial scale independently of the configuration of the gas cells surrounding the black hole and ii) wind mass is explicitly injected along with momentum and energy, compensating for mass expulsion.

Another potential concern with nearest neighbour injection in Lagrangian codes is that injection is anisotropic \citep[e.g.][]{Zubovas:16}, effectively mass-weighted and may occur along preferred directions (e.g. the disc plane). \citet{Hopkins:18} illustrate how the failure to ensure statistical isotropy and conservation of mass, momentum and energy in supernova feedback, for instance, generates artificial torques that can drastically alter the morphology of the simulated galaxies. Our model explicitly conserves mass, momentum and energy and ensures statistical isotropy, thus overcoming all these issues.

\subsubsection{AGN winds in Eulerian codes}
In Eulerian codes, AGN feedback is also typically modelled via injection of thermal or kinetic energy. In their `quasar-mode' implementations, \citet{Teyssier:11}, \citet{Dubois:12b} and \citet{Biernacki:17} inject energy in thermal form into all cells contained within a sphere of fixed radius $\lesssim 4 \Delta x_{\rm min}$, where $\Delta x_{\rm min}$ is the width of the smallest cells in their simulations. 

Unlike typical implementations in Lagrangian codes, and more closely to our model, the injection region is now spatially fixed.
Many of the same questions raised above apply to such AGN feedback implementations, however.
For instance, the scale at which the energy-driven phase begins should depend on the density of the ambient medium and the AGN luminosity (Eq.~\ref{eq_rfree}), while thermal injection implicitly assumes that energy-driving starts at the resolution scale, unlike in our model. 
Similarly, the temperature and density of heated gas depends on the size of the injection region and the enclosed mass (which is now time-dependent) and cannot straightforwardly be made to match the values predicted by our model.

Eulerian codes, however, present the advantage that energy and momentum can be distributed isotropically more straightforwardly, depending on whether and how injected quantities are weighted, and can naturally overcome the issues pointed out in the previous section. Cartesian grids, however, can lead to artefacts in spherical solutions, a problem which our {\sc AREPO} model overcomes through the use of spherically-symmetric cell layers aligned with the radial direction.
With high resolution, perhaps achieved through adaptive mesh refinement, it would be possible to implement our model in Eulerian codes. A number of cells within a thin, but well-sampled, ring could be selected and the fluxes across the inner boundary of the cells in these rings fixed as in our model.

\subsection{How star formation is suppressed}
In Section~\ref{sec:impact}, we see that a single AGN feedback mechanism can influence the star formation history through multiple channels.
Such multi-faceted effects on the star formation history have been reported for other mechanisms, such as for radiation pressure on dust \citep{Costa:18a, Costa:18b}, and also in simulations which adopt VSK and VMK recipes for AGN feedback \citep[e.g.][]{Barai:18,Zinger:20}.
Here we show that physical, small-scale AGN-driven winds, (i) eject and destroy star-forming gas from the galactic nucleus and (ii) expel halo gas.
Removal of dense gas causes rapid suppression in the star formation rate by factors $\lesssim 3$, whereas ejection from the gaseous halo operates on longer $\sim 100 \, \rm Myr$ timescales, by speeding up the decline in the halo gas inflow rate.

In no simulation, even those with extreme AGN luminosities $L_{\rm AGN} \gtrsim 10^{47} \, \rm erg \, s^{-1}$ do we find rapid, thorough star formation quenching.
In our disc galaxy simulations, complete quenching would require the destruction of the gas disc, which is implausible since even spherical outflows become collimated by the ambient gas density field and escape through paths of least resistance \citep[see also][]{Gabor:14,Costa:14}.
In Section~\ref{sec:discs}, direct gas ejection becomes relatively unimportant once the nuclear gas reservoir is depleted and star formation shifts to the extended disc component.
These results indicate that rapid quenching due to gas ejection through small-scale AGN winds is likely only in systems where the star formation region is highly compact, concentrated around the galactic nucleus and approximately spheroidal. Potential sites are thus high-redshift, compact, star-forming galaxies \citep[e.g.][]{Barro:13,Straatman:15}.

Some theoretical work, however, suggests that ejection even in such extreme circumstances may not result in long-term quenching.
\citet{Dubois:13} perform `zoom-in' simulations of a $5 \times 10^{11} \, \rm M_\odot$ halo at $z > 6$, indeed finding that AGN feedback quenches star formation effectively in the innermost $50 \, \rm pc$, with star formation levels of $\approx 10 \, \rm M_\odot \, yr^{-1}$ persisting within $\approx 3.5 \, \rm kpc$.
In the cosmological simulations of \citet{Costa:18b}, where AGN feedback is investigated in a remarkably compact ($R \lesssim 500 \, \rm pc$), massive galaxy hosted in a $\sim 10^{12} \, \rm M_\odot$ halo at $z \, > \, 6$, both radiation pressure on dust and AGN winds, modelled through continuous thermal energy injection, fail to completely halt star formation even if a bright central quasar is active for $100 \, \rm Myr$. 
Narrow, dense filaments of cold gas continuously replenish the central galaxy and, given their small solid angle, are resilient to even powerful AGN-driven outflows.
\citet{Bourne:15} and \citet{Curtis:16} further show that cold gas ejection and star formation suppression become increasingly inefficient as the resolution increases.

The available channels for star formation suppression are only as sophisticated as the star formation model adopted in the simulation. 
In most simulations, star formation depends mainly on the gas density and, due to insufficient resolution, most current simulations cannot resolve the low volume-filling cold phase of the interstellar medium. Small-scale, high-density gas clouds may be disrupted by an AGN wind without ejection from the galaxy \citep[e.g.][]{Hopkins:10} and the injection of solenoidal turbulence may counter cloud collapse \citep[e.g.][]{Federrath:12}. 
Many such new channels may be uncovered by applying physical AGN feedback models to high-resolution studies of the interstellar medium.

\subsection{Limitations of our model}
\label{sec_plasma}
The main simplifying assumption made in this study is that the small-scale wind is continuous, smooth and has a well-defined velocity $v_{\rm w}$ and momentum flux $\dot{P}_{\rm w}$.
Observational evidence \citep[e.g.][]{Gofford:15}, for example, indicates that the speed of accretion disc winds may scale weakly with the AGN luminosity as $v_{\rm w} \propto L_{\rm AGN}^{1/2}$.
In addition, theoretical work suggests that the speed of the wind depends on the scale within the accretion disc from which it is driven \citep[e.g.][]{Yuan:14}. 
Equally, the speed is predicted to depend on solid angle and is typically highest along the edges of the accretion disc, and somewhat lower at low inclinations \citep[e.g.][]{Nomura:17}.
Simulations of radiatively-inefficient accretion discs \citep[e.g][]{Sadowski:13}, for instance, predict two distinct outflow components: (i) a collimated, jet and (ii) a wide-angle, sub-relativistic wind.
In principle, our model is able to accommodate these more complex wind structures. For instance, we could choose to let vary $v_{\rm w}$ vary with $L_{\rm AGN}$ and with the solid angle in accordance with AGN disc wind simulations.

There are otherwise various other, likely important, missing physical ingredients in our model. 
One question we investigated in this paper is whether inverse Compton scattering from AGN photons can act as an efficient cooling mechanism for shocked wind close to the AGN (Section~\ref{sec:shockedwind}).
\citet{King:03}, for example, assumes that the wind thermalises within the Compton cooling radius and that the protons and electrons within shocked wind plasma rapidly reach thermal equilibrium.  
\citet{Faucher-Giguere:12}, however, argue that the Coulomb equilibration timescales in the shocked wind phase can be significant and that inverse Compton scattering should not efficiently cool the shocked wind.
In Section~\ref{sec:shockedwind}, we assumed that protons and electrons reach equilibrium instantaneously and nevertheless found that Compton cooling is efficient only for very high AGN luminosities and high ambient densities, for homogeneous media. Prolonging the equilibration timescales would only suppress Compton cooling even further reducing the parameter space in which momentum-driven solutions can occur.

Other physical processes which may be important include acceleration of cosmic rays in both reverse and forward shocks and AGN radiation.
Non-thermal pressure from cosmic rays may potentially enhance the momentum deposition of the outflow \citep[e.g.][]{Diesing:18} and, if transported into star-forming regions, potentially counter cloud collapse, helping to regulate star formation.
Momentum input by AGN radiation pressure on dust may compete with the momentum generated during the energy-driven phase \citep[e.g.][]{Costa:18b}, particularly if the small-scale wind velocity is low, while photo-ionisation, photo-heating and X-ray heating undoubtedly shape the thermodynamic state and, hence, the observability of outflowing gas.

In this study, we have focussed on how to accurately model the effect of a small-scale AGN wind on its surrounding medium. 
The source of power for the small-scale wind was understood to stem from an accreting black hole, but accretion was not modelled. 
The spherical boundary which we used to inject a wind can, in the future, be employed to measure inflow rates towards the galactic nucleus. These inflow rates could, for instance, be used as input for the growth rate of unresolved black hole accretion discs, as modelled e.g. in \citet{Fiacconi:18} and \citet{Bustamante:19}.
It will also be critical to couple the AGN luminosity to the black hole accretion rate in order to address whether black hole growth becomes self-regulated, to predict AGN lifetimes and to more accurately evaluate the impact of winds on their immediate environment.

\section{Conclusions}
\label{sec:conclusions}
The interaction of small-scale AGN winds with their host galaxies and their large-scale environment proceeds through physical processes that occur on an extreme range of scales, starting at $< 10^{-2} \, \rm pc$, accretion disc scales and extending out to the $\sim 100 \, \rm kpc$ scales of galactic haloes. 
In order to render this problem tractable, \citet{King:03}, \citet{Faucher-Giguere:12} and \citet{Zubovas:12} postulate the existence of a small-scale wind emanating from the nucleus and develop the analytical theory of the outflows that result from the collision between the small-scale wind and the surrounding medium.

Based on these analytical foundations, the two key assumptions of this study then are (i) that AGN drive winds with well-defined, terminal velocities, energy and momentum fluxes at small, unresolved scales and (ii) that the processes responsible for driving small-scale winds can be decoupled from those powering large-scale outflows. The latter is valid as long as the free-expansion radius ($\sim 1 \-- 100 \, \rm pc$) of the wind significantly exceeds the radius at which it is initially driven. While the generation of the small-scale wind cannot be captured in galaxy formation simulations, the scale at which it interacts significantly with the surrounding medium can, in fact, often be resolved.

We model a small-scale AGN wind by prescribing mass, momentum and energy fluxes across a fixed, spherical boundary, which we construct using two rigid, spherical layers of {\sc AREPO} cells. The main parameters describing the small-scale wind are its geometry, its speed and the solid-angle integrated momentum flux, all of which can be selected to obey a variety of observational and theoretical constraints, for different types of winds.

We test our model by analysing the propagation of AGN winds through homogeneous media. We show that our model predicts an initial free-expansion phase, which is later superseded by an energy-driven phase at the correct, analytically-derived radius. Our model reproduces the dynamical evolution of both the free-expansion and energy-driven regimes accurately. The classical structure of the large-scale outflows with its four flow zones (free-streaming wind, shocked wind, shocked and unshocked ambient media), the density and temperature of the different outflow phases and the location and gas phases at which radiative cooling is important all match analytical expectations very accurately. 

Since our model correctly captures the free-expansion of the small-scale wind, it predicts when the outflow thermalises and where it settles into the classical four-zone structure which is assumed in many analytical studies to hold from arbitrarily small radii. For homogenous media, at least, we find that the wind typically thermalises outside the Compton cooling radius and therefore find that momentum-driven solutions, while not impossible, do not always arise. 

We demonstrate that our model possesses good convergence properties down to the typical resolution of hydrodynamic simulations of galaxy formation. 
If the free-expansion radius is not resolved, the initially kinetic energy-dominated wind thermalises just after injection, launching an energy-driven outflow directly.
In particular, we find that the radial momentum, kinetic and thermal energy content of the outflow changes only by factors $\lesssim 4$ over variations in mass resolution of more than 3 orders of magnitude.

In order to test our new model in a more typical setup, we apply it to simulations of an isolated disc galaxy embedded in a galactic halo with $M_{\rm 200} \, = \, 10^{12} \, \rm M_\odot$, focussing on the ability of small-scale winds with speed $v_{\rm w} \, = \, 5000 \, \rm km \, s^{-1}$ and $v_{\rm w} \, = \, 30000 \, \rm km \, s^{-1}$ to power galactic outflows. The winds typically thermalise at scales $\lesssim 50 \, \rm pc$ and quickly evolve into energy-driven bubbles which propagate along the disc rotation axis.
The transition from a small-scale wind into a powerful super-wind, however, only occurs if the pressure gradient generated by the energy-driven bubble significantly exceeds the gravitational potential gradient.
This condition introduces a minimum AGN luminosity, above which the pressure of the energy-driven bubble becomes sufficient to power a large-scale outflow.

At a given small-scale wind speed, the mean speed of the large-scale outflow scales with the AGN luminosity as $v_{\rm out} \propto L_{\rm AGN}^{0.165}$. For $v_{\rm w} \, = \, 5000 \, \rm km \, s^{-1}$, it ranges from $\approx 300 \, \rm km \, s^{-1}$ at $L \, = \, 10^{45} \, \rm erg \, s^{-1}$ to $\approx 1400 \, \rm km \, s^{-1}$ at $L \, = \, 5 \times 10^{47} \, \rm erg \, s^{-1}$. If $v_{\rm w} \, = \, 30000 \, \rm km \, s^{-1}$, the mean outflow speed is higher by about $400 \, \rm km \, s^{-1}$ at any given AGN luminosity.
At the highest luminosities, where the work done by the confining pressure of ambient gas on the outflow is negligible, the large-scale outflows attain momentum fluxes $> L_{\rm AGN}/c$, for $v_{\rm w} \, = \, 30000 \, \rm km \, s^{-1}$, and $\approx L_{\rm AGN}/c$, for $v_{\rm w} \, = \, 5000 \, \rm km \, s^{-1}$. Momentum fluxes $\approx L_{\rm AGN}/c$ are indeed expected if the energy carried by the small-scale wind represents a small enough fraction of the AGN bolometric luminosity, even if it thermalises and develops into an energy-driven outflow. At high luminosities, the outflow kinetic luminosities approach the theoretical maximum $(L_{\rm AGN}/2) (v_{\rm w}/c)$, corresponding to $5 \% L_{\rm AGN}$ for the high-velocity wind case and $0.8 \% L_{\rm AGN}$ for the moderate-velocity wind. At intermediate and low AGN luminosities, when the work done by outflows as they expand into their surroundings constitutes a larger fraction of the available energy, the kinetic luminosities can drop by more than an order of magnitude.

When present, large-scale outflows affect their host galaxies via two main channels: (i) removal and destruction of high-density gas in the nucleus, which operates immediately when the AGN outburst begins, and (ii) suppression of halo gas accretion, which is more gradual and important when star formation activity moves to the outskirts of the galaxy and ejection becomes less efficient. Even as the star formation rate drops, the total star-forming gas mass of the galaxy increases. The star formation suppression is possible, because AGN winds, which only couple directly to the innermost few $100 \, \rm pc$, efficiently remove the densest and hence most star-forming gas from the galactic nucleus. After $350 \, \rm Myr$ of intermittent AGN wind activity, we find reductions in the total stellar mass of about $(4 \-- 10) \times 10^9 \, \rm M_\odot$ with respect to a simulation with no AGN feedback, where the stellar mass is $\approx 1.77 \times 10^{10} \, \rm M_\odot$. The magnitude of the stellar mass reduction increases with the speed of the small-scale wind and the AGN luminosity.

Our new model allows us to predict the generation of large-scale outflows based on the properties of small-scale winds in a physically validated and meaningful way.
It opens up the possibility to much more rigorously study the impact of AGN-driven winds on accretion flows and black hole self-regulation, quantify their effect on the interstellar medium and its ability to form stars and establish whether small-scale winds driven from the immediate vicinity of AGN can shape the evolution of galaxy populations.

\section*{Acknowledgements}
We thank Kastytis Zubovas for a prompt, constructive and thorough referee report. TC gratefully acknowledges Ildar Khabibulin and Francesco Tombesi for enlightening discussions and Martin Haehnelt for providing many helpful comments on the manuscript.

\section*{Data Availability}
The data underlying this article will be shared on reasonable request to the corresponding author.

\appendix
\section{Energy-driven outflow dynamics}
\label{sec:app:energy-driven}
In the energy-driven phase, the cooling time of the shocked wind phase is long compared to the outflow timescale.
If the ambient medium has spatially-constant density and is spherically symmetric, the thermal pressure of the internal bubble composed of shocked wind will sweep-up a shell composed of shocked ambient gas.
Its dynamics is captured by the same equation of motion as in the classic Sedov explosion, with the difference that energy injection into the internal bubble occurs continuously, or for a prolonged period, rather than in a short burst.

We consider a bubble with pressure P, volume V and internal energy $PV / (\gamma - 1) \, = \, (3/2) PV$ for $\gamma \, = \, 5/3$ expanding through a medium with constant density.
We neglect gravity, assume the pressure of the ambient medium to be $\ll P$ and that the ambient medium is static. These assumptions do often break down in more realistic applications (as illustrated in e.g. Section~\ref{sec:discs}). However, they allow us to derive analytical solutions for the dynamics of the outflowing shell and to borrow from literature results addressing collisions between winds and homogeneous media \citep[e.g.][]{Weaver:77} that are used here to validate our numerical model.

The thermal energy is assumed to be the dominant component, which applies because the bubble in fact forms through shock-heating of the supersonic AGN wind.
If thermal energy is added to the bubble at a rate of $\epsilon L_{\rm AGN}$, for some efficiency $\epsilon$, then integrating the energy flux density (Eq.~\ref{eqs_conservation_e}) over a spherical shell of radius $R_{\rm sh}$ gives
\begin{equation}
\frac{3}{2} \frac{d}{dt} \left( \frac{4 \pi}{3} P R_{\rm sh}^3 \right) \, = \, \epsilon L_{\rm AGN} - 4 \pi R^2 P \dot{R}_{\rm sh} \, ,
\end{equation}
which can be simplified to
\begin{equation}
2 \pi R_{\rm sh}^3 \dot{P} + 10\pi R_{\rm sh}^2 \dot{R}_{\rm sh} P - \epsilon L_{\rm AGN} \, = \, 0 \, .
\label{eq_appendix_energy}
\end{equation}
The momentum equation of the outflowing shell is simply
\begin{equation}
\frac{d}{dt} \left( \frac{4 \pi}{3} \rho_{\rm 0} R_{\rm sh}^3 \dot{R}_{\rm sh} \right)  \, = \, 4 \pi R_{\rm sh}^2 P \, ,
\label{eq_appendix_pressure}
\end{equation}
where $\rho_{\rm 0}$ is the density of the ambient medium through which the shell propagates.
Replacing the pressure $P$ in Eq.~\ref{eq_appendix_energy} with the expression found in Eq.~\ref{eq_appendix_pressure} gives the equation of motion
\begin{equation}
\frac{2\pi}{3}\dddot{R} R^4 + 8\pi \ddot{R} \dot{R} R^3 + 10 \pi \dot{R}^3 R^2 - \frac{\epsilon L_{\rm AGN}}{\rho_{\rm 0}} \, = \, 0 \, .
\label{eq_appendix_motion}
\end{equation}
Eq.~\ref{eq_appendix_motion} can be solved by looking for a power law solution of the form $R_{\rm sh}(t) \propto t^\alpha$, which gives $\alpha \, = \, 3/5$, i.e. $R_{\rm sh}(t) \propto t^{3/5}$ and $\dot{R}_{\rm sh}(t) \propto t^{-2/5}$. The full solution reads
\begin{eqnarray}
\label{eq_solution}
R_{\rm sh} (t) \, & = & \, \left( \frac{125}{154 \pi} \frac{\epsilon L_{\rm AGN}}{\rho_{\rm 0}} \right)^{1/5} t^{3/5} + R(t \, = \, 0) \, \\
                       \, & \approx & \, \left(\frac{\epsilon}{0.05} \right)^{1/5} \left( \frac{L_{\rm AGN}}{10^{45} \, \rm erg \, s^{-1}} \right)^{1/5} \left( \frac{n_{\rm H}}{\mathrm{cm^{-3}}} \right)^{-1/5} \left( \frac{t}{\mathrm{Myr}} \right)^{3/5} \, \rm kpc \, , \nonumber
\end{eqnarray}
where $R(t \, = \, 0)$ was taken to be zero in the last step.

The result in Eq.~\ref{eq_solution} only holds in the case of a homogeneous ambient medium of fixed density and the equation of motion will differ depending on the assumed density profile.
A general equation of motion for energy-driven shells and its solution for isothermal, NFW and Hernquist profiles are presented in \citet{Zubovas:12b}, while solutions for general power law profiles can be found in Appendices A and B of \citet{Faucher-Giguere:12}.
Equally, the equation of motion (Eq.~\ref{eq_appendix_motion}) and its solution are only valid in the energy-driven limit and do not apply for the early free-expansion phase or for a potential momentum-driven phase.
Finally, we consider the appropriate value of the efficiency $\epsilon$ appearing in Eq.~\ref{eq_appendix_motion} and in its solution.
Detailed numerical calculations \citep{Weaver:77} show that the kinetic energy of the shocked wind phase is negligible.
For this reason, $\epsilon$ can be equated to the kinetic efficiency of the AGN wind (Eq.~\ref{eq_efficiency}), i.e. $\epsilon \approx \, \frac{\tau \beta}{2}$.

\section{Additional numerical tests}
\label{sec:app:numerical}

\begin{figure*}
	\includegraphics[width=0.475\textwidth]{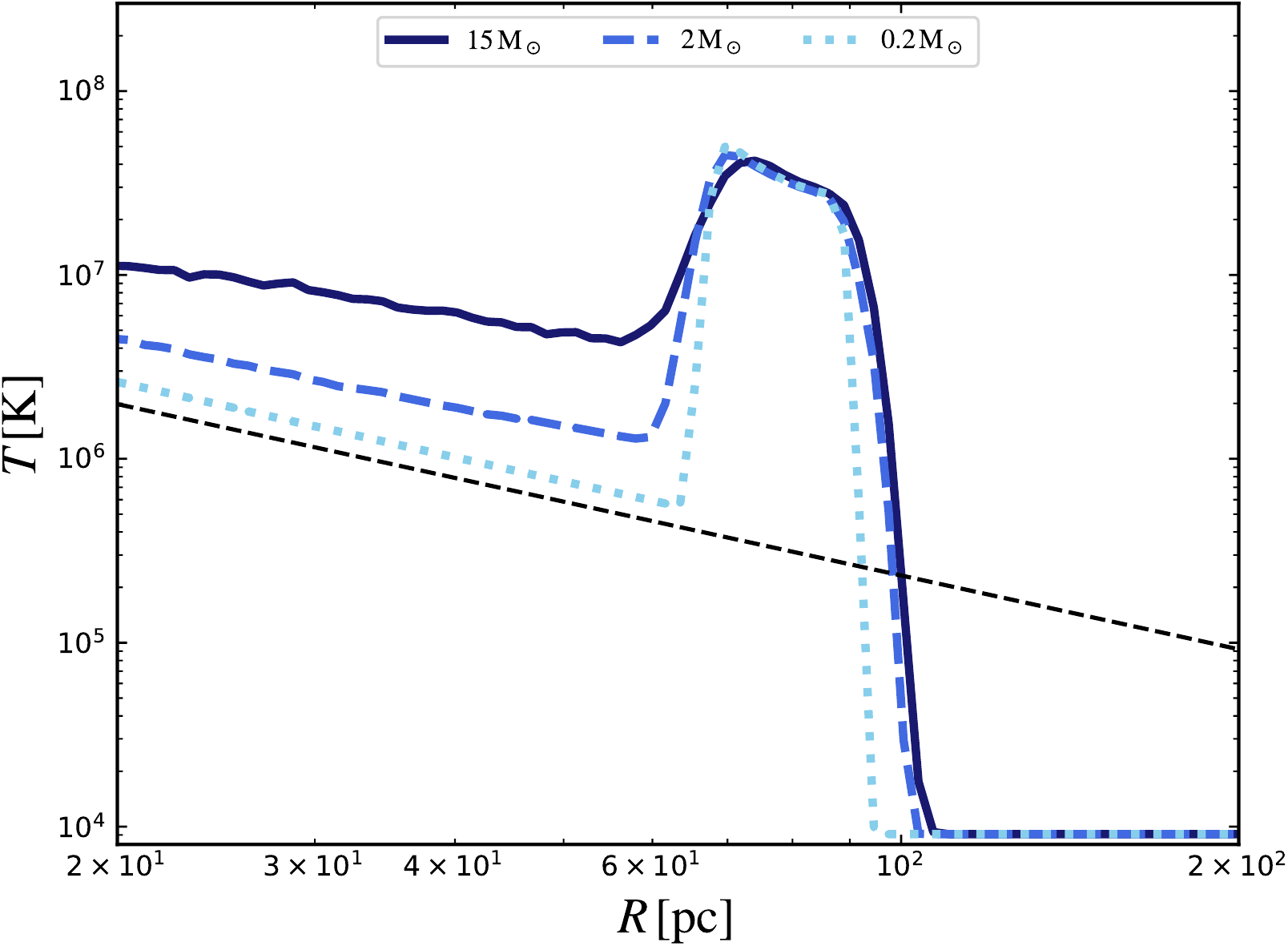}
	\includegraphics[width=0.475\textwidth]{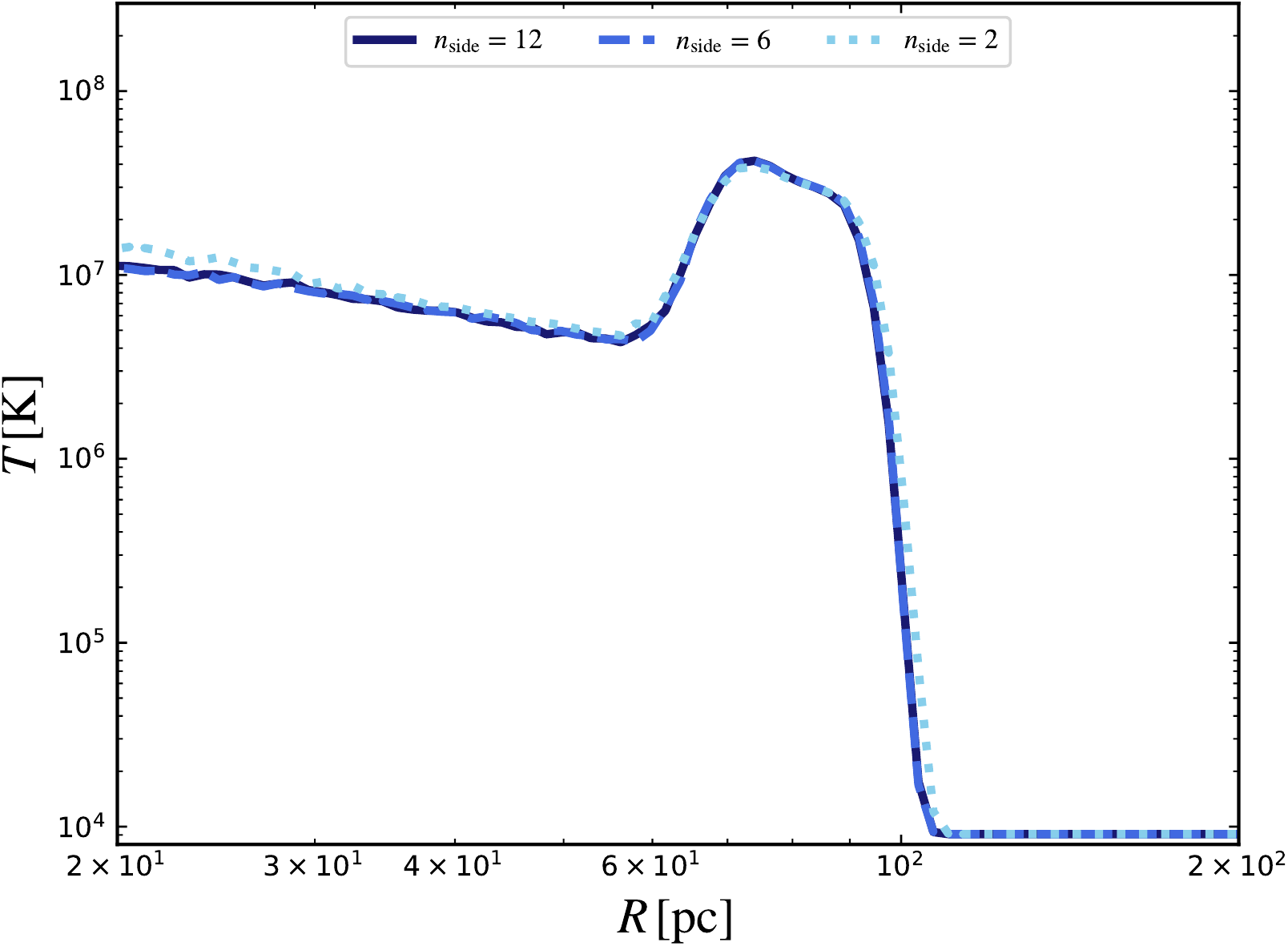}
	\caption{Left: Temperature profiles for simulations performed at different mass resolutions. While the position of the temperature peak is mostly insensitive to resolution, the normalisation of the temperature profile decreases as $m_{\rm target}$ drops. For sufficiently low $m_{\rm target}$, the temperature profile matches the expected temperature profile of the wind. Right: Dependence of the temperature profile on the number with which the wind injection boundary cell layers are sampled.}\
	\label{fig_convergence}
\end{figure*}

In Section~\ref{sec:transitionrad}, we showed that the normalisation of the wind temperature profile can be higher than expected based on the choice of $T_{\rm w}$.
Here we present various numerical tests that show that this issue (i) does not affect the dynamics of the outflow and (ii) becomes less important with increasing resolution.

We perform a number of simulations following the propagation of small-scale winds through a homogeneous medium with $n_{\rm H} \, = \, 1 \rm cm^{-3}$. We use $r_{\rm sp} \, = \, 8 \, \rm pc$, $\beta \, = \, 0.01$, $L_{\rm AGN} \, = \, 10^{45} \, \rm erg \, s^{-1}$ and $T_{\rm W} \, = \, 5 \times 10^6 \, \rm K$.
We explore various mass resolution values, varying $m_{\rm target} \, = \, 15 \, \rm M_\odot$ to $m_{\rm target} \, = \, 0.2 \, \rm M_\odot$.

In the left-hand panel Fig.~\ref{fig_convergence}, we test how the profile depends on the overall mass resolution of the simulations.
As resolution increases, (i) the position of the shocks move slightly towards smaller radii, (ii) the shocks become sharper and (iii) the normalisation of the temperature profile for the adiabatic section of the flow drops. We also see that, for sufficiently high resolution, the wind temperature profile converges on the expected value (dashed, grey line).  
Since the winds in our simulations are highly supersonic, the choice of $T_{\rm W}$ is unimportant for the dynamics of the emerging large-scale outflows.
In future studies employing our new model, the wind temperature, however, will have to be interpreted cautiously and with this caveat in mind.
Besides increasing the numerical resolution, the problem we have identified here is likely to be relieved by using higher-order hydrodynamic codes.

The temperature profile is, otherwise, only weakly sensitive to the number of resolution elements in the wind injection sphere.
In the right-hand panel of Fig.~\ref{fig_convergence}, we show temperature profiles for spherical boundaries generated with $n_{\rm side}$ ranging from $n_{\rm side} \, = \, 2$ to $n_{\rm side} \, = \, 12$. We find small temperature decrements as $n_{\rm side}$ increases from small values $n_{\rm side} \lesssim 4$, but also see that the profiles saturate for larger $n_{\rm side}$.
This saturation occurs when the number of pixels in the wind sphere starts exceeding that of the number of resolution elements in the ambient medium with which interacts directly.
For the same reason, choosing a low $n_{\rm side}$ leads to clear departures from spherical symmetry in the wind's properties.
There is, however, no advantage in increasing $n_{\rm side}$ indefinitely, as the incoming fluxes from a large number of small cells simply merge into the bigger, conventional {\sc AREPO} cells.
Optimally, $n_{\rm side}$ is chosen such that there is roughly one conventional {\sc AREPO} cell for every wind injection boundary cell.



\bibliographystyle{mnras}
\bibliography{lit} 

\begin{thebibliography}{}
\makeatletter
\relax
\def\mn@urlcharsother{\let\do\@makeother \do\$\do\&\do\#\do\^\do\_\do\%\do\~}
\def\mn@doi{\begingroup\mn@urlcharsother \@ifnextchar [ {\mn@doi@}
  {\mn@doi@[]}}
\def\mn@doi@[#1]#2{\def\@tempa{#1}\ifx\@tempa\@empty \href
  {http://dx.doi.org/#2} {doi:#2}\else \href {http://dx.doi.org/#2} {#1}\fi
  \endgroup}
\def\mn@eprint#1#2{\mn@eprint@#1:#2::\@nil}
\def\mn@eprint@arXiv#1{\href {http://arxiv.org/abs/#1} {{\tt arXiv:#1}}}
\def\mn@eprint@dblp#1{\href {http://dblp.uni-trier.de/rec/bibtex/#1.xml}
  {dblp:#1}}
\def\mn@eprint@#1:#2:#3:#4\@nil{\def\@tempa {#1}\def\@tempb {#2}\def\@tempc
  {#3}\ifx \@tempc \@empty \let \@tempc \@tempb \let \@tempb \@tempa \fi \ifx
  \@tempb \@empty \def\@tempb {arXiv}\fi \@ifundefined
  {mn@eprint@\@tempb}{\@tempb:\@tempc}{\expandafter \expandafter \csname
  mn@eprint@\@tempb\endcsname \expandafter{\@tempc}}}

\bibitem[\protect\citeauthoryear{{Angl{\'e}s-Alc{\'a}zar}, {Dav{\'e}},
  {Faucher-Gigu{\`e}re}, {{\"O}zel}  \& {Hopkins}}{{Angl{\'e}s-Alc{\'a}zar}
  et~al.}{2017}]{Angles-Alcazar:17}
{Angl{\'e}s-Alc{\'a}zar} D.,  {Dav{\'e}} R.,  {Faucher-Gigu{\`e}re} C.-A.,
  {{\"O}zel} F.,   {Hopkins} P.~F.,  2017, \mn@doi [\mnras]
  {10.1093/mnras/stw2565}, \href
  {https://ui.adsabs.harvard.edu/abs/2017MNRAS.464.2840A} {464, 2840}

\bibitem[\protect\citeauthoryear{{Barai}, {Murante}, {Borgani}, {Gaspari},
  {Granato}, {Monaco}  \& {Ragone-Figueroa}}{{Barai} et~al.}{2016}]{Barai:16}
{Barai} P.,  {Murante} G.,  {Borgani} S.,  {Gaspari} M.,  {Granato} G.~L.,
  {Monaco} P.,   {Ragone-Figueroa} C.,  2016, \mn@doi [\mnras]
  {10.1093/mnras/stw1389}, \href
  {https://ui.adsabs.harvard.edu/abs/2016MNRAS.461.1548B} {461, 1548}

\bibitem[\protect\citeauthoryear{{Barai}, {Gallerani}, {Pallottini}, {Ferrara},
  {Marconi}, {Cicone}, {Maiolino}  \& {Carniani}}{{Barai}
  et~al.}{2018}]{Barai:18}
{Barai} P.,  {Gallerani} S.,  {Pallottini} A.,  {Ferrara} A.,  {Marconi} A.,
  {Cicone} C.,  {Maiolino} R.,   {Carniani} S.,  2018, \mn@doi [\mnras]
  {10.1093/mnras/stx2563}, \href
  {https://ui.adsabs.harvard.edu/abs/2018MNRAS.473.4003B} {473, 4003}

\bibitem[\protect\citeauthoryear{{Barro} et~al.,}{{Barro}
  et~al.}{2013}]{Barro:13}
{Barro} G.,  et~al., 2013, \mn@doi [\apj] {10.1088/0004-637X/765/2/104}, \href
  {https://ui.adsabs.harvard.edu/abs/2013ApJ...765..104B} {765, 104}

\bibitem[\protect\citeauthoryear{{Biernacki} \& {Teyssier}}{{Biernacki} \&
  {Teyssier}}{2018}]{Biernacki:18}
{Biernacki} P.,  {Teyssier} R.,  2018, \mn@doi [\mnras] {10.1093/mnras/sty216},
  \href {https://ui.adsabs.harvard.edu/abs/2018MNRAS.475.5688B} {475, 5688}

\bibitem[\protect\citeauthoryear{{Biernacki}, {Teyssier}  \&
  {Bleuler}}{{Biernacki} et~al.}{2017}]{Biernacki:17}
{Biernacki} P.,  {Teyssier} R.,   {Bleuler} A.,  2017, \mn@doi [\mnras]
  {10.1093/mnras/stx845}, \href
  {https://ui.adsabs.harvard.edu/abs/2017MNRAS.469..295B} {469, 295}

\bibitem[\protect\citeauthoryear{{Booth} \& {Schaye}}{{Booth} \&
  {Schaye}}{2009}]{Booth:09}
{Booth} C.~M.,  {Schaye} J.,  2009, \mn@doi [\mnras]
  {10.1111/j.1365-2966.2009.15043.x}, \href
  {https://ui.adsabs.harvard.edu/abs/2009MNRAS.398...53B} {398, 53}

\bibitem[\protect\citeauthoryear{{Bourne}, {Zubovas}  \& {Nayakshin}}{{Bourne}
  et~al.}{2015}]{Bourne:15}
{Bourne} M.~A.,  {Zubovas} K.,   {Nayakshin} S.,  2015, \mn@doi [\mnras]
  {10.1093/mnras/stv1730}, \href
  {https://ui.adsabs.harvard.edu/abs/2015MNRAS.453.1829B} {453, 1829}

\bibitem[\protect\citeauthoryear{{Bourne}, {Sijacki}  \& {Puchwein}}{{Bourne}
  et~al.}{2019}]{Bourne:19}
{Bourne} M.~A.,  {Sijacki} D.,   {Puchwein} E.,  2019, \mn@doi [\mnras]
  {10.1093/mnras/stz2604}, \href
  {https://ui.adsabs.harvard.edu/abs/2019MNRAS.490..343B} {490, 343}

\bibitem[\protect\citeauthoryear{{Bower}, {Benson}, {Malbon}, {Helly}, {Frenk},
  {Baugh}, {Cole}  \& {Lacey}}{{Bower} et~al.}{2006}]{Bower:06}
{Bower} R.~G.,  {Benson} A.~J.,  {Malbon} R.,  {Helly} J.~C.,  {Frenk} C.~S.,
  {Baugh} C.~M.,  {Cole} S.,   {Lacey} C.~G.,  2006, \mn@doi [\mnras]
  {10.1111/j.1365-2966.2006.10519.x}, \href
  {https://ui.adsabs.harvard.edu/abs/2006MNRAS.370..645B} {370, 645}

\bibitem[\protect\citeauthoryear{{Braito} et~al.,}{{Braito}
  et~al.}{2018}]{Braito:18}
{Braito} V.,  et~al., 2018, \mn@doi [\mnras] {10.1093/mnras/sty1697}, \href
  {https://ui.adsabs.harvard.edu/abs/2018MNRAS.479.3592B} {479, 3592}

\bibitem[\protect\citeauthoryear{{Bustamante} \& {Springel}}{{Bustamante} \&
  {Springel}}{2019}]{Bustamante:19}
{Bustamante} S.,  {Springel} V.,  2019, \mn@doi [\mnras]
  {10.1093/mnras/stz2836}, \href
  {https://ui.adsabs.harvard.edu/abs/2019MNRAS.490.4133B} {490, 4133}

\bibitem[\protect\citeauthoryear{{Cappi} et~al.,}{{Cappi}
  et~al.}{2009}]{Cappi:09}
{Cappi} M.,  et~al., 2009, \mn@doi [\aap] {10.1051/0004-6361/200912137}, \href
  {https://ui.adsabs.harvard.edu/abs/2009A&A...504..401C} {504, 401}

\bibitem[\protect\citeauthoryear{{Castor}, {McCray}  \& {Weaver}}{{Castor}
  et~al.}{1975}]{Castor:75}
{Castor} J.,  {McCray} R.,   {Weaver} R.,  1975, \mn@doi [\apjl]
  {10.1086/181908}, \href
  {https://ui.adsabs.harvard.edu/abs/1975ApJ...200L.107C} {200, L107}

\bibitem[\protect\citeauthoryear{{Choi}, {Ostriker}, {Naab}  \&
  {Johansson}}{{Choi} et~al.}{2012}]{Choi:12}
{Choi} E.,  {Ostriker} J.~P.,  {Naab} T.,   {Johansson} P.~H.,  2012, \mn@doi
  [\apj] {10.1088/0004-637X/754/2/125}, \href
  {https://ui.adsabs.harvard.edu/abs/2012ApJ...754..125C} {754, 125}

\bibitem[\protect\citeauthoryear{{Choi}, {Somerville}, {Ostriker}, {Naab}  \&
  {Hirschmann}}{{Choi} et~al.}{2018}]{Choi:18}
{Choi} E.,  {Somerville} R.~S.,  {Ostriker} J.~P.,  {Naab} T.,   {Hirschmann}
  M.,  2018, \mn@doi [\apj] {10.3847/1538-4357/aae076}, \href
  {https://ui.adsabs.harvard.edu/abs/2018ApJ...866...91C} {866, 91}

\bibitem[\protect\citeauthoryear{{Churazov}, {Br{\"u}ggen}, {Kaiser},
  {B{\"o}hringer}  \& {Forman}}{{Churazov} et~al.}{2001}]{Churazov:01}
{Churazov} E.,  {Br{\"u}ggen} M.,  {Kaiser} C.~R.,  {B{\"o}hringer} H.,
  {Forman} W.,  2001, \mn@doi [\apj] {10.1086/321357}, \href
  {https://ui.adsabs.harvard.edu/abs/2001ApJ...554..261C} {554, 261}

\bibitem[\protect\citeauthoryear{{Cicone} et~al.,}{{Cicone}
  et~al.}{2014}]{Cicone:14}
{Cicone} C.,  et~al., 2014, \mn@doi [\aap] {10.1051/0004-6361/201322464}, \href
  {https://ui.adsabs.harvard.edu/abs/2014A&A...562A..21C} {562, A21}

\bibitem[\protect\citeauthoryear{{Cicone} et~al.,}{{Cicone}
  et~al.}{2015}]{Cicone:15}
{Cicone} C.,  et~al., 2015, \mn@doi [\aap] {10.1051/0004-6361/201424980}, \href
  {https://ui.adsabs.harvard.edu/abs/2015A&A...574A..14C} {574, A14}

\bibitem[\protect\citeauthoryear{{Contopoulos} \& {Lovelace}}{{Contopoulos} \&
  {Lovelace}}{1994}]{Contopoulos:94}
{Contopoulos} J.,  {Lovelace} R.~V.~E.,  1994, \mn@doi [\apj] {10.1086/174307},
  \href {https://ui.adsabs.harvard.edu/abs/1994ApJ...429..139C} {429, 139}

\bibitem[\protect\citeauthoryear{{Costa}, {Sijacki}  \& {Haehnelt}}{{Costa}
  et~al.}{2014}]{Costa:14}
{Costa} T.,  {Sijacki} D.,   {Haehnelt} M.~G.,  2014, \mn@doi [\mnras]
  {10.1093/mnras/stu1632}, \href
  {http://adsabs.harvard.edu/abs/2014MNRAS.444.2355C} {444, 2355}

\bibitem[\protect\citeauthoryear{{Costa}, {Sijacki}  \& {Haehnelt}}{{Costa}
  et~al.}{2015}]{Costa:15}
{Costa} T.,  {Sijacki} D.,   {Haehnelt} M.~G.,  2015, \mn@doi [\mnras]
  {10.1093/mnrasl/slu193}, \href
  {http://adsabs.harvard.edu/abs/2015MNRAS.448L..30C} {448, L30}

\bibitem[\protect\citeauthoryear{{Costa}, {Rosdahl}, {Sijacki}  \&
  {Haehnelt}}{{Costa} et~al.}{2018a}]{Costa:18a}
{Costa} T.,  {Rosdahl} J.,  {Sijacki} D.,   {Haehnelt} M.~G.,  2018a, \mn@doi
  [\mnras] {10.1093/mnras/stx2598}, \href
  {https://ui.adsabs.harvard.edu/abs/2018MNRAS.473.4197C} {473, 4197}

\bibitem[\protect\citeauthoryear{{Costa}, {Rosdahl}, {Sijacki}  \&
  {Haehnelt}}{{Costa} et~al.}{2018b}]{Costa:18b}
{Costa} T.,  {Rosdahl} J.,  {Sijacki} D.,   {Haehnelt} M.~G.,  2018b, \mn@doi
  [\mnras] {10.1093/mnras/sty1514}, \href
  {http://adsabs.harvard.edu/abs/2018MNRAS.479.2079C} {479, 2079}

\bibitem[\protect\citeauthoryear{{Curtis} \& {Sijacki}}{{Curtis} \&
  {Sijacki}}{2015}]{Curtis:15}
{Curtis} M.,  {Sijacki} D.,  2015, \mn@doi [\mnras] {10.1093/mnras/stv2246},
  \href {https://ui.adsabs.harvard.edu/abs/2015MNRAS.454.3445C} {454, 3445}

\bibitem[\protect\citeauthoryear{{Curtis} \& {Sijacki}}{{Curtis} \&
  {Sijacki}}{2016}]{Curtis:16}
{Curtis} M.,  {Sijacki} D.,  2016, \mn@doi [\mnras] {10.1093/mnrasl/slv199},
  \href {http://adsabs.harvard.edu/abs/2016MNRAS.457L..34C} {457, L34}

\bibitem[\protect\citeauthoryear{{Dav{\'e}}, {Angl{\'e}s-Alc{\'a}zar},
  {Narayanan}, {Li}, {Rafieferantsoa}  \& {Appleby}}{{Dav{\'e}}
  et~al.}{2019}]{Dave:19}
{Dav{\'e}} R.,  {Angl{\'e}s-Alc{\'a}zar} D.,  {Narayanan} D.,  {Li} Q.,
  {Rafieferantsoa} M.~H.,   {Appleby} S.,  2019, \mn@doi [\mnras]
  {10.1093/mnras/stz937}, \href
  {https://ui.adsabs.harvard.edu/abs/2019MNRAS.486.2827D} {486, 2827}

\bibitem[\protect\citeauthoryear{{Di Matteo}, {Springel}  \& {Hernquist}}{{Di
  Matteo} et~al.}{2005}]{DiMatteo:05}
{Di Matteo} T.,  {Springel} V.,   {Hernquist} L.,  2005, \mn@doi [\nat]
  {10.1038/nature03335}, \href
  {https://ui.adsabs.harvard.edu/abs/2005Natur.433..604D} {433, 604}

\bibitem[\protect\citeauthoryear{{Diesing} \& {Caprioli}}{{Diesing} \&
  {Caprioli}}{2018}]{Diesing:18}
{Diesing} R.,  {Caprioli} D.,  2018, \mn@doi [\prl]
  {10.1103/PhysRevLett.121.091101}, \href
  {https://ui.adsabs.harvard.edu/abs/2018PhRvL.121i1101D} {121, 091101}

\bibitem[\protect\citeauthoryear{{Dubois}, {Devriendt}, {Slyz}  \&
  {Teyssier}}{{Dubois} et~al.}{2012}]{Dubois:12b}
{Dubois} Y.,  {Devriendt} J.,  {Slyz} A.,   {Teyssier} R.,  2012, \mn@doi
  [\mnras] {10.1111/j.1365-2966.2011.20236.x}, \href
  {https://ui.adsabs.harvard.edu/abs/2012MNRAS.420.2662D} {420, 2662}

\bibitem[\protect\citeauthoryear{{Dubois}, {Gavazzi}, {Peirani}  \&
  {Silk}}{{Dubois} et~al.}{2013}]{Dubois:13}
{Dubois} Y.,  {Gavazzi} R.,  {Peirani} S.,   {Silk} J.,  2013, \mn@doi [\mnras]
  {10.1093/mnras/stt997}, \href
  {https://ui.adsabs.harvard.edu/abs/2013MNRAS.433.3297D} {433, 3297}

\bibitem[\protect\citeauthoryear{{Dubois}, {Peirani}, {Pichon}, {Devriendt},
  {Gavazzi}, {Welker}  \& {Volonteri}}{{Dubois} et~al.}{2016}]{Dubois:16}
{Dubois} Y.,  {Peirani} S.,  {Pichon} C.,  {Devriendt} J.,  {Gavazzi} R.,
  {Welker} C.,   {Volonteri} M.,  2016, \mn@doi [\mnras]
  {10.1093/mnras/stw2265}, \href
  {https://ui.adsabs.harvard.edu/abs/2016MNRAS.463.3948D} {463, 3948}

\bibitem[\protect\citeauthoryear{{Fabian}}{{Fabian}}{1999}]{Fabian:99}
{Fabian} A.~C.,  1999, \mn@doi [\mnras] {10.1046/j.1365-8711.1999.03017.x},
  \href {http://adsabs.harvard.edu/abs/1999MNRAS.308L..39F} {308, L39}

\bibitem[\protect\citeauthoryear{{Fabian}}{{Fabian}}{2012}]{Fabian:12}
{Fabian} A.~C.,  2012, \mn@doi [\araa] {10.1146/annurev-astro-081811-125521},
  \href {https://ui.adsabs.harvard.edu/abs/2012ARA&A..50..455F} {50, 455}

\bibitem[\protect\citeauthoryear{{Fabian} et~al.,}{{Fabian}
  et~al.}{2011}]{Fabian:11}
{Fabian} A.~C.,  et~al., 2011, \mn@doi [\mnras]
  {10.1111/j.1365-2966.2011.19402.x}, \href
  {https://ui.adsabs.harvard.edu/abs/2011MNRAS.418.2154F} {418, 2154}

\bibitem[\protect\citeauthoryear{{Faucher-Gigu{\`e}re} \&
  {Quataert}}{{Faucher-Gigu{\`e}re} \& {Quataert}}{2012}]{Faucher-Giguere:12}
{Faucher-Gigu{\`e}re} C.-A.,  {Quataert} E.,  2012, \mn@doi [\mnras]
  {10.1111/j.1365-2966.2012.21512.x}, \href
  {http://adsabs.harvard.edu/abs/2012MNRAS.425..605F} {425, 605}

\bibitem[\protect\citeauthoryear{{Faucher-Gigu{\`e}re}, {Lidz}, {Zaldarriaga}
  \& {Hernquist}}{{Faucher-Gigu{\`e}re} et~al.}{2009}]{Faucher-Giguere:09}
{Faucher-Gigu{\`e}re} C.-A.,  {Lidz} A.,  {Zaldarriaga} M.,   {Hernquist} L.,
  2009, \mn@doi [\apj] {10.1088/0004-637X/703/2/1416}, \href
  {http://adsabs.harvard.edu/abs/2009ApJ...703.1416F} {703, 1416}

\bibitem[\protect\citeauthoryear{{Federrath} \& {Klessen}}{{Federrath} \&
  {Klessen}}{2012}]{Federrath:12}
{Federrath} C.,  {Klessen} R.~S.,  2012, \mn@doi [\apj]
  {10.1088/0004-637X/761/2/156}, \href
  {https://ui.adsabs.harvard.edu/abs/2012ApJ...761..156F} {761, 156}

\bibitem[\protect\citeauthoryear{{Fiacconi}, {Sijacki}  \&
  {Pringle}}{{Fiacconi} et~al.}{2018}]{Fiacconi:18}
{Fiacconi} D.,  {Sijacki} D.,   {Pringle} J.~E.,  2018, \mn@doi [\mnras]
  {10.1093/mnras/sty893}, \href
  {https://ui.adsabs.harvard.edu/abs/2018MNRAS.477.3807F} {477, 3807}

\bibitem[\protect\citeauthoryear{{Fiore} et~al.,}{{Fiore}
  et~al.}{2017}]{Fiore:17}
{Fiore} F.,  et~al., 2017, \mn@doi [\aap] {10.1051/0004-6361/201629478}, \href
  {https://ui.adsabs.harvard.edu/abs/2017A&A...601A.143F} {601, A143}

\bibitem[\protect\citeauthoryear{{Fluetsch} et~al.,}{{Fluetsch}
  et~al.}{2019}]{Fluetsch:19}
{Fluetsch} A.,  et~al., 2019, \mn@doi [\mnras] {10.1093/mnras/sty3449}, \href
  {https://ui.adsabs.harvard.edu/abs/2019MNRAS.483.4586F} {483, 4586}

\bibitem[\protect\citeauthoryear{{Forman} et~al.,}{{Forman}
  et~al.}{2007}]{Forman:07}
{Forman} W.,  et~al., 2007, \mn@doi [\apj] {10.1086/519480}, \href
  {https://ui.adsabs.harvard.edu/abs/2007ApJ...665.1057F} {665, 1057}

\bibitem[\protect\citeauthoryear{{F{\"o}rster Schreiber} et~al.,}{{F{\"o}rster
  Schreiber} et~al.}{2014}]{Foerster-Schreiber:14}
{F{\"o}rster Schreiber} N.~M.,  et~al., 2014, \mn@doi [\apj]
  {10.1088/0004-637X/787/1/38}, \href
  {https://ui.adsabs.harvard.edu/abs/2014ApJ...787...38F} {787, 38}

\bibitem[\protect\citeauthoryear{{Gabor} \& {Bournaud}}{{Gabor} \&
  {Bournaud}}{2014}]{Gabor:14}
{Gabor} J.~M.,  {Bournaud} F.,  2014, \mn@doi [\mnras] {10.1093/mnras/stu677},
  \href {https://ui.adsabs.harvard.edu/abs/2014MNRAS.441.1615G} {441, 1615}

\bibitem[\protect\citeauthoryear{{Gaspari}, {Brighenti}  \& {Temi}}{{Gaspari}
  et~al.}{2012}]{Gaspari:12}
{Gaspari} M.,  {Brighenti} F.,   {Temi} P.,  2012, \mn@doi [\mnras]
  {10.1111/j.1365-2966.2012.21183.x}, \href
  {https://ui.adsabs.harvard.edu/abs/2012MNRAS.424..190G} {424, 190}

\bibitem[\protect\citeauthoryear{{Gebhardt} et~al.,}{{Gebhardt}
  et~al.}{2000}]{Gebhardt:00}
{Gebhardt} K.,  et~al., 2000, \mn@doi [\apjl] {10.1086/312840}, \href
  {https://ui.adsabs.harvard.edu/abs/2000ApJ...539L..13G} {539, L13}

\bibitem[\protect\citeauthoryear{{Gofford}, {Reeves}, {McLaughlin}, {Braito},
  {Turner}, {Tombesi}  \& {Cappi}}{{Gofford} et~al.}{2015}]{Gofford:15}
{Gofford} J.,  {Reeves} J.~N.,  {McLaughlin} D.~E.,  {Braito} V.,  {Turner}
  T.~J.,  {Tombesi} F.,   {Cappi} M.,  2015, \mn@doi [\mnras]
  {10.1093/mnras/stv1207}, \href
  {https://ui.adsabs.harvard.edu/abs/2015MNRAS.451.4169G} {451, 4169}

\bibitem[\protect\citeauthoryear{{G{\'o}rski}, {Hivon}, {Banday}, {Wand elt},
  {Hansen}, {Reinecke}  \& {Bartelmann}}{{G{\'o}rski} et~al.}{2005}]{Gorski:05}
{G{\'o}rski} K.~M.,  {Hivon} E.,  {Banday} A.~J.,  {Wand elt} B.~D.,  {Hansen}
  F.~K.,  {Reinecke} M.,   {Bartelmann} M.,  2005, \mn@doi [\apj]
  {10.1086/427976}, \href
  {https://ui.adsabs.harvard.edu/abs/2005ApJ...622..759G} {622, 759}

\bibitem[\protect\citeauthoryear{{Gull}}{{Gull}}{1973}]{Gull:73}
{Gull} S.~F.,  1973, \mn@doi [\mnras] {10.1093/mnras/161.1.47}, \href
  {https://ui.adsabs.harvard.edu/abs/1973MNRAS.161...47G} {161, 47}

\bibitem[\protect\citeauthoryear{{H{\"a}ring} \& {Rix}}{{H{\"a}ring} \&
  {Rix}}{2004}]{Haring:04}
{H{\"a}ring} N.,  {Rix} H.-W.,  2004, \mn@doi [\apjl] {10.1086/383567}, \href
  {https://ui.adsabs.harvard.edu/abs/2004ApJ...604L..89H} {604, L89}

\bibitem[\protect\citeauthoryear{{Harrison}, {Alexander}, {Mullaney}  \&
  {Swinbank}}{{Harrison} et~al.}{2014}]{Harrison:14}
{Harrison} C.~M.,  {Alexander} D.~M.,  {Mullaney} J.~R.,   {Swinbank} A.~M.,
  2014, \mn@doi [\mnras] {10.1093/mnras/stu515}, \href
  {https://ui.adsabs.harvard.edu/abs/2014MNRAS.441.3306H} {441, 3306}

\bibitem[\protect\citeauthoryear{{Harrison}, {Costa}, {Tadhunter},
  {Fl{\"u}tsch}, {Kakkad}, {Perna}  \& {Vietri}}{{Harrison}
  et~al.}{2018}]{Harrison:18}
{Harrison} C.~M.,  {Costa} T.,  {Tadhunter} C.~N.,  {Fl{\"u}tsch} A.,  {Kakkad}
  D.,  {Perna} M.,   {Vietri} G.,  2018, \mn@doi [Nature Astronomy]
  {10.1038/s41550-018-0403-6}, \href
  {https://ui.adsabs.harvard.edu/abs/2018NatAs...2..198H} {2, 198}

\bibitem[\protect\citeauthoryear{{Hartwig}, {Volonteri}  \&
  {Dashyan}}{{Hartwig} et~al.}{2018}]{Hartwig:18}
{Hartwig} T.,  {Volonteri} M.,   {Dashyan} G.,  2018, \mn@doi [\mnras]
  {10.1093/mnras/sty229}, \href
  {https://ui.adsabs.harvard.edu/abs/2018MNRAS.476.2288H} {476, 2288}

\bibitem[\protect\citeauthoryear{{Herrera-Camus} et~al.,}{{Herrera-Camus}
  et~al.}{2019}]{Herrera-Camus:19}
{Herrera-Camus} R.,  et~al., 2019, \mn@doi [\apj] {10.3847/1538-4357/aaf6a7},
  \href {https://ui.adsabs.harvard.edu/abs/2019ApJ...871...37H} {871, 37}

\bibitem[\protect\citeauthoryear{{Hopkins} \& {Elvis}}{{Hopkins} \&
  {Elvis}}{2010}]{Hopkins:10}
{Hopkins} P.~F.,  {Elvis} M.,  2010, \mn@doi [\mnras]
  {10.1111/j.1365-2966.2009.15643.x}, \href
  {https://ui.adsabs.harvard.edu/abs/2010MNRAS.401....7H} {401, 7}

\bibitem[\protect\citeauthoryear{{Hopkins}, {Torrey}, {Faucher-Gigu{\`e}re},
  {Quataert}  \& {Murray}}{{Hopkins} et~al.}{2016}]{Hopkins:16}
{Hopkins} P.~F.,  {Torrey} P.,  {Faucher-Gigu{\`e}re} C.-A.,  {Quataert} E.,
  {Murray} N.,  2016, \mn@doi [\mnras] {10.1093/mnras/stw289}, \href
  {https://ui.adsabs.harvard.edu/abs/2016MNRAS.458..816H} {458, 816}

\bibitem[\protect\citeauthoryear{{Hopkins} et~al.,}{{Hopkins}
  et~al.}{2018}]{Hopkins:18}
{Hopkins} P.~F.,  et~al., 2018, \mn@doi [\mnras] {10.1093/mnras/sty674}, \href
  {https://ui.adsabs.harvard.edu/abs/2018MNRAS.477.1578H} {477, 1578}

\bibitem[\protect\citeauthoryear{{Husemann}, {Scharw{\"a}chter}, {Bennert},
  {Mainieri}, {Woo}  \& {Kakkad}}{{Husemann} et~al.}{2016}]{Husemann:16}
{Husemann} B.,  {Scharw{\"a}chter} J.,  {Bennert} V.~N.,  {Mainieri} V.,  {Woo}
  J.~H.,   {Kakkad} D.,  2016, \mn@doi [\aap] {10.1051/0004-6361/201527992},
  \href {https://ui.adsabs.harvard.edu/abs/2016A&A...594A..44H} {594, A44}

\bibitem[\protect\citeauthoryear{{Jarvis} et~al.,}{{Jarvis}
  et~al.}{2019}]{Jarvis:19}
{Jarvis} M.~E.,  et~al., 2019, \mn@doi [\mnras] {10.1093/mnras/stz556}, \href
  {https://ui.adsabs.harvard.edu/abs/2019MNRAS.485.2710J} {485, 2710}

\bibitem[\protect\citeauthoryear{{King}}{{King}}{2003}]{King:03}
{King} A.,  2003, \mn@doi [\apjl] {10.1086/379143}, \href
  {http://adsabs.harvard.edu/abs/2003ApJ...596L..27K} {596, L27}

\bibitem[\protect\citeauthoryear{{King}}{{King}}{2005}]{King:05}
{King} A.,  2005, \mn@doi [\apjl] {10.1086/499430}, \href
  {http://adsabs.harvard.edu/abs/2005ApJ...635L.121K} {635, L121}

\bibitem[\protect\citeauthoryear{{King} \& {Pounds}}{{King} \&
  {Pounds}}{2015}]{King:15}
{King} A.,  {Pounds} K.,  2015, \mn@doi [\araa]
  {10.1146/annurev-astro-082214-122316}, \href
  {https://ui.adsabs.harvard.edu/abs/2015ARA&A..53..115K} {53, 115}

\bibitem[\protect\citeauthoryear{{Kormendy} \& {Ho}}{{Kormendy} \&
  {Ho}}{2013}]{Kormendy:13}
{Kormendy} J.,  {Ho} L.~C.,  2013, \mn@doi [\araa]
  {10.1146/annurev-astro-082708-101811}, \href
  {https://ui.adsabs.harvard.edu/abs/2013ARA&A..51..511K} {51, 511}

\bibitem[\protect\citeauthoryear{{Lansbury}, {Banerji}, {Fabian}  \&
  {Temple}}{{Lansbury} et~al.}{2020}]{Lansbury:19}
{Lansbury} G.~B.,  {Banerji} M.,  {Fabian} A.~C.,   {Temple} M.~J.,  2020,
  \mn@doi [\mnras] {10.1093/mnras/staa1220}, \href
  {https://ui.adsabs.harvard.edu/abs/2020MNRAS.tmp.1350L} {}

\bibitem[\protect\citeauthoryear{{Maiolino} et~al.,}{{Maiolino}
  et~al.}{2012}]{Maiolino:12}
{Maiolino} R.,  et~al., 2012, \mn@doi [\mnras]
  {10.1111/j.1745-3933.2012.01303.x}, \href
  {https://ui.adsabs.harvard.edu/abs/2012MNRAS.425L..66M} {425, L66}

\bibitem[\protect\citeauthoryear{{McConnell} \& {Ma}}{{McConnell} \&
  {Ma}}{2013}]{McConnell:13}
{McConnell} N.~J.,  {Ma} C.-P.,  2013, \mn@doi [\apj]
  {10.1088/0004-637X/764/2/184}, \href
  {https://ui.adsabs.harvard.edu/abs/2013ApJ...764..184M} {764, 184}

\bibitem[\protect\citeauthoryear{{McNamara} et~al.,}{{McNamara}
  et~al.}{2000}]{McNamara:00}
{McNamara} B.~R.,  et~al., 2000, \mn@doi [\apjl] {10.1086/312662}, \href
  {https://ui.adsabs.harvard.edu/abs/2000ApJ...534L.135M} {534, L135}

\bibitem[\protect\citeauthoryear{{Morganti}, {Fogasy}, {Paragi}, {Oosterloo}
  \& {Orienti}}{{Morganti} et~al.}{2013}]{Morganti:13}
{Morganti} R.,  {Fogasy} J.,  {Paragi} Z.,  {Oosterloo} T.,   {Orienti} M.,
  2013, \mn@doi [Science] {10.1126/science.1240436}, \href
  {https://ui.adsabs.harvard.edu/abs/2013Sci...341.1082M} {341, 1082}

\bibitem[\protect\citeauthoryear{{Murray}, {Quataert}  \& {Thompson}}{{Murray}
  et~al.}{2005}]{Murray:05}
{Murray} N.,  {Quataert} E.,   {Thompson} T.~A.,  2005, \mn@doi [\apj]
  {10.1086/426067}, \href {http://adsabs.harvard.edu/abs/2005ApJ...618..569M}
  {618, 569}

\bibitem[\protect\citeauthoryear{{Narayan} \& {Yi}}{{Narayan} \&
  {Yi}}{1994}]{Narayan:94}
{Narayan} R.,  {Yi} I.,  1994, \mn@doi [\apjl] {10.1086/187381}, \href
  {https://ui.adsabs.harvard.edu/abs/1994ApJ...428L..13N} {428, L13}

\bibitem[\protect\citeauthoryear{{Nardini} et~al.,}{{Nardini}
  et~al.}{2015}]{Nardini:15}
{Nardini} E.,  et~al., 2015, \mn@doi [Science] {10.1126/science.1259202}, \href
  {https://ui.adsabs.harvard.edu/abs/2015Sci...347..860N} {347, 860}

\bibitem[\protect\citeauthoryear{{Navarro}, {Frenk}  \& {White}}{{Navarro}
  et~al.}{1997}]{Navarro:97}
{Navarro} J.~F.,  {Frenk} C.~S.,   {White} S. D.~M.,  1997, \mn@doi [\apj]
  {10.1086/304888}, \href
  {https://ui.adsabs.harvard.edu/abs/1997ApJ...490..493N} {490, 493}

\bibitem[\protect\citeauthoryear{{Nayakshin} \& {Power}}{{Nayakshin} \&
  {Power}}{2010}]{Nayakshin:10}
{Nayakshin} S.,  {Power} C.,  2010, \mn@doi [\mnras]
  {10.1111/j.1365-2966.2009.15946.x}, \href
  {https://ui.adsabs.harvard.edu/abs/2010MNRAS.402..789N} {402, 789}

\bibitem[\protect\citeauthoryear{{Nayakshin} \& {Zubovas}}{{Nayakshin} \&
  {Zubovas}}{2012}]{Nayakshin:12}
{Nayakshin} S.,  {Zubovas} K.,  2012, \mn@doi [\mnras]
  {10.1111/j.1365-2966.2012.21950.x}, \href
  {https://ui.adsabs.harvard.edu/abs/2012MNRAS.427..372N} {427, 372}

\bibitem[\protect\citeauthoryear{{Nelson} et~al.,}{{Nelson}
  et~al.}{2019}]{Nelson:19}
{Nelson} D.,  et~al., 2019, \mn@doi [\mnras] {10.1093/mnras/stz2306}, \href
  {https://ui.adsabs.harvard.edu/abs/2019MNRAS.490.3234N} {490, 3234}

\bibitem[\protect\citeauthoryear{{Nesvadba} et~al.,}{{Nesvadba}
  et~al.}{2010}]{Nesvadba:10}
{Nesvadba} N.~P.~H.,  et~al., 2010, \mn@doi [\aap]
  {10.1051/0004-6361/200913333}, \href
  {https://ui.adsabs.harvard.edu/abs/2010A&A...521A..65N} {521, A65}

\bibitem[\protect\citeauthoryear{{Nims}, {Quataert}  \&
  {Faucher-Gigu{\`e}re}}{{Nims} et~al.}{2015}]{Nims:15}
{Nims} J.,  {Quataert} E.,   {Faucher-Gigu{\`e}re} C.-A.,  2015, \mn@doi
  [\mnras] {10.1093/mnras/stu2648}, \href
  {https://ui.adsabs.harvard.edu/abs/2015MNRAS.447.3612N} {447, 3612}

\bibitem[\protect\citeauthoryear{{Nomura} \& {Ohsuga}}{{Nomura} \&
  {Ohsuga}}{2017}]{Nomura:17}
{Nomura} M.,  {Ohsuga} K.,  2017, \mn@doi [\mnras] {10.1093/mnras/stw2877},
  \href {https://ui.adsabs.harvard.edu/abs/2017MNRAS.465.2873N} {465, 2873}

\bibitem[\protect\citeauthoryear{{Nomura}, {Ohsuga}, {Takahashi}, {Wada}  \&
  {Yoshida}}{{Nomura} et~al.}{2016}]{Nomura:16}
{Nomura} M.,  {Ohsuga} K.,  {Takahashi} H.~R.,  {Wada} K.,   {Yoshida} T.,
  2016, \mn@doi [\pasj] {10.1093/pasj/psv124}, \href
  {https://ui.adsabs.harvard.edu/abs/2016PASJ...68...16N} {68, 16}

\bibitem[\protect\citeauthoryear{{Oppenheimer}, {Segers}, {Schaye}, {Richings}
  \& {Crain}}{{Oppenheimer} et~al.}{2018}]{Oppenheimer:18}
{Oppenheimer} B.~D.,  {Segers} M.,  {Schaye} J.,  {Richings} A.~J.,   {Crain}
  R.~A.,  2018, \mn@doi [\mnras] {10.1093/mnras/stx2967}, \href
  {https://ui.adsabs.harvard.edu/abs/2018MNRAS.474.4740O} {474, 4740}

\bibitem[\protect\citeauthoryear{{Ostriker}, {Choi}, {Ciotti}, {Novak}  \&
  {Proga}}{{Ostriker} et~al.}{2010}]{Ostriker:10}
{Ostriker} J.~P.,  {Choi} E.,  {Ciotti} L.,  {Novak} G.~S.,   {Proga} D.,
  2010, \mn@doi [\apj] {10.1088/0004-637X/722/1/642}, \href
  {https://ui.adsabs.harvard.edu/abs/2010ApJ...722..642O} {722, 642}

\bibitem[\protect\citeauthoryear{{Pakmor}, {Springel}, {Bauer}, {Mocz},
  {Munoz}, {Ohlmann}, {Schaal}  \& {Zhu}}{{Pakmor} et~al.}{2016}]{Pakmor:16}
{Pakmor} R.,  {Springel} V.,  {Bauer} A.,  {Mocz} P.,  {Munoz} D.~J.,
  {Ohlmann} S.~T.,  {Schaal} K.,   {Zhu} C.,  2016, \mn@doi [\mnras]
  {10.1093/mnras/stv2380}, \href
  {https://ui.adsabs.harvard.edu/abs/2016MNRAS.455.1134P} {455, 1134}

\bibitem[\protect\citeauthoryear{{Peirani} et~al.,}{{Peirani}
  et~al.}{2017}]{Peirani:17}
{Peirani} S.,  et~al., 2017, \mn@doi [\mnras] {10.1093/mnras/stx2099}, \href
  {https://ui.adsabs.harvard.edu/abs/2017MNRAS.472.2153P} {472, 2153}

\bibitem[\protect\citeauthoryear{{Perna}, {Lanzuisi}, {Brusa}, {Mignoli}  \&
  {Cresci}}{{Perna} et~al.}{2017}]{Perna:17}
{Perna} M.,  {Lanzuisi} G.,  {Brusa} M.,  {Mignoli} M.,   {Cresci} G.,  2017,
  \mn@doi [\aap] {10.1051/0004-6361/201630369}, \href
  {https://ui.adsabs.harvard.edu/abs/2017A&A...603A..99P} {603, A99}

\bibitem[\protect\citeauthoryear{{Pinto} et~al.,}{{Pinto}
  et~al.}{2018}]{Pinto:18}
{Pinto} C.,  et~al., 2018, \mn@doi [\mnras] {10.1093/mnras/sty231}, \href
  {https://ui.adsabs.harvard.edu/abs/2018MNRAS.476.1021P} {476, 1021}

\bibitem[\protect\citeauthoryear{{Pounds} \& {Reeves}}{{Pounds} \&
  {Reeves}}{2009}]{Pounds:09}
{Pounds} K.~A.,  {Reeves} J.~N.,  2009, \mn@doi [\mnras]
  {10.1111/j.1365-2966.2009.14971.x}, \href
  {https://ui.adsabs.harvard.edu/abs/2009MNRAS.397..249P} {397, 249}

\bibitem[\protect\citeauthoryear{{Pounds}, {Reeves}, {King}, {Page}, {O'Brien}
  \& {Turner}}{{Pounds} et~al.}{2003}]{Pounds:03}
{Pounds} K.~A.,  {Reeves} J.~N.,  {King} A.~R.,  {Page} K.~L.,  {O'Brien}
  P.~T.,   {Turner} M.~J.~L.,  2003, \mn@doi [\mnras]
  {10.1046/j.1365-8711.2003.07006.x}, \href
  {https://ui.adsabs.harvard.edu/abs/2003MNRAS.345..705P} {345, 705}

\bibitem[\protect\citeauthoryear{{Prasad}, {Sharma}  \& {Babul}}{{Prasad}
  et~al.}{2017}]{Prasad:17}
{Prasad} D.,  {Sharma} P.,   {Babul} A.,  2017, \mn@doi [\mnras]
  {10.1093/mnras/stx1698}, \href
  {https://ui.adsabs.harvard.edu/abs/2017MNRAS.471.1531P} {471, 1531}

\bibitem[\protect\citeauthoryear{{Proga}, {Stone}  \& {Kallman}}{{Proga}
  et~al.}{2000}]{Proga:00}
{Proga} D.,  {Stone} J.~M.,   {Kallman} T.~R.,  2000, \mn@doi [\apj]
  {10.1086/317154}, \href
  {https://ui.adsabs.harvard.edu/abs/2000ApJ...543..686P} {543, 686}

\bibitem[\protect\citeauthoryear{{Raimundo}, {Fabian}, {Bauer}, {Alexand er},
  {Brandt}, {Luo}, {Vasudevan}  \& {Xue}}{{Raimundo}
  et~al.}{2010}]{Raimundo:10}
{Raimundo} S.~I.,  {Fabian} A.~C.,  {Bauer} F.~E.,  {Alexand er} D.~M.,
  {Brandt} W.~N.,  {Luo} B.,  {Vasudevan} R.~V.,   {Xue} Y.~Q.,  2010, \mn@doi
  [\mnras] {10.1111/j.1365-2966.2010.17234.x}, \href
  {https://ui.adsabs.harvard.edu/abs/2010MNRAS.408.1714R} {408, 1714}

\bibitem[\protect\citeauthoryear{{Ricci} et~al.,}{{Ricci}
  et~al.}{2017}]{Ricci:17}
{Ricci} C.,  et~al., 2017, \mn@doi [\nat] {10.1038/nature23906}, \href
  {https://ui.adsabs.harvard.edu/abs/2017Natur.549..488R} {549, 488}

\bibitem[\protect\citeauthoryear{{Richings} \&
  {Faucher-Gigu{\`e}re}}{{Richings} \&
  {Faucher-Gigu{\`e}re}}{2018a}]{Richings:18}
{Richings} A.~J.,  {Faucher-Gigu{\`e}re} C.-A.,  2018a, \mn@doi [\mnras]
  {10.1093/mnras/stx3014}, \href
  {https://ui.adsabs.harvard.edu/abs/2018MNRAS.474.3673R} {474, 3673}

\bibitem[\protect\citeauthoryear{{Richings} \&
  {Faucher-Gigu{\`e}re}}{{Richings} \&
  {Faucher-Gigu{\`e}re}}{2018b}]{Richings:18b}
{Richings} A.~J.,  {Faucher-Gigu{\`e}re} C.-A.,  2018b, \mn@doi [\mnras]
  {10.1093/mnras/sty1285}, \href
  {https://ui.adsabs.harvard.edu/abs/2018MNRAS.478.3100R} {478, 3100}

\bibitem[\protect\citeauthoryear{{Risaliti} \& {Elvis}}{{Risaliti} \&
  {Elvis}}{2010}]{Risaliti:10}
{Risaliti} G.,  {Elvis} M.,  2010, \mn@doi [\aap]
  {10.1051/0004-6361/200912579}, \href
  {https://ui.adsabs.harvard.edu/abs/2010A&A...516A..89R} {516, A89}

\bibitem[\protect\citeauthoryear{{Roth}, {Kasen}, {Hopkins}  \&
  {Quataert}}{{Roth} et~al.}{2012}]{Roth:12}
{Roth} N.,  {Kasen} D.,  {Hopkins} P.~F.,   {Quataert} E.,  2012, \mn@doi
  [\apj] {10.1088/0004-637X/759/1/36}, \href
  {https://ui.adsabs.harvard.edu/abs/2012ApJ...759...36R} {759, 36}

\bibitem[\protect\citeauthoryear{{Sazonov} \& {Sunyaev}}{{Sazonov} \&
  {Sunyaev}}{2001}]{Sazonov:01}
{Sazonov} S.~Y.,  {Sunyaev} R.~A.,  2001, \mn@doi [Astronomy Letters]
  {10.1134/1.1388915}, \href
  {https://ui.adsabs.harvard.edu/abs/2001AstL...27..481S} {27, 481}

\bibitem[\protect\citeauthoryear{{Sazonov}, {Ostriker}  \& {Sunyaev}}{{Sazonov}
  et~al.}{2004}]{Sazonov:04}
{Sazonov} S.~Y.,  {Ostriker} J.~P.,   {Sunyaev} R.~A.,  2004, \mn@doi [\mnras]
  {10.1111/j.1365-2966.2004.07184.x}, \href
  {https://ui.adsabs.harvard.edu/abs/2004MNRAS.347..144S} {347, 144}

\bibitem[\protect\citeauthoryear{{Schaye} et~al.,}{{Schaye}
  et~al.}{2015}]{Schaye:15}
{Schaye} J.,  et~al., 2015, \mn@doi [\mnras] {10.1093/mnras/stu2058}, \href
  {https://ui.adsabs.harvard.edu/abs/2015MNRAS.446..521S} {446, 521}

\bibitem[\protect\citeauthoryear{{Schiano}}{{Schiano}}{1985}]{Schiano:85}
{Schiano} A.~V.~R.,  1985, \mn@doi [\apj] {10.1086/163680}, \href
  {https://ui.adsabs.harvard.edu/abs/1985ApJ...299...24S} {299, 24}

\bibitem[\protect\citeauthoryear{{Segers}, {Schaye}, {Bower}, {Crain},
  {Schaller}  \& {Theuns}}{{Segers} et~al.}{2016}]{Segers:16}
{Segers} M.~C.,  {Schaye} J.,  {Bower} R.~G.,  {Crain} R.~A.,  {Schaller} M.,
  {Theuns} T.,  2016, \mn@doi [\mnras] {10.1093/mnrasl/slw111}, \href
  {https://ui.adsabs.harvard.edu/abs/2016MNRAS.461L.102S} {461, L102}

\bibitem[\protect\citeauthoryear{{Serafinelli}, {Tombesi}, {Vagnetti},
  {Piconcelli}, {Gaspari}  \& {Saturni}}{{Serafinelli}
  et~al.}{2019}]{Serafinelli:19}
{Serafinelli} R.,  {Tombesi} F.,  {Vagnetti} F.,  {Piconcelli} E.,  {Gaspari}
  M.,   {Saturni} F.~G.,  2019, \mn@doi [\aap] {10.1051/0004-6361/201935275},
  \href {https://ui.adsabs.harvard.edu/abs/2019A&A...627A.121S} {627, A121}

\bibitem[\protect\citeauthoryear{{Shakura} \& {Sunyaev}}{{Shakura} \&
  {Sunyaev}}{1973}]{Shakura:73}
{Shakura} N.~I.,  {Sunyaev} R.~A.,  1973, \aap, \href
  {https://ui.adsabs.harvard.edu/abs/1973A&A....24..337S} {500, 33}

\bibitem[\protect\citeauthoryear{{Sijacki}, {Springel}, {Di Matteo}  \&
  {Hernquist}}{{Sijacki} et~al.}{2007}]{Sijacki:07}
{Sijacki} D.,  {Springel} V.,  {Di Matteo} T.,   {Hernquist} L.,  2007, \mn@doi
  [\mnras] {10.1111/j.1365-2966.2007.12153.x}, \href
  {https://ui.adsabs.harvard.edu/abs/2007MNRAS.380..877S} {380, 877}

\bibitem[\protect\citeauthoryear{{Sijacki}, {Vogelsberger}, {Genel},
  {Springel}, {Torrey}, {Snyder}, {Nelson}  \& {Hernquist}}{{Sijacki}
  et~al.}{2015}]{Sijacki:15}
{Sijacki} D.,  {Vogelsberger} M.,  {Genel} S.,  {Springel} V.,  {Torrey} P.,
  {Snyder} G.~F.,  {Nelson} D.,   {Hernquist} L.,  2015, \mn@doi [\mnras]
  {10.1093/mnras/stv1340}, \href
  {https://ui.adsabs.harvard.edu/abs/2015MNRAS.452..575S} {452, 575}

\bibitem[\protect\citeauthoryear{{Silk} \& {Rees}}{{Silk} \&
  {Rees}}{1998}]{Silk:98}
{Silk} J.,  {Rees} M.~J.,  1998, \aap, \href
  {http://adsabs.harvard.edu/abs/1998A%26A...331L...1S} {331, L1}

\bibitem[\protect\citeauthoryear{{Sirressi} et~al.,}{{Sirressi}
  et~al.}{2019}]{Sirressi:19}
{Sirressi} M.,  et~al., 2019, \mn@doi [\mnras] {10.1093/mnras/stz2249}, \href
  {https://ui.adsabs.harvard.edu/abs/2019MNRAS.489.1927S} {489, 1927}

\bibitem[\protect\citeauthoryear{{S{\k{a}}dowski}, {Narayan}, {Penna}  \&
  {Zhu}}{{S{\k{a}}dowski} et~al.}{2013}]{Sadowski:13}
{S{\k{a}}dowski} A.,  {Narayan} R.,  {Penna} R.,   {Zhu} Y.,  2013, \mn@doi
  [\mnras] {10.1093/mnras/stt1881}, \href
  {https://ui.adsabs.harvard.edu/abs/2013MNRAS.436.3856S} {436, 3856}

\bibitem[\protect\citeauthoryear{{S{\k{a}}dowski}, {Lasota}, {Abramowicz}  \&
  {Narayan}}{{S{\k{a}}dowski} et~al.}{2016}]{Sadowski:16}
{S{\k{a}}dowski} A.,  {Lasota} J.-P.,  {Abramowicz} M.~A.,   {Narayan} R.,
  2016, \mn@doi [\mnras] {10.1093/mnras/stv2854}, \href
  {https://ui.adsabs.harvard.edu/abs/2016MNRAS.456.3915S} {456, 3915}

\bibitem[\protect\citeauthoryear{{Soker}}{{Soker}}{2016}]{Soker:16}
{Soker} N.,  2016, \mn@doi [\nar] {10.1016/j.newar.2016.08.002}, \href
  {https://ui.adsabs.harvard.edu/abs/2016NewAR..75....1S} {75, 1}

\bibitem[\protect\citeauthoryear{{Springel}}{{Springel}}{2010}]{Springel:10}
{Springel} V.,  2010, \mn@doi [\mnras] {10.1111/j.1365-2966.2009.15715.x},
  \href {https://ui.adsabs.harvard.edu/abs/2010MNRAS.401..791S} {401, 791}

\bibitem[\protect\citeauthoryear{{Springel} \& {Hernquist}}{{Springel} \&
  {Hernquist}}{2003}]{Springel:03}
{Springel} V.,  {Hernquist} L.,  2003, \mn@doi [\mnras]
  {10.1046/j.1365-8711.2003.06206.x}, \href
  {https://ui.adsabs.harvard.edu/abs/2003MNRAS.339..289S} {339, 289}

\bibitem[\protect\citeauthoryear{{Springel}, {Di Matteo}  \&
  {Hernquist}}{{Springel} et~al.}{2005}]{Springel:05}
{Springel} V.,  {Di Matteo} T.,   {Hernquist} L.,  2005, \mn@doi [\apjl]
  {10.1086/428772}, \href
  {https://ui.adsabs.harvard.edu/abs/2005ApJ...620L..79S} {620, L79}

\bibitem[\protect\citeauthoryear{{Straatman} et~al.,}{{Straatman}
  et~al.}{2015}]{Straatman:15}
{Straatman} C. M.~S.,  et~al., 2015, \mn@doi [\apjl]
  {10.1088/2041-8205/808/1/L29}, \href
  {https://ui.adsabs.harvard.edu/abs/2015ApJ...808L..29S} {808, L29}

\bibitem[\protect\citeauthoryear{{Sturm} et~al.,}{{Sturm}
  et~al.}{2011}]{Sturm:11}
{Sturm} E.,  et~al., 2011, \mn@doi [\apjl] {10.1088/2041-8205/733/1/L16}, \href
  {https://ui.adsabs.harvard.edu/abs/2011ApJ...733L..16S} {733, L16}

\bibitem[\protect\citeauthoryear{{Tadhunter}, {Morganti}, {Rose}, {Oonk}  \&
  {Oosterloo}}{{Tadhunter} et~al.}{2014}]{Tadhunter:14}
{Tadhunter} C.,  {Morganti} R.,  {Rose} M.,  {Oonk} J.~B.~R.,   {Oosterloo} T.,
   2014, \mn@doi [\nat] {10.1038/nature13520}, \href
  {https://ui.adsabs.harvard.edu/abs/2014Natur.511..440T} {511, 440}

\bibitem[\protect\citeauthoryear{{Taylor} \& {Kobayashi}}{{Taylor} \&
  {Kobayashi}}{2015}]{Taylor:15}
{Taylor} P.,  {Kobayashi} C.,  2015, \mn@doi [\mnras] {10.1093/mnras/stv139},
  \href {https://ui.adsabs.harvard.edu/abs/2015MNRAS.448.1835T} {448, 1835}

\bibitem[\protect\citeauthoryear{{Teyssier}, {Moore}, {Martizzi}, {Dubois}  \&
  {Mayer}}{{Teyssier} et~al.}{2011}]{Teyssier:11}
{Teyssier} R.,  {Moore} B.,  {Martizzi} D.,  {Dubois} Y.,   {Mayer} L.,  2011,
  \mn@doi [\mnras] {10.1111/j.1365-2966.2011.18399.x}, \href
  {https://ui.adsabs.harvard.edu/abs/2011MNRAS.414..195T} {414, 195}

\bibitem[\protect\citeauthoryear{{Tombesi}, {Cappi}, {Reeves}, {Palumbo},
  {Braito}  \& {Dadina}}{{Tombesi} et~al.}{2011}]{Tombesi:11}
{Tombesi} F.,  {Cappi} M.,  {Reeves} J.~N.,  {Palumbo} G.~G.~C.,  {Braito} V.,
   {Dadina} M.,  2011, \mn@doi [\apj] {10.1088/0004-637X/742/1/44}, \href
  {https://ui.adsabs.harvard.edu/abs/2011ApJ...742...44T} {742, 44}

\bibitem[\protect\citeauthoryear{{Tombesi}, {Cappi}, {Reeves}  \&
  {Braito}}{{Tombesi} et~al.}{2012}]{Tombesi:12}
{Tombesi} F.,  {Cappi} M.,  {Reeves} J.~N.,   {Braito} V.,  2012, \mn@doi
  [\mnras] {10.1111/j.1745-3933.2012.01221.x}, \href
  {https://ui.adsabs.harvard.edu/abs/2012MNRAS.422L...1T} {422, L1}

\bibitem[\protect\citeauthoryear{{Tombesi}, {Cappi}, {Reeves}, {Nemmen},
  {Braito}, {Gaspari}  \& {Reynolds}}{{Tombesi} et~al.}{2013}]{Tombesi:13}
{Tombesi} F.,  {Cappi} M.,  {Reeves} J.~N.,  {Nemmen} R.~S.,  {Braito} V.,
  {Gaspari} M.,   {Reynolds} C.~S.,  2013, \mn@doi [\mnras]
  {10.1093/mnras/sts692}, \href
  {https://ui.adsabs.harvard.edu/abs/2013MNRAS.430.1102T} {430, 1102}

\bibitem[\protect\citeauthoryear{{Tombesi}, {Mel{\'e}ndez}, {Veilleux},
  {Reeves}, {Gonz{\'a}lez-Alfonso}  \& {Reynolds}}{{Tombesi}
  et~al.}{2015}]{Tombesi:15}
{Tombesi} F.,  {Mel{\'e}ndez} M.,  {Veilleux} S.,  {Reeves} J.~N.,
  {Gonz{\'a}lez-Alfonso} E.,   {Reynolds} C.~S.,  2015, \mn@doi [\nat]
  {10.1038/nature14261}, \href
  {https://ui.adsabs.harvard.edu/abs/2015Natur.519..436T} {519, 436}

\bibitem[\protect\citeauthoryear{{Veilleux} et~al.,}{{Veilleux}
  et~al.}{2013}]{Veilleux:13}
{Veilleux} S.,  et~al., 2013, \mn@doi [\apj] {10.1088/0004-637X/776/1/27},
  \href {https://ui.adsabs.harvard.edu/abs/2013ApJ...776...27V} {776, 27}

\bibitem[\protect\citeauthoryear{{Veilleux}, {Bolatto}, {Tombesi},
  {Mel{\'e}ndez}, {Sturm}, {Gonz{\'a}lez-Alfonso}, {Fischer}  \&
  {Rupke}}{{Veilleux} et~al.}{2017}]{Veilleux:17}
{Veilleux} S.,  {Bolatto} A.,  {Tombesi} F.,  {Mel{\'e}ndez} M.,  {Sturm} E.,
  {Gonz{\'a}lez-Alfonso} E.,  {Fischer} J.,   {Rupke} D.~S.~N.,  2017, \mn@doi
  [\apj] {10.3847/1538-4357/aa767d}, \href
  {https://ui.adsabs.harvard.edu/abs/2017ApJ...843...18V} {843, 18}

\bibitem[\protect\citeauthoryear{{Veilleux}, {Maiolino}, {Bolatto}  \&
  {Aalto}}{{Veilleux} et~al.}{2020}]{Veilleux:20}
{Veilleux} S.,  {Maiolino} R.,  {Bolatto} A.~D.,   {Aalto} S.,  2020, \mn@doi
  [\aapr] {10.1007/s00159-019-0121-9}, \href
  {https://ui.adsabs.harvard.edu/abs/2020A&ARv..28....2V} {28, 2}

\bibitem[\protect\citeauthoryear{{Vishniac}}{{Vishniac}}{1983}]{Vishniac:83}
{Vishniac} E.~T.,  1983, \mn@doi [\apj] {10.1086/161433}, \href
  {https://ui.adsabs.harvard.edu/abs/1983ApJ...274..152V} {274, 152}

\bibitem[\protect\citeauthoryear{{Vogelsberger}, {Genel}, {Sijacki}, {Torrey},
  {Springel}  \& {Hernquist}}{{Vogelsberger} et~al.}{2013}]{Vogelsberger:13}
{Vogelsberger} M.,  {Genel} S.,  {Sijacki} D.,  {Torrey} P.,  {Springel} V.,
  {Hernquist} L.,  2013, \mn@doi [\mnras] {10.1093/mnras/stt1789}, \href
  {https://ui.adsabs.harvard.edu/abs/2013MNRAS.436.3031V} {436, 3031}

\bibitem[\protect\citeauthoryear{{Volonteri}, {Dubois}, {Pichon}  \&
  {Devriendt}}{{Volonteri} et~al.}{2016}]{Volonteri:16}
{Volonteri} M.,  {Dubois} Y.,  {Pichon} C.,   {Devriendt} J.,  2016, \mn@doi
  [\mnras] {10.1093/mnras/stw1123}, \href
  {https://ui.adsabs.harvard.edu/abs/2016MNRAS.460.2979V} {460, 2979}

\bibitem[\protect\citeauthoryear{{Wagner}, {Umemura}  \& {Bicknell}}{{Wagner}
  et~al.}{2013}]{Wagner:13}
{Wagner} A.~Y.,  {Umemura} M.,   {Bicknell} G.~V.,  2013, \mn@doi [\apjl]
  {10.1088/2041-8205/763/1/L18}, \href
  {https://ui.adsabs.harvard.edu/abs/2013ApJ...763L..18W} {763, L18}

\bibitem[\protect\citeauthoryear{{Weaver}, {McCray}, {Castor}, {Shapiro}  \&
  {Moore}}{{Weaver} et~al.}{1977}]{Weaver:77}
{Weaver} R.,  {McCray} R.,  {Castor} J.,  {Shapiro} P.,   {Moore} R.,  1977,
  \mn@doi [\apj] {10.1086/155692}, \href
  {https://ui.adsabs.harvard.edu/abs/1977ApJ...218..377W} {218, 377}

\bibitem[\protect\citeauthoryear{{Weinberger}, {Ehlert}, {Pfrommer}, {Pakmor}
  \& {Springel}}{{Weinberger} et~al.}{2017}]{Weinberger:17}
{Weinberger} R.,  {Ehlert} K.,  {Pfrommer} C.,  {Pakmor} R.,   {Springel} V.,
  2017, \mn@doi [\mnras] {10.1093/mnras/stx1409}, \href
  {https://ui.adsabs.harvard.edu/abs/2017MNRAS.470.4530W} {470, 4530}

\bibitem[\protect\citeauthoryear{{Weinberger} et~al.,}{{Weinberger}
  et~al.}{2018}]{Weinberger:18}
{Weinberger} R.,  et~al., 2018, \mn@doi [\mnras] {10.1093/mnras/sty1733}, \href
  {https://ui.adsabs.harvard.edu/abs/2018MNRAS.479.4056W} {479, 4056}

\bibitem[\protect\citeauthoryear{{Weinberger}, {Springel}  \&
  {Pakmor}}{{Weinberger} et~al.}{2019}]{Weinberger:19}
{Weinberger} R.,  {Springel} V.,   {Pakmor} R.,  2019, \apjs \, (submitted),
  \href {https://ui.adsabs.harvard.edu/abs/2019arXiv190904667W} {p.
  arXiv:1909.04667}

\bibitem[\protect\citeauthoryear{{Wiersma}, {Schaye}  \& {Smith}}{{Wiersma}
  et~al.}{2009}]{Wiersma:09}
{Wiersma} R. P.~C.,  {Schaye} J.,   {Smith} B.~D.,  2009, \mn@doi [\mnras]
  {10.1111/j.1365-2966.2008.14191.x}, \href
  {https://ui.adsabs.harvard.edu/abs/2009MNRAS.393...99W} {393, 99}

\bibitem[\protect\citeauthoryear{{Wurster} \& {Thacker}}{{Wurster} \&
  {Thacker}}{2013}]{Wurster:13}
{Wurster} J.,  {Thacker} R.~J.,  2013, \mn@doi [\mnras] {10.1093/mnras/stt346},
  \href {https://ui.adsabs.harvard.edu/abs/2013MNRAS.431.2513W} {431, 2513}

\bibitem[\protect\citeauthoryear{{Wylezalek}, {Flores}, {Zakamska}, {Greene}
  \& {Riffel}}{{Wylezalek} et~al.}{2020}]{Wylezalek:20}
{Wylezalek} D.,  {Flores} A.~M.,  {Zakamska} N.~L.,  {Greene} J.~E.,   {Riffel}
  R.~A.,  2020, \mn@doi [\mnras] {10.1093/mnras/staa062}, \href
  {https://ui.adsabs.harvard.edu/abs/2020MNRAS.492.4680W} {492, 4680}

\bibitem[\protect\citeauthoryear{{Yang} \& {Reynolds}}{{Yang} \&
  {Reynolds}}{2016}]{Yang:16}
{Yang} H. Y.~K.,  {Reynolds} C.~S.,  2016, \mn@doi [\apj]
  {10.3847/0004-637X/818/2/181}, \href
  {https://ui.adsabs.harvard.edu/abs/2016ApJ...818..181Y} {818, 181}

\bibitem[\protect\citeauthoryear{{Yuan} \& {Narayan}}{{Yuan} \&
  {Narayan}}{2014}]{Yuan:14}
{Yuan} F.,  {Narayan} R.,  2014, \mn@doi [\araa]
  {10.1146/annurev-astro-082812-141003}, \href
  {https://ui.adsabs.harvard.edu/abs/2014ARA&A..52..529Y} {52, 529}

\bibitem[\protect\citeauthoryear{{Yuan}, {Bu}  \& {Wu}}{{Yuan}
  et~al.}{2012}]{Yuan:12}
{Yuan} F.,  {Bu} D.,   {Wu} M.,  2012, \mn@doi [\apj]
  {10.1088/0004-637X/761/2/130}, \href
  {https://ui.adsabs.harvard.edu/abs/2012ApJ...761..130Y} {761, 130}

\bibitem[\protect\citeauthoryear{{Yuan}, {Gan}, {Narayan}, {Sadowski}, {Bu}  \&
  {Bai}}{{Yuan} et~al.}{2015}]{Yuan:15}
{Yuan} F.,  {Gan} Z.,  {Narayan} R.,  {Sadowski} A.,  {Bu} D.,   {Bai} X.-N.,
  2015, \mn@doi [\apj] {10.1088/0004-637X/804/2/101}, \href
  {https://ui.adsabs.harvard.edu/abs/2015ApJ...804..101Y} {804, 101}

\bibitem[\protect\citeauthoryear{{Zakamska} et~al.,}{{Zakamska}
  et~al.}{2016}]{Zakamska:16}
{Zakamska} N.~L.,  et~al., 2016, \mn@doi [\mnras] {10.1093/mnras/stw718}, \href
  {https://ui.adsabs.harvard.edu/abs/2016MNRAS.459.3144Z} {459, 3144}

\bibitem[\protect\citeauthoryear{{Zhang}, {Churazov}  \&
  {Schekochihin}}{{Zhang} et~al.}{2018}]{Zhang:18}
{Zhang} C.,  {Churazov} E.,   {Schekochihin} A.~A.,  2018, \mn@doi [\mnras]
  {10.1093/mnras/sty1269}, \href
  {https://ui.adsabs.harvard.edu/abs/2018MNRAS.478.4785Z} {478, 4785}

\bibitem[\protect\citeauthoryear{{Zhuravleva} et~al.,}{{Zhuravleva}
  et~al.}{2014}]{Zhuravleva:14}
{Zhuravleva} I.,  et~al., 2014, \mn@doi [\nat] {10.1038/nature13830}, \href
  {https://ui.adsabs.harvard.edu/abs/2014Natur.515...85Z} {515, 85}

\bibitem[\protect\citeauthoryear{{Zinger} et~al.,}{{Zinger}
  et~al.}{2020}]{Zinger:20}
{Zinger} E.,  et~al., 2020, \mnras \, (submitted), \href
  {https://ui.adsabs.harvard.edu/abs/2020arXiv200406132Z} {p. arXiv:2004.06132}

\bibitem[\protect\citeauthoryear{{Zubovas} \& {King}}{{Zubovas} \&
  {King}}{2012a}]{Zubovas:12b}
{Zubovas} K.,  {King} A.~R.,  2012a, \mn@doi [\mnras]
  {10.1111/j.1365-2966.2012.21845.x}, \href
  {https://ui.adsabs.harvard.edu/abs/2012MNRAS.426.2751Z} {426, 2751}

\bibitem[\protect\citeauthoryear{{Zubovas} \& {King}}{{Zubovas} \&
  {King}}{2012b}]{Zubovas:12}
{Zubovas} K.,  {King} A.,  2012b, \mn@doi [\apjl]
  {10.1088/2041-8205/745/2/L34}, \href
  {http://adsabs.harvard.edu/abs/2012ApJ...745L..34Z} {745, L34}

\bibitem[\protect\citeauthoryear{{Zubovas} \& {King}}{{Zubovas} \&
  {King}}{2014}]{Zubovas:14}
{Zubovas} K.,  {King} A.~R.,  2014, \mn@doi [\mnras] {10.1093/mnras/stt2472},
  \href {https://ui.adsabs.harvard.edu/abs/2014MNRAS.439..400Z} {439, 400}

\bibitem[\protect\citeauthoryear{{Zubovas} \& {Nayakshin}}{{Zubovas} \&
  {Nayakshin}}{2012}]{Zubovas:12c}
{Zubovas} K.,  {Nayakshin} S.,  2012, \mn@doi [\mnras]
  {10.1111/j.1365-2966.2012.21250.x}, \href
  {https://ui.adsabs.harvard.edu/abs/2012MNRAS.424..666Z} {424, 666}

\bibitem[\protect\citeauthoryear{{Zubovas}, {Bourne}  \& {Nayakshin}}{{Zubovas}
  et~al.}{2016}]{Zubovas:16}
{Zubovas} K.,  {Bourne} M.~A.,   {Nayakshin} S.,  2016, \mn@doi [\mnras]
  {10.1093/mnras/stv2971}, \href
  {https://ui.adsabs.harvard.edu/abs/2016MNRAS.457..496Z} {457, 496}

\bibitem[\protect\citeauthoryear{{van de Voort}, {Schaye}, {Booth}  \& {Dalla
  Vecchia}}{{van de Voort} et~al.}{2011}]{vandeVoort:11}
{van de Voort} F.,  {Schaye} J.,  {Booth} C.~M.,   {Dalla Vecchia} C.,  2011,
  \mn@doi [\mnras] {10.1111/j.1365-2966.2011.18896.x}, \href
  {https://ui.adsabs.harvard.edu/abs/2011MNRAS.415.2782V} {415, 2782}

\bibitem[\protect\citeauthoryear{{van der Vlugt} \& {Costa}}{{van der Vlugt} \&
  {Costa}}{2019}]{vanderVlugt:19}
{van der Vlugt} D.,  {Costa} T.,  2019, \mn@doi [\mnras]
  {10.1093/mnras/stz2944}, \href
  {https://ui.adsabs.harvard.edu/abs/2019MNRAS.490.4918V} {490, 4918}

\makeatother
\end{thebibliography}


\bsp	
\label{lastpage}
\end{document}